\documentclass[12pt]{article}
\usepackage{amssymb,amsmath,amsbsy}
\usepackage{mathrsfs}
\usepackage{dsfont}
\usepackage{geometry}  
\usepackage{graphicx}
\usepackage{booktabs}
\usepackage{axodraw}
\usepackage{cite}

\def\theequation{\arabic{section}.\arabic{equation}}

\newenvironment{Eqnarray}%
     {\arraycolsep 0.24em\begin{eqnarray}}{\end{eqnarray}}
\def\beq{\begin{equation}}
\def\eeq{\end{equation}}
\def\slashchar#1{\setbox0=\hbox{$#1$}           
   \dimen0=\wd0                                 
   \setbox1=\hbox{/} \dimen1=\wd1               
   \ifdim\dimen0>\dimen1                        
      \rlap{\hbox to \dimen0{\hfil/\hfil}}      
      #1                                        
   \else                                        
      \rlap{\hbox to \dimen1{\hfil$#1$\hfil}}   
      /                                         
   \fi}                                        %
\newcommand{\bea}{\begin{Eqnarray}}
\newcommand{\pslash}{\slashchar{p}}
\newcommand{\eea}{\end{Eqnarray}}
\newcommand{\ra}{\rightarrow}
\newcommand{\Hc}{\mathrm{H.c.}}

\newcommand{\snr}{\tilde\nu_h}
\newcommand{\snl}{\tilde\nu_\ell}
\newcommand{\ynu}{Y_{\nu}}
\newcommand{\ynuh}{ Y_{\nu}}
\newcommand{\re}{\mathrm{Re}}
\newcommand{\im}{\mathrm{Im}}
\newcommand{\znu}{\mathcal{Z}_{\tilde\nu}}
\newcommand{\snu}{\tilde{\nu}}
\newcommand{\Gsnu}{\Gamma_{\tilde{\nu}}}
\newcommand{\snup}{\widetilde{\nu}^{(+)}}
\newcommand{\snum}{\widetilde{\nu}^{(-)}}
\newcommand{\snupm}{\widetilde{\nu}^{(\pm}}

\newcommand{\ket}[1]{|#1\rangle}
\newcommand{\braket}[2]{\langle #1|#2\rangle}
\newcommand{\wt}{\widetilde}
\newcommand{\wh}{\widehat}
\newcommand{\tr}{\rm tr}
\newcommand{\BR}{{\rm BR}}
\newcommand{\lsim}{\raisebox{-0.13cm}{~\shortstack{$<$\\[-0.07cm] $\sim$}}~}
\newcommand{\gsim}{\raisebox{-0.13cm}{~\shortstack{$>$\\[-0.07cm] $\sim$}}~}

\def\phm{\phantom{-}}
\def\phs{\phantom{*}}
\def\vev#1{\left\langle #1\right\rangle}
\def\ifmath#1{\relax\ifmmode #1\else $#1$\fi}
\def\half{\ifmath{{\textstyle{\frac{1}{2}}}}}
\def\ls#1{\ifmath{_{\lower1.5pt\hbox{$\scriptstyle #1$}}}}
\def\lsup#1{^{\lower 3pt\hbox{$\scriptstyle#1$}}}
\def\eq#1{eq.~(\ref{#1})}
\def\Eq#1{Eq.~(\ref{#1})}
\def\Eqs#1#2{Eqs.~(\ref{#1}) and (\ref{#2})}
\def\eqs#1#2{eqs.~(\ref{#1}) and (\ref{#2})}
\def\eqss#1#2#3{eqs.~(\ref{#1}), (\ref{#2}) and (\ref{#3})}
\def\eqst#1#2{eqs.~(\ref{#1})--(\ref{#2})}

\def\Ref#1{ref.~\cite{#1}}
\def\Rref#1{Ref.~\cite{#1}}
\def\Refs#1#2{refs.~\cite{#1} and \cite{#2}}
\def\fig#1{fig.~\ref{#1}}

\def\sect#1{Section \ref{#1}}

\begin{document}
\tolerance=100000

\setcounter{page}{0}
\thispagestyle{empty}

\begin{flushright}
IPPP-07-16\\[-1mm]
CPT-07-32\\[-1mm]
SCIPP-07/10\\[-1mm]
arXiv:0707.3718 [hep-ph]\\[-1mm]
\end{flushright}

\bigskip

\begin{center}
{\Large \bf Seesaw mechanism in the sneutrino sector }\\[0.3cm]
{\Large \bf and its consequences}\\[1.7cm]

{{\large Athanasios Dedes}$^{\, a,d}$, {\large Howard E.~Haber}$^{\, b}$
{\large and Janusz Rosiek}$^{\, a,c}$}\\[0.5cm]
{\it $^a$Institute for Particle Physics Phenomenology, University of
Durham, DH1 3LE, UK}\\[3mm]
{\it $^b$Santa Cruz Institute for Particle Physics, University of
California, Santa Cruz CA 95064}\\[3mm]
{\it $^c$Institute of Theoretical Physics, University of Warsaw, Ho{\.z}a
69, 00-681 Warsaw, Poland}\\[3mm]
{\it $^d$Division of Theoretical Physics, University of Ioannina, Ioannina,
GR 45110,  Greece}

\end{center}

\vspace*{0.8cm}\centerline{\bf ABSTRACT}
\vspace{0.5cm}
\noindent{\small
The seesaw-extended MSSM provides a framework in which the observed
light neutrino masses and mixing angles can be generated in the
context of a natural theory for the TeV-scale.  Sneutrino-mixing
phenomena provide valuable tools for connecting the physics of
neutrinos and supersymmetry.  We examine the theoretical structure of
the seesaw-extended MSSM, retaining the full complexity of three
generations of neutrinos and sneutrinos.  In this general framework,
new flavor-changing and CP-violating sneutrino processes are allowed,
and are parameterized in terms of two $3\times 3$ matrices that
respectively preserve and violate lepton number. The elements of these
matrices can be bounded by analyzing the rate for rare flavor-changing
decays of charged leptons and the one-loop contribution to neutrino
masses.  In the former case, new contributions arise in the seesaw
extended model which are not present in the ordinary MSSM.  In the
latter case, sneutrino--antisneutrino mixing generates the leading
correction at one-loop to neutrino masses, and could provide the
origin of the observed texture of the light neutrino mass matrix.
Finally, we derive general formulae for sneutrino--antisneutrino
oscillations and sneutrino flavor-oscillations.  Unfortunately,
neither oscillation phenomena is likely to be observable at future
colliders.}

\vspace*{\fill}
\newpage

\setcounter{equation}{0}
\section{Introduction}
\label{sec:intro}

The Standard Model of particle physics provides a remarkable
description of the fundamental interactions of elementary particles at
energy scales of order 100 GeV and below.  Precision tests at LEP, the
Tevatron and other lower energy colliders have detected no significant
deviations from the predictions of observed electroweak
phenomena~\cite{lepewwg}.  Although the scalar sector responsible for
electroweak symmetry breaking has not yet been discovered, the
precision electroweak data is consistent with the Standard Model
including a scalar Higgs boson of mass $114~{\rm GeV}<m_h<182~{\rm
GeV}$ at 95\% CL.  Despite its successes, the Standard Model is widely
acknowledged to be only a low-energy effective theory, to be
superseded (most likely at the TeV energy scale) by a more fundamental
theory that can explain the puzzling large hierarchy between the
energy scale that governs electroweak symmetry-breaking and the Planck
scale~\cite{hierarchy}.

Numerous proposals for a more fundamental theory that supersedes the
Standard Model have been advanced over the last thirty
years~\cite{bsm}.  Low-energy supersymmetric theories (in which
supersymmetry breaking effects of order the TeV scale are ultimately
responsible for electroweak symmetry breaking) are perhaps the most
well-studied framework for TeV-scale physics beyond the Standard Model
\cite{lowenergysusy,susybooks,haberkane}.   The simplest 
supersymmetric extension consists
of the particle content of the two-Higgs-doublet extension of the
Standard Model and its supersymmetric partners.   In addition to the
supersymmetric interactions of the particle supermultiplets, one adds
the most general set of soft-supersymmetry-breaking terms, which
parameterizes the unknown dynamics responsible for supersymmetry
breaking~\cite{Rosiek,pdghaber}.   The resulting minimal supersymmetric
Standard Model (MSSM) yields a rich phenomenology of new superpartners
and interactions, which if present in nature is poised for discovery
at the Tevatron and/or Large Hadron Collider (LHC).

Although no significant deviations from Standard Model predictions
have been observed at colliders, there is of course one definitive set
of observations that are in conflict with (the minimal version of) the
Standard Model---the observation of neutrino mixing and its
implications for neutrino masses~\cite{Gonzalez-Garcia:2007ib}.  Since
neutrinos are strictly massless in the Standard Model, the latter must
be modified in order to incorporate the observed phenomena of neutrino
oscillations.  The simplest approach is to introduce a gauge invariant
dimension-five operator~\cite{refdim5}%
\footnote{Following \Refs{Rosiek}{haberkane}, we
employ a convention where $\epsilon_{12}=-1=-\epsilon_{21}$.
\label{fnone}}
\beq
\label{dim5}
\mathscr{L}_{5}=-\frac{f_{IK}}{\Lambda}(\epsilon_{ij}L^I_i H_j)
(\epsilon_{k\ell}L^K_k H_\ell)+{\rm H.c.}\,,
\eeq
where $H_j$ is the complex Higgs doublet and $L_i^I\equiv
(\nu_L^I\,,\,\ell_L^I)$ is the SU(2)-doublet of two-component lepton
fields,\footnote{To translate the two-component spinor product $L_i^I
L_k^K$ into four-component spinor notation, see~\ref{app:fermion}.}
where $I$ and $K$ label the three generations.

After electroweak symmetry breaking, the neutral component of the
doublet Higgs field acquires a vacuum expectation value, and a
Majorana mass matrix for the neutrinos is generated.  The
dimension-five term [\eq{dim5}] is generated by new physics beyond the
Standard Model at the scale $\Lambda$.  Current bounds on light
neutrino masses suggest that $v^2/\Lambda\lsim 1~{\rm
eV}$~\cite{PDG,numass}, or $\Lambda\gsim 10^{13}~{\rm GeV}$.  A
possible realization of \eq{dim5} is based on the seesaw mechanism,
which was independently discovered by a number of different
authors~\cite{seesaw,numassmodel}.  In the seesaw extension of the
Standard Model~\cite{numassmodel}, one simply adds SU(2)$\times$U(1)
gauge singlet neutrino fields $\nu^{cI}_L$ and writes down the most
general renormalizable couplings of $\nu^{cI}_L$ to the Standard Model
fields:
\beq
\mathscr{L}_{\rm seesaw}= -\epsilon_{ij} Y_\nu^{IJ}H_i L^I_j
\nu_L^{c\,J}-\half M^{IJ}\nu_L^{c\,I}\nu_L^{c\,J}+{\rm H.c.}
\eeq
If $\|M\|\gg v$, then at energy scales below $M$ a dimension-five
operator of the form given by \eq{dim5} is generated.

The MSSM is a minimal extension of the Standard Model.  Nevertheless,
there is a potential source for lepton-number violation and hence
neutrino masses.  Unlike the Standard Model, it is possible to
construct renormalizable operators that violate lepton number and
baryon number~\cite{susylviolation}.  In their most generic forms,
such operators would lead to extremely fast proton decay in conflict
with the observations.  The traditional solution is to introduce a
discrete symmetry called R parity~\cite{rparity} that distinguishes
Standard Model particles and their superpartners.  In the
R-parity-conserving (RPC) MSSM, neutrinos are massless just as in the
Standard Model.  Thus, one way to incorporate massive neutrinos in the
RPC-MSSM is to formulate a minimal supersymmetric extension of the
seesaw-extended Standard Model~\cite{susyseesaw,Howie,Farzan,Chun,herrero}. 
An alternative approach is to choose a different discrete symmetry that
preserves baryon number but violates lepton number~\cite{z3}.  In such
an R-parity-violating (RPV) MSSM, a $\mathds{Z}_3$ baryon triality
guarantees that baryon number is conserved by the renormalizable
operators of the model (hence preventing fast proton decay).  This
approach has the advantage that no new fields beyond those of the MSSM
need to be introduced.  However, certain RPV (lepton-number-violating)
couplings must be taken to be quite small in order to explain the
scale of neutrino masses~\cite{Banks:1995by,rpv,Rimmer}.

In this paper, we shall consider the minimal supersymmetric extension
of the seesaw-extended Standard 
Model~\cite{susyseesaw,Howie,Farzan,Chun,herrero}.
In this model, neutrino masses and mixing are governed by the same
seesaw mechanism originally introduced into the (non-supersymmetric)
Standard Model.  In the supersymmetry-extended model, new
lepton-violating phenomena enter due to additional effective
lepton-violating operators generated by soft-supersymmetry-breaking.
Such effects govern the behavior of the neutrino superpartners---the
sneutrinos.  Thus, the supersymmetric seesaw model provides new
sources for lepton-number-violating phenomena.  For example,
sneutrinos and antisneutrinos can mix due to effective $\Delta L=2$
operators~\cite{Howie,snumix}.  Although such mixing effects are
expected to be quite small, there are some scenarios in which
sneutrino mixing phenomena could be observed in future collider
experiments~\cite{Howie,snumixpheno}.  Sneutrino mixing also
contributes a significant one-loop correction to neutrino masses and
could be partially responsible for the observed pattern of neutrino
masses and mixing~\cite{Howie,Rimmer,mixmass}.  The supersymmetric
seesaw can also introduce lepton-flavor-violation and CP-violating
effects due to the non-trivial flavor structure of the seesaw
interactions~\cite{Farzan,Chun,Farzan2}.  Such phenomena are
exhibited in the flavor oscillations of the charged 
sleptons~\cite{ArkaniHamed:1997km} and the sneutrinos, respectively.
Moreover, new one-loop
processes contribute to $\ell^{\,I}\to\ell^{\,J}\gamma$ and electric
dipole moments, and provide interesting constraints on the model
parameters.

In Section~\ref{sec:lagr}, we introduce the Lagrangian for the
three-generation supersymmetric seesaw model, focusing on the
interaction of the lepton and Higgs superfields.  
Our notation for fermion fields are described in
\ref{app:fermion}.  In
Section~\ref{sec:massmat}, we derive the mass matrices for neutrinos
and squared-mass matrices for the sneutrinos.  In the limit of $M\gg
v$, one can use perturbation theory to obtain accurate analytical
expressions for the diagonalization of the effective mass and
squared-mass matrices for the light and heavy neutral fermion and
scalar states, respectively.  
The origin of a non-decoupling contribution to sneutrino
masses noted in Section~\ref{sec:massmat} is provided 
in \ref{app:nondecoupling}.
In Section~\ref{sec:lc}, we examine the
constraints on the lepton-number conserving parameters of the model
due to the observed $g-2$ of the muon, the (unobserved) electric
dipole moment of the electron, and the unobserved radiative decays of
charged leptons.  In Section~\ref{sec:lv}, constraints on the
lepton-number violating parameters of the model are obtained based on
observed neutrino mass and mixing data.  The general theory and
phenomenology of sneutrino oscillations and mixing are addressed in
Section~\ref{sec:osc}.  Our conclusions are given in
Section~\ref{sec:conc}.  Although the neutrino are most easily treated
as two-component spinor fields, it is convenient to present the
Feynman rules of the model using four-component spinor notation.
In~\ref{app:fermion}, we demonstrate how to translate between
two-component and four-component spinor notation in the interaction
Lagrangian.  
The relevant Feynman rules needed for the computations of
this paper are listed in~\ref{app:feyrul}.  Finally, some order of
magnitude estimates for the contributions to one-loop neutrino masses
(relevant for the discussion of \sect{sec:self}) are provided
in~\ref{app:oneloopnu}.

\setcounter{equation}{0}
\section{Lagrangian and the scalar potential}
\label{sec:lagr}

In this section, we examine the terms of the Lagrangian that
contribute to the masses and the non-gauge interactions of the
neutrinos and sneutrinos.  That is, we focus on terms that involve the
charged leptons, neutrinos, charged sleptons, sneutrinos and the Higgs
fields.  The relevant superfields (denoted with hats above the
corresponding field symbol) are specified in Table~\ref{tab:fields}.
\renewcommand{\arraystretch}{1.3}
\setlength{\tabcolsep}{0.01in}
\begin{table}[htb]
\centering
\caption{}
\vskip6pt
\begin{tabular}{cccc}
&          &             & Fermionic \\[-5pt]
Superfield& \phantom{XX}hypercharge &
\phantom{xxxxx}Boson Fields\phantom{xxxxx}& Partners \\
\hline
$\wh L^I$& $-1$ &
$\wt L^I_j\,\equiv\,(\widetilde\nu^I_L\,,\,\widetilde \ell^{\,I}_L)$  \hfill&
            $(\nu^I_L\,,\,\ell^{\,I}_L)$  \\[5pt]
$\wh R^I$& $+2$ &
$\wt R^I\,\equiv\,(\wt \ell^{\,I}_R)^*\hphantom{(\nu,)}$ \hfill&
$\ell^{\,cI}_L$ \\[5pt]
$\wh N^I$& $\phm 0$ &
$\wt N^I\,\equiv\,(\wt\nu^I_R)^*\hphantom{(\nu,)}$
\hfill&  $\nu^{cI}_L$
\\[5pt]
$\wh H^1$ & $-1$ &
$H^1_j\equiv(H^1_1\,,\,H^{1}_2)$ \hfill & $(\wt H^{1}_1\,,\,\wt H^{1}_2)$
\\[5pt]
$\wh H^2$ & $+1$ &
$H^2_j\equiv(H^{2}_1\,,\,H^{2}_2)$ \hfill &
$(\wt H^{2}_1\,,\,\wt H^{2}_2)$
\\ \noalign{\vskip8pt}
\hline
\end{tabular}
\label{tab:fields}
\end{table}

The electric charge (in units of $e$) is given by $Q=T_3+Y/2$, where
$Y$ is the hypercharge specified above.  The index $j$ labels
components of the SU(2) doublets with $T_3=\pm 1/2$ for $j=1,2$
respectively (and $T_3=0$ for the SU(2) singlets).  The fermionic
partners can be viewed either as two-component fermion fields or the
left-handed projections of four-component fermion fields, as explained
in~\ref{app:fermion}.  The index $I=1,2,3$ labels three possible
generations of charged lepton and neutrino superfields.  The notation
for the scalar field components of the hypercharge-zero superfield is
motivated by the fact that in the lepton-number-conserving limit,
$\widehat R$ and $\widehat N$ possess the same lepton number (which is
opposite in sign to that of $\widehat L$).  Consequently, $\wt\nu_L$
and $\wt\nu_R$ possess identical lepton numbers [cf.  \eq{lepton}].

The most general (renormalizable) form of the superpotential involving
the lepton and Higgs superfields in the R-parity-conserving extended
MSSM is given by:
\beq \label{superpot}
W = \epsilon_{ij}(\mu \wh H^1_i \wh H^2_j - Y_\ell^{IJ} \wh H^1_i
\wh L^I_j \wh R^J +
Y_{\nu}^{IJ} \wh H^2_i \wh L^I_j \wh N^J) + \half M^{IJ} \wh N^I \wh N^J \,,
\eeq
where $Y_\ell$ and $Y_{\nu}$ are complex $3\times 3$ matrices, $M$ is
a complex symmetric $3\times 3$ matrix and $\mu$ is a complex
parameter.\footnote{With the convention for $\epsilon_{ij}$ as
specified in footnote \ref{fnone}, it is convenient to insert an extra
minus sign in front of $Y_\ell$ in \eq{superpot}.  This ensures that
in a basis where $Y_\ell$ is a real positive diagonal matrix, the
charged lepton masses are also positive.  Note that this convention
differs from the one adopted in \Ref{Rosiek}.}
In addition, there are soft-supersymmetry-breaking terms that involve
the scalar field components of the above superfields.  Before writing
these terms explicitly, it is convenient to perform field
redefinitions of the (charged and neutral) lepton superfields:
\beq \label{redefsf}
\wh L^I\to V_L^{IJ} \wh L^J\,,\qquad \wh R^I\to V_R^{IJ} \wh R^J\,,\qquad
\wh N^I\to V_N^{IJ} \wh N^J\,,
\eeq
where $V_L$, $V_R$ and $V_N$ are $3\times 3$ unitary matrices.  Note
that the kinetic energy terms (and the couplings of the lepton
superfields to the gauge fields) are invariant under the above unitary
transformations.  However, the coefficients of the terms of the
superpotential are modified:
\beq
Y_\ell\to V_L^T Y_\ell V_R\,,\qquad
Y_\nu\to V_L^T Y_\nu V_N\,,\qquad
M\to V_N^T M V_N\,.
\eeq
We shall choose $V_L$, $V_R$ and $V_N$ such that:
\bea
V_L^T Y_\ell V_R &=& {\rm diag}
(Y_e\,,\,Y_\mu\,,\,Y_\tau)\,,\label{svd}\\
V_N^T M V_N &=& {\rm diag}(M_1\,,\,M_2\,,\,M_3)\,,
\label{takagi}
\eea
where the elements of the two diagonal matrices above are real and
non-negative.  It is always possible to find unitary matrices $V_L$
and $V_R$ such that \eq{svd} is satisfied---this is the singular value
decomposition of an arbitrary complex matrix \cite{linalg}.  Likewise,
it is always possible to find a unitary matrix $V_N$ such that
\eq{takagi} holds---this is the Takagi-diagonalization of an arbitrary
complex symmetric matrix \cite{linalg,takagi,chkz}.  Thus, the
redefinition of the lepton superfields [\eq{redefsf}] implies that one
can assume from the beginning without loss of generality that $Y_\ell$
and $M$ are real non-negative diagonal matrices.\footnote{After
electroweak symmetry breaking, \eq{svd} corresponds to working in a
basis in which the charged lepton mass matrices are (real)
non-negative and diagonal.}  Note that the (transformed) $Y_\nu$ is in
general an arbitrary complex $3\times 3$ matrix.

We next introduce the most general set of R-parity-conserving
soft-supersymmetry (SUSY)-breaking terms (following the usual rules
of~\cite{gg}) involving the slepton, sneutrino and Higgs fields:
\bea
V_{\rm SOFT} &=& m_{H_1}^2 H^{1*}_i H^1_i
+ m_{H_2}^2 H^{2*}_i H^2_i
+(m_L^2)^{IJ} \wt L^{I*}_i \wt L^J_i
+(m_R^2)^{IJ} \wt R^{I*} \wt R^J
+(m_N^2)^{IJ} \wt N^{I*} \wt N^J
\nonumber \\ [5pt]
&&\hspace{-0.2in} -\left[(m_B^2)^{IJ}\wt N^I \wt N^J +
\epsilon_{ij}\left(m_{12}^2 H^1_i H^2_j + A_\ell^{IJ} H^1_i
\wt L^I_j \wt R^J + A_\nu^{IJ} H^2_i \wt L^I_j \wt N^J\right)
+ \mathrm{H.c.}\right], \label{lsoft}
\eea
where $m_L^2$, $m_R^2$ and $m_N^2$ are hermitian matrices, $m_B^2$ is
a complex symmetric matrix and $A_\ell$ and $A_\nu$ are complex
matrices.  In general, these $3\times 3$ matrices do not take a
simplified form in the basis defined by \eqs{svd}{takagi}.  The total
scalar potential is made up of three contributions: the $F$-terms,
which are derived from \eq{superpot}, the $D$-terms, which arise from
the gauge interactions, and and the soft SUSY-breaking terms, which
have been specified in \eq{lsoft}.  The total scalar potential is then
given by:
\beq \label{pottotal}
V=V_F+V_D+V_{\rm SOFT}\,, \qquad {\rm where} \qquad V_F\equiv
\sum_i \left|\frac{\partial W}{\partial\phi_i}\right|^2
\eeq
and the sum over $i$ is taken over all scalar components of the
corresponding superfields.

The Yukawa couplings of the leptons and the Higgs fields and the
corresponding fermion mass terms are derived from \eq{superpot} using
the well-known formula~\cite{haberkane,Rosiek}:
\beq \label{susyyuk}
-\mathscr{L}_{\rm mass}-\mathscr{L}_{\rm Yuk}=
\half\sum_{ij}\left[\frac{\partial^2 W[\phi]}{\partial\phi_i\partial\phi_j}
\psi_i\psi_j+\Hc\right]\,,
\eeq
where the $\psi_i$ are the two-component fermion field superpartners
of the corresponding $\phi_i$, and $W[\phi]$ is the superpotential
function with superfields replaced by their scalar components.  After
electroweak symmetry breaking, the neutral Higgs fields acquire vacuum
expectation values,\footnote{We define the overall phases of the
neutral Higgs fields, $H_1^1$ and $H_2^2$, such that the corresponding
vacuum expectation values $v_{1,2}/\sqrt{2}$ are real and positive.}
\beq
\vev{H^1_1} = \frac{v_1}{\sqrt{2}}\,,\qquad\qquad
\vev{H^2_2} = \frac{v_2}{\sqrt{2}}\,,
\eeq
where $v^2\equiv v_1^2+v_2^2=(246~\rm{GeV})^2$ and $\tan\beta\equiv
v_2/v_1$.  Inserting the Higgs field vacuum expectation values into
\eqs{pottotal}{susyyuk}, one can isolate the terms of the Lagrangian
that are quadratic in the scalar fields and fermion fields,
respectively.  These terms yield squared-mass matrices for the charged
sleptons and sneutrinos and mass matrices for the charged leptons and
neutrinos.  In the basis defined by \eq{svd}, the charged lepton mass
matrix is diagonal, with diagonal elements $m_{\ell^I}=v_1
Y_\ell^I/\sqrt{2}$.

In general, the diagonalization of these mass matrices cannot be
performed analytically, and one must resort to numerical techniques.
However, the large hierarchy between neutrino masses and charged
lepton masses strongly suggests that the parameters $M_I\gg v$, in
which case an analytic perturbative diagonalization permits one to
isolate the light (s)neutrino sector and integrate out the superheavy
(s)neutrino sector, whose particle masses are of order the $M_I$.
This procedure was carried out for the CP-conserving one-generation
model in \Ref{Howie}.  In Section~\ref{sec:massmat}, we shall
generalize this analysis to the most general (potentially
CP-violating) three-generation model.

First, we clarify the expected magnitudes of the parameters of the
model:
\begin{itemize}
\item[1.]
We assume that the Yukawa couplings $Y^{IJ}_\nu$ satisfy:%
\footnote{The Euclidean matrix norm is defined by
$\|A\|\equiv\left[\tr(A^\dagger A)\right]\lsup{1/2}=\left[\sum_{i,j}
|a_{ij}|^2\right]\lsup{1/2}$, for a matrix $A$ whose matrix elements
are given by $a_{ij}$.}
\beq \label{assume1}
\|Y_\nu\|\lsim\mathcal{O}(1)\,.
\eeq

\item[2.] The Majorana mass $M$ is much heavier than the electroweak
scale (seesaw mechanism~\cite{seesaw})
\bea \label{assume3}
 \|M\|\gg v \;.
\eea

\item[3.] Although $\mu$ is a supersymmetric parameter, we require it
to be of a similar order to the low-energy supersymmetry-breaking
scale, $M_{\rm SUSY}$~\cite{pol}:
\beq \label{assumemu}
\mu\sim M_{\rm SUSY}\,.
\eeq

\item[4.]  The non-singlet soft SUSY-breaking squared-masses are of
a similar order to the supersymmetry-breaking scale:
\bea \label{assume2}
\|m^2_L\| \sim \|m^2_R\| \sim M_{\rm SUSY}^2 \;.
\eea

\item[5.] The parameters $m^2_B$ and $A_{\nu}$ are unconnected to
electroweak symmetry breaking at tree-level.  However, 
these parameters generate a mass-splitting between sneutrinos and
antisneutrinos.  The latter contributes via loop
corrections to neutrino mass splittings, which are experimentally
constrained.  One expects that~\cite{bfarzan}:
\beq \label{assume4}
\|A_{\nu}\| \lsim M_{\rm SUSY}\,,\qquad\qquad 
\|m_B^2\|\lsim M_{\rm SUSY}\|M\|\,,
\eeq
although these parameters could conceivably be larger by as much as a
factor of $10^3$~\cite{Howie}.  Large $A_{\nu}$ also leads also to
large corrections to charged slepton masses.  Thus, to avoid unnatural
fine-tuning in order to prevent charged slepton masses from being
larger than about 1 TeV, one again expects that $A_\nu$ cannot be much
larger than the supersymmetry-breaking scale.
The impact of the one-loop effects of $m_B^2$ 
on charged lepton radiative decays and 
the Higgs mass parameters also yield constraints and imply
that the bound on $m_B^2$ given by \eq{assume4} cannot be
significantly relaxed.

\item[6.] The singlet soft SUSY-breaking parameter $m^2_N$ is also
unconnected to electroweak symmetry breaking at tree-level.  However,
the one-loop corrections to the Higgs mass parameters depend
quadratically on $m_N^2$, so to avoid unnatural
fine-tuning of the electroweak symmetry breaking scale, one
expects that $m_N^2$ cannot be much larger than $(1~{\rm TeV})^2$.
This expectation is confirmed in \ref{app:nondecoupling}, in which case
\bea \label{assume5}
\|m^2_N\|\lsim M_{\rm SUSY}^2 \;.
\eea
If significant fine-tuning of the electroweak scale is allowed
(as in the split-super\-symmetry \cite{split} approach), then the
constraints on $m_N^2$ are significantly relaxed. The one-loop effects of
$m_N^2$ on \textit{physical} observables are rather mild, even as 
$\|m^2_N\|$ approaches $\|M^2\|$.  For example, in \Ref{cao}, the
one-loop corrections to Higgs masses in the seesaw-extended MSSM are
found to be large and negative if $\|m^2_L\|\,,\,\|m^2_N\|\sim\|M^2\|$.
However, these corrections become negligible once these
soft-SUSY-breaking masses are taken somewhat below the seesaw scale.

Thus, we shall present results in this paper that allow
for the possibility that:
\bea \label{assume5p}
  \|m^2_N\|\sim \|M^2\| \;.
\eea
If \eq{assume5p} holds, then remnants of the heavy 
neutrino/sneutrino sector can survive in the effective theory of the
light sneutrinos.  The origin of this non-decoupling effect is
explored in \ref{app:nondecoupling}.
\end{itemize}

Although naturalness demands that the scale of low-energy
supersymmetry-breaking, $M_{\rm SUSY}$, should be (roughly) of
$\mathcal{O}(v)$, the absence of observed supersymmetric phenomena
(and a light CP-even Higgs boson) suggest that $M_{\rm SUSY}$ may be
somewhat larger, of order 1~TeV.
Nevertheless, in \eqst{assumemu}{assume5},
one could substitute $M_{\rm SUSY}$ with $v$; the results of this
paper are consistent with either choice.

\setcounter{equation}{0}
\section{The (s)neutrino (squared-)mass matrices}
\label{sec:massmat}

In this section, we examine in detail the neutrino mass matrix and the
sneutrino squared-mass matrix.  In a three-generation model, the
neutrino mass matrix is a $6\times 6$ complex symmetric matrix, which
can be written in block (partitioned) form in terms of $3\times 3$
matrix blocks.  The sneutrino squared-mass matrix is a $12\times 12$
hermitian matrix, which can be written in block (partitioned) form in
terms of $6\times 6$ matrix blocks.  Each of these $6\times 6$
matrices can be further partitioned in terms of $3\times 3$ matrix
blocks.  In order to accommodate the proliferation of matrices of
dimension 3, 6 and 12, we adopt a notational device that allows the
reader to instantly discern the dimension of a given matrix.  Thus, we
use a boldface capital letter ($\boldsymbol{M}$) to denote a $12\times
12$ matrix, a calligraphic letter ($\mathcal M$) to denote a $6\times
6$ matrix, and a Latin letter ($M$ or $m$) to denote a $3\times 3$
matrix.  Latin letters will also be used to denote (scalar) mass
parameters, with appropriate identifying subscript or superscript
labels to distinguish these from the $3\times 3$ matrices introduced
in Sections~\ref{sec:lagr} and~\ref{sec:massmat}.  Following the
conventions of Section~\ref{sec:lagr}, we shall employ subscript and
superscript upper case Latin indices $I$, $J$, $K$ as generation
labels that run from 1 to 3.  Lower case Latin indices $i$, $j$, $k$
are employed for other purposes, either as SU(2) gauge indices or as
labels representing the six light sneutrino mass eigenstates.  Other
subscripts appearing in this section will be used to distinguish among
different matrix quantities.

\subsection{The neutrino mass matrices}
\label{sec:numass}

Working in a basis where $M$ is a diagonal matrix [cf.~\eq{takagi}],
we begin by analyzing the neutrino mass matrix.  The resulting terms
quadratic in the neutrino fields are given in terms of two-component
fermion fields\footnote{In~\ref{app:fermion}, we show how to rewrite
\eq{eq:mterms} in terms of four-component neutrino fields.
However, the two-component formalism is more economical, so we adopt
this notation in what follows.} by:
\beq
-\mathscr{L}_{m_\nu} = \half\left(v_2\sqrt{2} \, Y_{\nu}^{IJ}
\nu_L^I \nu_L^{cJ} \ + \ M^{IJ} \nu_L^{cI} \nu_L^{cJ} + \Hc\right)
= \half\, (\nu_L^T \quad \nu_L^{cT})\,\mathcal{M}_\nu
\left(\begin{array}{c}\nu_L \\ \nu_L^c \\ \end{array}\right)+ \Hc
\label{eq:mterms}
\eeq
The neutrino mass matrix $\mathcal{M}_\nu$ is a $6\times 6$ complex
symmetric matrix given in block form by:
\beq \label{eq:vmass}
\mathcal{M}_\nu\equiv\left(\begin{array}{cc}
0 & \,\,\, m_D \\
m_D^T & \,\,\, M
\end{array}\right)\,,
\eeq
where the $3\times 3$ complex matrix
\beq \label{diracmm}
m_D\equiv v_2\ynuh/\sqrt{2}
\eeq
generalizes the neutrino Dirac mass term of the one-generation model
[cf.~\eq{fn4}].

Provided that $\|M\|\gg \|m_D\|$ [as suggested by \eq{assume3}],
$\mathcal{M}_\nu$ is of a seesaw type~\cite{seesaw}.  The neutrino
mass matrix can be Takagi block-diagonalized \cite{herrero,Rimmer,chkz} as
follows.  Introduce the $6\times 6$ (approximate) unitary matrix:
\beq \label{udef}
\mathcal{U}=\left(\begin{array}{cc} \mathds{1} -\half m_D^*M^{-2}m_D^T &
\quad m_D^*M^{-1} \\ -M^{-1}m_D^T & \quad \mathds{1} -\half
M^{-1}m_D^T m_D^* M^{-1}
\end{array}\right)\,,
\eeq
where $\mathds{1}$ is the $3\times 3$ identity matrix.

One can check that:
\beq
\mathcal{U}^\dagger \mathcal{U}=
\begin{pmatrix} \mathds{1}+\mathcal{O}(m_D^4 M^{-4}) & 0
\\ 0 & \mathds{1}+\mathcal{O}(m_D^4 M^{-4})\end{pmatrix}\,.
\eeq
We define transformed (light and heavy) neutrino states $\nu_\ell$ and
$\nu^c_h$ by:
\beq \label{nutransform}
\begin{pmatrix} \nu_L \\ \nu_L^c \end{pmatrix}=
\mathcal{U}\begin{pmatrix}\nu_\ell \\ \nu^c_h \end{pmatrix}\,.
\eeq
By straightforward matrix multiplication, one can verify that
\beq \label{nublockdiag}
\mathcal{U}^T\mathcal{M}_\nu \,\mathcal{U}=
\begin{pmatrix} -m_D M^{-1}m_D^T+\mathcal{O}(m_D^4 M^{-3}) & \mathcal{O}(m_D^3
M^{-2})\\
\mathcal{O}(m_D^3 M^{-2}) & M+\half(M^{-1}m_D^\dagger m_D+m_D^T m_D^* M^{-1})
+\mathcal{O}(m_D^4 M^{-3})\end{pmatrix}\!.
\eeq

At this stage, we can identify an effective (complex symmetric) mass
matrix $M_{\nu_\ell}$ for the three light (left-handed) neutrinos with
respect to the $\{\nu_\ell\}$-basis:
\beq
M_{\nu_\ell} \simeq  -m_D M^{-1} m_D^T \,.
\label{eq:mnu}
\eeq

To identify the physical light neutrino states, we must perform a
Takagi-diagonalization of $M_{\nu_\ell}$.  This is accomplished by
introducing the unitary MNS matrix\cite{PMNS}, $U_{\rm MNS}$,
via
\beq \label{MNS}
\nu_{\ell} ^I=U_{\rm MNS}^{IJ}\, (\nu_\ell^J)^{\rm phys}\,,
\eeq
where the $(\nu_\ell^J)^{\rm phys}$ [$J=1,2,3$] denote the physical
light neutrino fields.  $U_{\rm MNS}$ is determined by the
Takagi-diagonalization of $M_{\nu_\ell}$:
\beq
U^T_{\rm MNS} M_{\nu_\ell} U_{\rm MNS} = {\rm
diag}(m_{\nu_{\ell 1}}\,,\,m_{\nu_{\ell 2}}\,,\,m_{\nu_{\ell 3}})\,,
\label{eq:vmphys}
\eeq
where the $m_{\nu_{\ell J}}$ are the (real non-negative) masses of the
light neutrino mass eigenstates.

For completeness, we examine the effective mass matrix of the heavy
neutrino states.  Although $M$ is diagonal by assumption, the lower
right-handed block in \eq{nublockdiag} is no longer diagonal due to
the second-order perturbative correction.  However, we do not have to
perform another Takagi-diagonalization, since the off-diagonal
elements are of $\mathcal{O}(m_D^2M^{-1})$, and would only affect the
physical (diagonal) masses at order $\mathcal{O}(m_D^4 M^{-3})$, which
we neglect.  The corresponding mixing angles would be of
$\mathcal{O}(m_D^2 M^{-2})$, which we also neglect here.  Thus, we
identify the physical heavy neutrino mass eigenstates to leading order
by:
\beq
(\nu_h^{c\,I})^{\rm{phys}}\simeq\nu^{cI}_h\,,
\eeq
with masses
\beq \label{heavynumass}
m_{\nu_{hI}}=M_I\left(1+\frac{1}{M_I^2}\sum_J |m\ls{D}^{JI}|^2\right)\,,
\eeq
where the $M_I$ are the diagonal elements of $M$ in our chosen basis.

\subsection{The sneutrino squared-mass matrices}
\label{sec:snumass}

We now turn to the sneutrino sector.  It is convenient to separate out
various pieces that comprise the $F$-term contributions to the scalar
potential [\eq{pottotal}]:
\beq
V_F\equiv V_\nu+V_\mu+V_{\rm other}\,,
\eeq
where $V_\nu\equiv\sum_{i=\widetilde L_1^I\,,\,\widetilde N^I}|
\partial W/\partial\phi_i|^2$ and $V_\mu\equiv |\partial W/\partial
H_2^2|^2$ ultimately contribute to the sneutrino squared-mass matrix,
whereas $V_{\rm other}$ (which involves derivatives of the
superpotential with respect to the other scalar fields) makes no
contributions to tree-level sneutrino masses.

As a pedagogical exercise, we first analyze the supersymmetric limit.
Although super\-symmetry-breaking is required in the MSSM to generate
electroweak symmetry breaking, one often finds supersymmetric-like
relations between the fermion and sfermion sectors in the limit of
$v_1=v_2$ and $\mu=0$, \textit{i.e.} for $V_\mu=V_D=0$.  Thus, in the
following computation the supersymmetric limit corresponds to taking
the total scalar potential [\eq{pottotal}] to be $V=V_\nu$.  To
analyze the contributions of $V_\nu$ to sneutrino masses, we can
employ the following trick.  Focus on the following two terms of the
superpotential:
\beq
W_{\nu}\equiv Y_{\nu}^{IJ}\widehat H_2^2\widehat L^I_1 \widehat N^J+
\half M^{IJ}N^I N^J=\half
\left(\widehat L_1^T \quad \widehat N^T\right)
\left(\begin{array}{cc}
0 &  \widehat H_2^2 \ynuh\\
\widehat H_2^2\ynuh^T &  M \\
\end{array}\right)
\left(\begin{array}{c}
\widehat L_1 \\
\widehat N \\
\end{array}\right)\,.
\eeq
Consistent with \eq{nutransform}, we redefine the neutrino superfields
as follows:
\beq \label{nusftransform}
\begin{pmatrix} \widehat L_1 \\ \widehat N \end{pmatrix}=
\mathcal{U}\begin{pmatrix}\widehat L_{1\ell} \\ \widehat N_h
\end{pmatrix}\,,
\eeq
where the unitary matrix $\mathcal{U}$ is given by \eq{udef}.
Defining the matrix ${H}\equiv \widehat H_2^2 Y_\nu$, the effect of
\eq{nusftransform} is to transform $W_\nu$ into\footnote{Strictly
speaking, this is not a permissible transformation, since $W$ must be
holomorphic in the superfields, whereas \eq{wnupr} is a function of
both $\widehat H_2^2$ and $\widehat H_2^{2\,*}$.  However, since we
ultimately set $H_2^2=v_2/\sqrt{2}$ and only take derivatives of
$W_\nu$ with respect to $\wt L_{1\ell}$ and $\wt N_h$, the procedure
outlined here yields correct results.}
\beq \label{wnupr}
W_\nu\simeq\half({H}M^{-1}{H}^T)^{IJ}\widehat L_{1\ell}^{I}
\widehat
L_{1\ell}^{J}+\half\left[M^{IJ}+\half(M^{-1}{H}^\dagger
{H}+{H}^T {H}^*M^{-1})^{IJ}\right]
\widehat N_h^I\widehat N_h^J+\mathcal{O}({H}^4M^{-3})\,,
\eeq
where there is an implicit sum over $I$ and $J$.  In deriving
\eq{wnupr}, we have used the fact that $M^{IJ}$ is a non-negative
diagonal matrix.  Setting $H_2^2=v_2/\sqrt{2}$ and using
\eq{pottotal}, we can directly make use of \eq{wnupr} to isolate the
contributions to the sneutrino squared-mass matrix that arise from
$V_\nu$:
\beq
-\mathscr{L}_{\rm mass}=\widetilde L^\dagger_{1\ell}
M^2_{\ell^\dagger\ell}\widetilde L_{1\ell}+\widetilde N^\dagger_{h}
M^2_{h^\dagger h}\widetilde N_h\,,
\eeq
where the $3\times 3$ hermitian matrices $M^2_{\ell^\dagger\ell}$ and
$M^2_{h^\dagger h}$ are given by:
\bea
M^2_{\ell^\dagger\ell}&=&m_D^*M^{-1}m_D^\dagger
m_DM^{-1}m_D^T+\mathcal{O}(m_D^6 M^{-4})\,,\label{susysnu1}\\
M^2_{h^\dagger h}&=& M^2+m_D^\dagger m_D+\half(Mm_D^T
m_D^*M^{-1}+M^{-1}m_D^Tm_D^*M) +\mathcal{O}(m_D^4
M^{-2})\,.\label{susysnu2}
\eea
Moreover, the effective light and heavy neutrino mass matrices,
$M_{\nu_\ell}$ and $M_{\nu_h}$, can also be derived by inserting
\eq{wnupr} into \eq{susyyuk}.    As expected, the resulting neutrino
mass matrices are related in a supersymmetric way to the sneutrino
squared-mass matrices obtained in \eqs{susysnu1}{susysnu2}:
\beq \label{susylimit}
M^2_{\ell^\dagger\ell}=M_{\nu_\ell}^\dagger M_{\nu_\ell}\,,\qquad
M^2_{h^\dagger h}=M_{\nu_h}^\dagger M_{\nu_h}\,.
\eeq
In particular, in the supersymmetric limit,
\beq
U_{\rm MNS}^T\,M^2_{\ell^\dagger\ell}\,U^*_{\rm MNS}=
{\rm diag}~(m_{\nu_{\ell 1}}^2\,,\,m_{\nu_{\ell 2}}^2\,,
\,m_{\nu_{\ell 3}}^2)\,,
\eeq
which implies that the light neutrino and sneutrino masses coincide.

We now turn to the complete calculation of the sneutrino mass matrix.
Although one could perform the computation with respect to the basis
of sneutrino states defined by \eq{nusftransform}, this basis is not
especially convenient.  This is due to the fact that the effective
squared-mass matrix of the light sneutrinos is dominated by
supersymmetry-breaking effects.  In particular, the supersymmetric
contribution of $\mathcal{O}(m_D^4 M^{-2})$ [cf.~\eq{susysnu1}] is
completely negligible relative to the supersymmetry-breaking
contributions.  Thus, there is no advantage to performing in the
sneutrino sector the same change of basis used to isolate the
effective mass matrix of the light neutrinos.  Hence we will write the
$12\times 12$ hermitian sneutrino squared-mass matrix in block form
as:
\beq \label{snumass2}
-\mathcal{L}_{\rm mass}=\half \left(\begin{array}{cc} \phi_L^\dagger &
\phi_N^\dagger\end{array}\right)\left(\begin{array}{cc} \mathcal{M}^2_{LL} &
\mathcal{M}^2_{LN} \\ (\mathcal{M}_{LN}^{2})^\dagger &
\mathcal{M}^2_{NN}\end{array}\right)\left(\begin{array} {c}\phi_L \\
\phi_N\end{array}\right)\,,
\eeq
where $\phi_L\equiv (\widetilde L_{1}\,,\,\widetilde L^*_{1})^T$ and
$\phi_N\equiv (\widetilde N\,,\,\widetilde N^*)^T$ are six-dimensional
vectors.  The $6\times 6$ hermitian matrices $\mathcal{M}^2_{LL}$,
$\mathcal{M}^2_{NN}$ and the $6\times 6$ complex matrix
$\mathcal{M}^2_{LN}$ can be written in block partitioned form as:
\beq \label{m2jk}
\mathcal{M}^2_{AB}\equiv \left(\begin{array}{cc} M^2\ls{A^\dagger B} &
M^{2\,*}\ls{A^T B} \\ M^2\ls{A^T B} &
M^{2\,*}\ls{A^\dagger B}\end{array}\right)\,,
\eeq
where the subscripts $A$ and $B$ can take on possible values $L$ and
$N$ [this labeling allows one to keep track of the origin of the
various matrix blocks].  The $M^2_{A^\dagger A}$ are $3\times 3$
hermitian matrices and the $M^2_{A^T A}$ are $3\times 3$ complex
symmetric matrices, for $A=L\,,\,N$.  There are no restrictions on the
$3\times 3$ complex matrices $M^2_{A^\dagger B}$ and $M^2_{A^T B}$ for
$A\neq B$.

Adding up the contributions of $V_\nu$, $V_\mu$, $V_D$ and $V_{\rm
SOFT}$ to the sneutrino masses yields:
\bea
M^2_{L^\dagger L}&= &m_L^2+\half M_Z^2\cos 2\beta
+m_D^* m_D^T\,,\label{msnu1}\\
M^2_{N^\dagger N}&=& M^2+m_N^2+m_D^\dagger m_D\,,\label{msnu2}\\
M^2_{L^\dagger N}&=& m_D^*M\,,\label{msnu3}\\
M^2_{L^T N}&=&-X_\nu m_D\,,\label{msnu4} \\
M^2_{N^T N} &=& -2m_B^2\,,\label{msnu5}\\
M^2_{L^T L} &=&0 \,,\label{msnu6}
\eea
where we have introduced the complex $3\times 3$ matrix parameter
$X_\nu$ by the following definition:
\beq \label{xnudef}
X_\nu m_D\equiv\frac{1}{\sqrt{2}}\left(v_2 A_\nu+\mu^* v_1 Y_\nu\right)\,.
\eeq
A quick check of the supersymmetric limit confirms the expected
relation between the neutrino mass matrix and the sneutrino
squared-mass matrix:
\beq
\mathcal{M}_\nu^\dagger\mathcal{M}_\nu=
\left(\begin{array}{cc} m_D^* m_D^T & m_D^* M\\ M m_D^\dagger
& M^2+m_D^\dagger m_D \end{array}\right)\,.
\eeq
As noted above, because of the dominance of supersymmetry-breaking
contributions to the light sneutrino masses, the diagonalization of
the light neutrino mass matrix and the light sneutrino squared-mass
matrix are completely independent.

Under the assumptions of \eqst{assume1}{assume5}, the $12\times 12$
sneutrino mass matrix, written in terms of $6\times 6$ matrix blocks
with estimated magnitudes,
\bea
\boldsymbol{M}^2_{\tilde\nu}\equiv\left(\begin{array}{cc}
\mathcal{M}_{LL}^2 & \,\,\, \mathcal{M}_{LN}^2 \\
\left(\mathcal{M}_{LN}^2\right)^\dagger & \,\,\, \mathcal{M}_{NN}^2 \\
\end{array}\right) =
\left(\begin{array}{cc} \mathcal{O}(v^2) & \,\,\,\mathcal{O}(v M) \\
\mathcal{O}(v M) &  \,\,\,\mathcal{O}(M^2)\\
\end{array}\right)\,,
\eea
also exhibits a seesaw type behavior, analogous to the seesaw type
mass matrix [\eq{eq:vmass}] of the neutrino sector.  Following the
standard procedure for diagonalizing such matrices (see \Ref{Rimmer}),
we introduce a $12\times 12$ unitary matrix:
\beq \label{boldvee}
\boldsymbol{V}=\left(\begin{array}{cc} \mathcal{I}-\half \mathcal{M}_{LN}^2
\mathcal{M}_{NN}^{-4}(\mathcal{M}_{LN}^{2})^\dagger
& \mathcal{M}_{LN}^2 \mathcal{M}_{NN}^{-2} \\ -\mathcal{M}_{NN}^{-2}
(\mathcal{M}_{LN}^{2})^\dagger &
\mathcal{I}-\half  \mathcal{M}_{NN}^{-2} (\mathcal{M}_{LN}^{2})^\dagger
\mathcal{M}_{LN}^2 \mathcal{M}_{NN}^{-2}
\end{array}\right)\,,
\eeq
where $\mathcal{I}$ is the $6\times 6$ identity matrix.  One can
easily compute:
\beq
\boldsymbol{V}^\dagger\boldsymbol{M}^2_{\tilde\nu}\,\,\boldsymbol{V}=
\left(\begin{array}{cc} \mathcal{M}_{LL}^2 - \mathcal{M}_{LN}^2
\mathcal{M}_{NN}^{-2} (\mathcal{M}_{LN}^{2})^\dagger
+\mathcal{O}(v^4 M^{-2}) & \quad\mathcal{O}(v^3 M^{-1}) \\
\mathcal{O}(v^3 M^{-1}) & \quad \mathcal{M}_{NN}^{2}
+\mathcal{O}(v^2)\end{array}\right)\,.
\eeq
Hence, the effective $6\times 6$ hermitian squared-mass matrix for the
light sneutrinos reads:
\bea
\mathcal{M}_{\snl}^2\equiv \mathcal{M}_{LL}^2
- \mathcal{M}_{LN}^2 \mathcal{M}_{NN}^{-2}
\left(\mathcal{M}_{LN}^2\right)^\dagger+\mathcal{O}(v^4
M^{-2}) \;, \label{snueff}
\eea
analogous to the light effective neutrino mass matrix of \eq{eq:mnu}.
Likewise, the effective $6\times 6$ hermitian squared-mass matrix for
the superheavy sneutrinos reads:
\beq \label{snuheavy}
\mathcal{M}_{\snr}^2\equiv \mathcal{M}_{NN}^{2}
+\half\left[\mathcal{M}_{NN}^{-2}
(\mathcal{M}_{LN}^{2})^\dagger \mathcal{M}_{LN}^2
+ (\mathcal{M}_{LN}^{2})^\dagger \mathcal{M}_{LN}^2
\mathcal{M}_{NN}^{-2}\right]+\mathcal{O}(v^4 M^{-2})\,,
\eeq
where for completeness, we have exhibited the $\mathcal{O}(v^2)$
corrections to the leading term.  As expected, the masses of half of
the sneutrino eigenstates are of order the electroweak symmetry
breaking scale, whereas the other half are superheavy, of order $M$.

Following the notation of Table~\ref{tab:fields}, the (complex)
sneutrino interaction eigenstates are denoted by: $\wt\nu_L\equiv\wt
L_1$ and $\wt\nu_R\equiv \wt N^*$.  The latter convention reflects the
fact that in the lepton-number conserving limit of $M^{IJ}=m_B^2=0$,
the lepton numbers of $\wt\nu_L$ and $\wt\nu_R$ are identical, as
previously noted.  (Of course, the limit of interest in this paper,
$\|M\|\gg v$, is very far from the lepton-number conserving limit.)
In analogy to $\nu_\ell$ and $\nu_h$, we define transformed (light and
heavy) sneutrino states $\tilde\nu_\ell$ and $\tilde\nu_h$ by:
\beq \label{snutransform}
\begin{pmatrix} \phi_L \\ \phi_N \end{pmatrix}=
\mathcal{\boldsymbol{V}}\begin{pmatrix} \phi_\ell \\ \phi_h \end{pmatrix}\,,
\eeq
where $\phi_\ell\equiv (\wt\nu_\ell\,,\,\wt\nu_\ell^*)^T$ and
$\phi_h\equiv (\wt\nu_h^*\,,\,\wt\nu_h)^T$ are six-dimensional
vectors.  Sneutrino--antisneutrino oscillations are a consequence of
the $\Delta L=2$ elements in the light and heavy sneutrino
squared-mass matrices $\mathcal{M}^2_{\wt\nu_\ell}$ and
$\mathcal{M}^2_{\wt\nu_h}$, and are governed by $M^2_{N^T N}$ and
$M^2_{L^\dagger N}$ (note that $M^2_{L^T L}$, which would also violate
lepton number by two units, is zero).

Using the form of $\mathcal{M}^2_{AB}$ ($A$, $B=L$ or $N$) given by
\eq{m2jk} with the $M^2_{AB}$ given in \eqst{msnu1}{msnu6}, the
effective $6\times 6$ hermitian squared-mass matrix for the light
sneutrinos [\eq{snueff}] is given by:
\beq \label{eq:snlmass}
\mathcal{M}_{\snl}^2\equiv\left(\begin{array}{cc}
M^2_{LC} & (M^2_{LV})^* \\ M^2_{LV} &
(M^2_{LC})^*\end{array}\right)\,,
\eeq
where the lepton-number-conserving (LC) and lepton-number-violating
(LV) matrix elements are given by:
\bea \label{mlc}
\!\!\!\!\!\!\!\! M^2_{LC}&\equiv&m_L^2+\half M_Z^2\cos 2\beta +m_D^*
m_D^T-m_D^*M(M^2+m_N^2)^{-1} M m_D^T+\mathcal{O}(v^4 M^{-2})\,,
\\[6pt]
\!\!\!\!\!\!\!\!
M^2_{LV}&\equiv&m_D M (M^2+m_N^{2\,*})^{-1}m_D^T X_\nu^T +X_\nu
m_D(M^2+m_N^{2})^{-1}M m_D^T \nonumber\\ &&\qquad\qquad -2m_D
M(M^2+m_N^{2\,*})^{-1} m_B^2 (M^2+m_N^2)^{-1}Mm_D^T +\mathcal{O}(v^4
M^{-2})\,,
\label{mlv}
\eea
under the assumption that $m_B^2$ and $m_N^2$ can be as large
as indicated in \eqs{assume4}{assume5p}.  Note that
$M^2_{LC}$ is a $3\times 3$ hermitian matrix, and
$M^2_{LV}$ is a $3\times 3$ complex symmetric matrix.  Moreover,
although $M$ is a diagonal matrix with real positive entries
[cf.~\eq{takagi}], $m_N^2$ can be any $3 \times 3$ hermitian matrix,
not necessarily diagonal nor real.
The $M\to\infty$ limit of \eqs{mlc}{mlv} is noteworthy.  
In this limit, $M_{LV}^2=0$ and
the lepton-number-violating effects completely decouple, as expected.  If in
addition $m_N^2=0$, then $M_{LC}^2=m_L^2+\half M_Z^2\cos 2\beta$,
which reproduces the well known $3\times 3$ light
sneutrino squared-mass matrix of the MSSM.  However, according to
\eq{assume5}, $m_N^2 M^{-2}\sim\mathcal{O}(1)$ is possible, in which
case $M_{LC}^2$ deviates from its MSSM value
by a quantity of $\mathcal{O}(v^2)$ \textit{even in the exact
decoupling limit of $M\to\infty$}.  The origin of this
non-decoupling behavior is explained in \ref{app:nondecoupling}.
As a result of this non-decoupling phenomenon,
remnants of the heavy sector of the
seesaw mechanism may survive in the effective theory of light
sneutrinos.  These non-decoupling effects
can be detected in principle through measurements of the sneutrino and
charged slepton properties.

The physical light sneutrino states can be identified by diagonalizing
$\mathcal{M}_{\snl}^2$.  Note that if $M^2_{LV}=0$, then the
eigenvalues\footnote{Under the assumption that R-parity is not
spontaneously broken, the (real) eigenvalues of the hermitian matrix
$M_{LC}^2$ are non-negative.} of $\mathcal{M}_{\snl}^2$ are doubly
degenerate, corresponding to the fact that the conserved lepton number
implies that the six light sneutrino states are comprised of three
sneutrino antisneutrino pairs.  If $M^2_{LV}\neq 0$, then lepton
number is violated and the sneutrinos and antisneutrinos can mix.
This mixing splits the degenerate pairs and yields (in general) six
non-degenerate light sneutrinos.  In particular, the resulting
sneutrino mass-eigenstates are self-conjugate real fields, which we
denote by $S_1$\,,\,$S_2\,,\,\ldots\,,\,S_6$.

To determine the $S_k$ in terms of the interaction sneutrino
eigenstates, one must compute the $6\times 6$ unitary matrix
$\mathcal{W}$ that diagonalizes $\mathcal{M}^2_{\tilde\nu_\ell}$:
\beq \label{wdiag}
\mathcal{W}^{\,\dagger} \mathcal{M}_{\snl}^2 \mathcal{W}=
{\rm diag}~(m\ls{S_1}^2\,,\,m\ls{S_2}^2
\,,\,\ldots\,,\,m\ls{S_6}^2)\,.
\eeq
Noting that $\Sigma \mathcal{M}_{\snl}^2\Sigma= \mathcal{M}_{\snl}^{2\,*}$,
where $\Sigma\equiv\left(\begin{smallmatrix} 0 & \,\,\mathds{1}
\\ \mathds{1} & \,\, 0 \end{smallmatrix}\right)$, it follows that
if $\mathcal{W}$ satisfies \eq{wdiag} then so does $\Sigma
\mathcal{W}^{\,*}$.    However, the unitary matrix that diagonalizes
$\mathcal{M}_{\snl}^2$ is unique up to a multiplication on the right
by a unitary matrix $\mathcal{U}_D$ that is arbitrary within a
subspace of degenerate eigenvalues and is otherwise diagonal.  Denote
the set of all such unitary matrices by $\mathcal{S}$.  Hence, one can
conclude that $\Sigma \mathcal{W}^*=\mathcal{W}\mathcal{U}_D$ for some
$\mathcal{U}_D\in\mathcal{S}$.  Since $\mathcal{W}$ is unitary,
$\mathcal{U}_D=\mathcal{W}^{\,\dagger}\Sigma \mathcal{W}^{\,*}$, and
it follows that $\mathcal{U}_D \mathcal{U}_D^*=\mathcal{I}$.  That is,
$\mathcal{U}_D$ must be a symmetric unitary matrix.  It then follows
that the matrix $\mathcal{W}^{\,\prime}\equiv \mathcal{W}U_D^{1/2}$
satisfies $\mathcal{W}^{\,\prime}=\Sigma \mathcal{W}^{\,\prime\,*}$.%
\footnote{We define $\mathcal{U}_D^{1/2}\in\mathcal{S}$
to be the unique square root of $\mathcal{U}_D$ that is symmetric and
unitary.  This is accomplished by noting that there exists a (unique)
real symmetric matrix $\mathcal{H}$ such that $\mathcal{U}_D
=\exp(i\mathcal{H})$.  Then, $\mathcal{U}_D^{1/2}
\equiv\exp(i\mathcal{H}/2)$.  Note that there is still some freedom
left in the choice of $\mathcal{W}^\prime$, which is unique up to a
multiplication on the right by a real orthogonal matrix that is
arbitrary within a degenerate subspace and is otherwise diagonal.}

Thus, without loss of generality, we may drop the primed superscripts
and impose the constraint $\mathcal{W}=\Sigma\mathcal{W}^*$ on the
diagonalizing matrix that satisfies \eq{wdiag}.  It then follows that
$\mathcal{W}$ has the following form:
\beq \label{wdef}
\mathcal{W}\equiv\left(\begin{array}{cc} X^{\phs}& \phm iY^{\phs} \\ X^* &
    -iY^*\end{array}\right)\,,
\eeq
where $X$ and $Y$ are $3\times 3$ complex matrices that satisfy:
\bea \label{xxyy}
&& XX^\dagger+YY^\dagger=\mathds{1}\,,\qquad\qquad\quad\,
XX^T=YY^T\,,\\
&& \re(X^\dagger X)=\re(Y^\dagger Y)=\half\,,\qquad\quad \im(X^\dagger Y)=0\,,
\label{xxyy2}
\eea
due to the unitarity of $\mathcal{W}$.  Consequently, the relation
between the sneutrino interaction-eigenstate fields $\wt\nu^I_\ell$
and the six self-conjugate sneutrino mass-eigenstate fields $S_k$ is
given by:
\beq \label{essdef}
\wt\nu_\ell^I= \sum_{k=1}^6\,\mathcal{W}^{Ik}S_k=
\sum_{K=1}^3 \left(X^{IK}S_K + iY^{IK}\,S_{K+3}\right)\,,
\qquad
(I=1,2,3)\,.
\eeq
One can then invert \eq{essdef} [using \eqs{xxyy}{xxyy2}] to obtain:
\beq \label{essinvert}
S_K=\sum_{I=1}^3\left(
X^{IK\,*}\wt\nu_\ell^I+X^{IK}(\wt\nu_\ell^I)^*\right)\,,\quad
S_{K+3}= -i \sum_{I=1}^3 \left(Y^{IK\,*}\wt\nu_\ell^I - Y^{IK}
(\wt\nu_\ell^I)^*\right)\,,\quad (K=1,2,3)\,.
\eeq
Indeed, the $S_k$ are self-conjugate real fields as noted above.

Since $M_{LC}^2\sim\mathcal{O}(v^2)$ and
$M_{LV}^2\sim\mathcal{O}(v^3M^{-1})$, the mass-splittings of the
would-be sneutrino-antisneutrino pairs are expected to be very small,
of order a typical neutrino mass.  To compute the magnitude of the
corresponding mass-splittings, we can employ perturbative techniques
to evaluate the eigenvalues of
$\mathcal{M}^2_{\tilde\nu_\ell}$~[\eq{eq:snlmass}].  First, we diagonalize the
sub-matrix $M^2_{LC}$:
\beq
\label{eq:di_def}
Q_0^\dagger M^2_{LC} Q_0=D\equiv {\rm diag}(d_1\,,\,d_2\,,\,d_3)\,,
\eeq
where $Q_0$ is a $3\times 3$ unitary matrix, and the eigenvalues $d_I$
are real.  Note that $Q_0$ is not unique.  In \sect{sec:llg}, we will
argue that the bounds on the radiative flavor-changing charged lepton
decay $\ell^J\to\ell^I\gamma$ imply that matrix $M^2_{LC}$ is very
close to a diagonal form.  In the limit of diagonal $M^2_{LC}$, we
shall take $Q_0=\mathds{1}$.  We can then determine the off-diagonal
elements of $Q_0$ by writing $M^2_{LC}\simeq {\rm
diag}(m_1^2\,,\,m_2^2\,,\,m_3^2)+m^2_{LC}$, where $m^2_{LC}$ is a
matrix made up of the off-diagonal elements of $M^2_{LC}$, and
$Q_0\simeq \mathds{1}+q_0$, where $q_0^\dagger = -q_0$.  By
assumption, the matrix elements of $m^2_{LC}$ are much smaller than
the $m_I^2$, and the matrix elements of $q_0$ are much smaller than
unity.  Thus treating~\eq{eq:di_def} to first order in the small
quantities, we can solve for the off-diagonal elements of $q_0$ in
terms of the elements of $m^2_{LC}$ and the $m_I^2$.  Since at first
order $m_I^2=d_I$, it follows that:
\beq \label{qzero}
(Q_0)_{IJ}\simeq \frac{(M^2_{LC})_{IJ}}{d_J-d_I}\,,\qquad I\neq J\,.
\eeq
The diagonal elements of $Q_0$ can then be determined to the same
order by using the unitarity of $Q_0$.  In the remainder of this
section, we will not make any assumption regarding the size of the
off-diagonal elements of $M^2_{LC}$, in which case \eq{qzero} does not
apply and $Q_0$ must be obtained numerically from \eq{eq:di_def}.

In the following, it will be convenient to define
\bea
Q = Q_0 T
\label{eq:qdef}
\eea
where $T$ is a $3\times 3$ diagonal matrix of phases given by:
\beq
T\equiv {\rm diag} \left(e^{-i\phi_1/2}\, , \,e^{-i\phi_2/2}\, ,
\,e^{-i\phi_3/2} \right)\,,\qquad \phi_J\equiv\arg
\left(Q_0^T M^2_{LV} Q_0\right)_{JJ}\,.
\label{eq:tdef}
\eeq
Note that the right hand side of \eq{eq:di_def} is unchanged when
$Q_0\to Q_0 T$, so that the unitary matrix $Q$ can also be used to
diagonalize $M^2_{LC}$.  It then follows that:
\beq \label{curlydee}
\mathscr{D}\equiv \left(\begin{array}{cc} D & \,\, B^*\\ B & \,\,
D^{\phantom{*}}\end{array}\right)\,=\,
\left(\begin{array}{cc} Q^\dagger & \,\, 0\\ 0 & \,\, Q^T\end{array}\right)
\left(\begin{array}{cc} M^2_{LC} & (M^2_{LV})^* \\
M^2_{LV} & (M^2_{LC})^*\end{array}\right)
\left(\begin{array}{cc} Q& \,\, 0\\ 0 & \,\, Q^*\end{array}\right)\,,
\eeq
where $B$ is the $3\times 3$ complex symmetric matrix
\beq \label{beedef}
B\equiv Q^T M_{LV}^2 Q\,.
\eeq
Due to the rephasing of $Q_0$ as specified by \eqs{eq:qdef}{eq:tdef},
the diagonal elements of $B$ are real and non-negative: $B_{JJ} =
|B_{JJ}|$.  This is the motivation for our choice of $Q$ in the
diagonalization of $M^2_{LC}$.  Note that if $M_{LC}^2$ is
approximately diagonal, then $Q_0\simeq\mathds{1}$, in which case
$\phi_J\simeq\arg[(M^2_{LV})_{JJ}]$.  Thus, unless the diagonal
elements of $M^2_{LV}$ are non-negative, $Q\simeq T\neq\mathds{1}$ in
this limiting case.

Even though $D \sim\mathcal{O}(v^2)$ and
$B\sim\mathcal{O}(v^3M^{-1})$, the unitary matrix that diagonalizes
$\mathscr{D}$ is not close to the identity matrix, due to the double
degeneracy of the diagonal elements.  In order to perform a
perturbative diagonalization of $\mathscr{D}$, we first introduce the
following $6\times 6$ unitary matrix $\mathcal{P}$, expressed in block
form as:
\beq \label{peedef}
\mathcal{P}\equiv\frac{1}{\sqrt{2}}\left(\begin{array}{cc}
\mathds{1} & \phm i\mathds{1} \\
\mathds{1} & -i\mathds{1}\end{array}\right)\,,
\eeq
A straightforward computation yields:
\beq
\mathcal{P}^\dagger\mathscr{D}\,\mathcal{P} =\left(\begin{array}{cc}
D+\re\,B & -\im\,B \\ -\im\,B & D-\re\,B\end{array}\right)\,,
\eeq
which is a $6\times 6$ real symmetric matrix.

If the elements of the diagonal matrix $D$ are
non-degenerate\footnote{In general, we would expect the $d_I$ (which
are the eigenvalues of $M_{LC}^2$) to be non-degenerate.  Even if the
parameters $m_L^2$ and $m_N^2$ were proportional to the identity
matrix at the high energy scale due to some flavor symmetry, this
latter symmetry would not be respected by the corresponding low-energy
parameters, due to flavor-violating effects that enter the
renormalization group running.  Moreover, the matrix $m_D$ is likely
to reflect some of the flavor-violating effects of the model.  Hence,
any (near) degeneracy among the $d_I$ would be purely accidental.}
such that $d_I-d_J\sim\mathcal{O}(v^2)$ for all $I\neq J$, then the
matrix $\mathcal{P}^\dagger\mathscr{D}\,\mathcal{P}$ can be
diagonalized by a real orthogonal matrix $\mathcal{R}$ that is close
to the identity:
\beq
\label{aredef} \mathcal{R}=
\left(\begin{array}{cc} \mathds{1}+\re\,R & \im\,R \\ \im\, R &
\mathds{1}-\re\,R\end{array}\right)+\mathcal{O}(v^2 M^{-2})\,,
\eeq
where the $3\times 3$ complex antisymmetric matrix $R$ is of order
$\mathcal{O}(vM^{-1})$:
\beq
R_{IJ}=-R_{JI}\equiv \frac{B^*_{IJ}}{d_J-d_I}\,,\qquad (I\neq J)\,.
\label{eq:rdef}
\eeq
One can check that:
\beq
\mathcal{R}^T\mathcal{P}^\dagger\mathscr{D}\,\mathcal{P}\mathcal{R}
={\rm diag}(m^2_{S_1}\,,\,m^2_{S_2}\,,\,\ldots\,,\,,m^2_{S_6})
+\mathcal{O}(v^4 M^{-2})\,,
\eeq
where the squared-masses of the light sneutrinos are given by:
\beq \label{essmass}
m^2_{S_J\,,\,S_{J+3}}=d_J\pm |B_{JJ}|+\mathcal{O}(v^4 M^{-2})
\,,\qquad (J=1,2,3)\,,
\eeq
and $m^2_{S_J}>m^2_{S_{J+3}}$.  Note that the perturbations due to the
off-diagonal elements of $B$ contribute only to the $\mathcal{O}(v^4
M^{-2})$ terms of the squared-masses.

Combining the results of \eqss{curlydee}{peedef}{aredef}, the light
sneutrino mixing matrix [defined in \eq{wdiag}] is given by:
\beq
\mathcal{W}=\frac{1}{\sqrt{2}}\left(\begin{array}{cc}
\!\!\!Q(\mathds{1}+R) & \,\,\, \! iQ(\mathds{1}-R) \\
\!\!\!\!\!\phm Q^* (\mathds{1}+R^*) &\,\,\, -iQ^* (\mathds{1}-R^*)
\end{array}\right) +\mathcal{O}(v^2M^{-2})\,.  \label{eq:xy}
\eeq
Comparing with \eq{wdef}, we identify:
\beq
X=\frac{1}{\sqrt{2}}Q(\mathds{1}+R) +\mathcal{O}(v^2M^{-2})\,,\quad
{\rm and} \quad
Y=\frac{1}{\sqrt{2}}Q(\mathds{1}-R)+\mathcal{O}(v^2M^{-2})\,.
\label{eq:xyqr}
\eeq
Inserting these results into \eqs{essdef}{essinvert} yields the
desired (approximate) relations between the sneutrino mass eigenstates
$S_k$ and the interaction eigenstates $\wt\nu_\ell^I$.

For completeness, we briefly examine the modifications to \eq{essmass}
if some of the $d_I$ are degenerate.  In this case, the diagonalizing
matrix $\mathcal{R}$ is not close to the identity matrix, and the
perturbative analysis above fails.  Consider the case of $d_I=d_J\neq
d_K$, where $\{I,J,K\}$ is some permutation of $\{1,2,3\}$.  The first
order shift in the eigenvalues of $\mathscr{D}$ will depend on
$B_{IJ}$ as well as on the diagonal elements of $B$.  However, the
perturbations due to $B_{IK}$ and $B_{JK}$ will only generate
second-order shifts to the eigenvalues, which we neglect here.  Thus,
it is sufficient to solve the characteristic equation of $\mathscr{D}$
in the limit of $d_I=d_J$ and $B_{IK}=B_{JK}=0$.  In this limit, the
characteristic polynomial factors into a product of two simpler
polynomial factors:\footnote{In the case of a near degeneracy where
$d_I-d_J\lsim\mathcal{O}(vM^{-1})$, the quartic polynomial factor of
the characteristic equation of $\mathscr{D}$ contains a term linear in
$\lambda-\half(d_I+d_J)$.  In this case, the resulting expressions for
$m^2_{S_I\,,\,S_{I+3}}$ and $m^2_{S_J\,,\,S_{J+3}}$ are significantly
more complicated than those presented in \eqs{degen1}{degen2}.}
\beq
\Bigl[(\lambda-d_K)^2-|B_{KK}|^2\Bigr]
\Bigl[(\lambda-d_I)^4-(\lambda-d_I)^2\left[
|B_{II}|^2+|B_{JJ}|^2+2|B_{IJ}|^2\right] 
+\bigl|B_{IJ}^2-B_{II}B_{JJ}\bigr|^2
\Bigr]\,.
\eeq
The resulting sneutrino squared-masses are:
\bea
m^2_{S_I\,,\,S_{I+3}}&\simeq& d_I\pm
\Bigl\{\half\left[|B_{II}|^2+|B_{JJ}|^2+2|B_{IJ}|^2+
\sqrt{\Delta}\,\right]\Bigr\}^{1/2}\,,\label{degen1} \\
m^2_{S_J\,,\,S_{J+3}}&\simeq& d_I\pm
\Bigl\{\half\left[|B_{II}|^2+|B_{JJ}|^2+2|B_{IJ}|^2-
\sqrt{\Delta}\,\right]\Bigr\}^{1/2}\,,\label{degen2} \\
m^2_{S_K\,,\,S_{K+3}}&\simeq & d_k\pm|B_{KK}|\,,\label{degen3}
\eea
where
\beq
\Delta\equiv \Bigl[|B_{II}|^2+|B_{JJ}|^2+2|B_{IJ}|^2\Bigr]^{2}
-4\,\bigl|B_{IJ}^2-B_{II}B_{JJ}\bigr|^{2}\,.
\eeq
The corresponding mixing matrix can be obtained by performing an exact
diagonalization within the two-dimensional degenerate subspace,
although we shall omit the details.

Finally, in the very unlikely scenario where $d_1=d_2=d_3\equiv d$,
all of the matrix elements of $B$ contribute to the first order shifts
of the eigenvalues of $\mathscr{D}$.  To determine these shifts, put
$\lambda=d+x$ in the characteristic equation of $\mathscr{D}$ to
obtain a sixth order polynomial in $x$.  No further perturbative
simplification is possible, since all the terms of this polynomial are
of the same order of magnitude.

As expected, the mass-splittings of the would-be
sneutrino--antisneutrino pairs are nonzero due to the presence of the
lepton-number violating matrix $M^2_{LV}$ [cf \eq{beedef}].  If we
denote the three sneutrino mass-splittings by $(\Delta
m_{\tilde\nu_\ell}) \ls{J} \equiv |m_{S_{J}}-m_{S_{J+3}}|$ (for
$J=1,2,3$), then in the non-degenerate case,
\beq \label{deltamnu}
(\Delta m_{\tilde\nu_\ell})\ls{J}\simeq \frac{|B_{JJ}|}{\sqrt{d_J}}\,.
\eeq
In the case of degenerate $d_I$, the mass-splittings $(\Delta
m_{\tilde\nu_\ell})\ls{J}$ also depend on the non-diagonal elements
of~$B$.

It is instructive to examine the above results in a simplified one
generation model.  In this case, $D\equiv M_{LC}^2$ and $B\equiv
M_{LV}^2$ are just numbers.  In particular, $m_N^2$ is a real
parameter and $\mathcal{M}_{\snl}^2$ is a $2\times 2$ hermitian
matrix, with eigenvalues
\bea
m_{S_1,S_2}^2 &=& M_{LC}^2\pm |M_{LV}^2| \nonumber \\
&=& m_L^2+\half M_Z^2\cos 2\beta+\frac{|m_D|^2 m_N^2}{M^2+m_N^2}\pm
\frac{2|m_D|^2 M}{M^2+m_N^2}\left|X_\nu-\frac{Mm_B^2}{M^2+m_N^2}\right|\,.
\eea
The corresponding sneutrino mass-splitting, $\Delta
m_{\tilde\nu_\ell}\equiv |m_{S_2}-m_{S_1}|$, is given by
\beq \label{eq:dmnu}
\frac{\Delta m_{\tilde\nu_\ell}}{m_{\nu_\ell}}=
\frac{2M^2}{m_{\tilde\nu_\ell}(M^2+m_N^2)}
\left|X_\nu-\frac{Mm_B^2}{M^2+m_N^2}\right|\,,
\eeq
where $m_{\nu_\ell}\equiv |m_D|^2/M$ is the mass of the light neutrino
and $m_{\tilde\nu_\ell}\equiv\half(m_{S_1}+m_{S_2})$ is the average
light sneutrino mass.  If $m_N\ll M$, then \eq{eq:dmnu} coincides with
the result given in \Ref{Howie} after taking into account a slight
difference in notation.\footnote{If we put $m_B^2\equiv -MB_N$ and
change the sign of $A_\nu$ (with the corresponding change in $X_\nu$
[cf.~\eq{xnudef}]), we recover the results of \Ref{Howie}.}

Assuming that $m_B^2\sim\mathcal{O}(vM)$, it follows that
both terms on the right hand side of \eq{eq:dmnu} are of the same
order, which implies that $\Delta m_{\tilde\nu_\ell}
\sim\mathcal{O}(m_{\nu_\ell})$.  However, as noted below \eq{assume4},
it is possible that $m_B^2$ could be as much as a factor of $10^3$ larger
than its naive estimate \cite{Howie}, in which case the sneutrino-antisneutrino
mass splitting could be three orders of magnitude larger than the
corresponding light neutrino mass.\footnote{A similarly
enhanced sneutrino-antisneutrino mass splitting also 
arises in the supersymmetric triplet seesaw model of \Ref{rossi}.}

The same set of manipulations described above can be carried out to
obtain the corresponding results for the effective $6\times 6$
hermitian squared-mass matrix for the heavy sneutrinos
[\eq{snuheavy}]:
\beq \label{heavysnumatrix}
\mathcal{M}_{\snr}^2\equiv \left(\begin{array}{cc} M^2_{H} & -2(m_B^2)^*
 \\ -2m_B^2 & \phm (M^2_{H})^*\end{array}\right)+\mathcal{O}(v^4 M^{-2})\,,
\eeq
where the $3\times 3$ hermitian matrix $M_H^2$ is defined by:
\beq \label{m2Hdef}
M^2_{H}\equiv M^2+m_N^2+m_D^\dagger m_D+\half(M^2+m_N^2)^{-1}Mm_D^T
m_D^* M +\half Mm_D^T m_D^* M(M^2+m_N^2)^{-1}\,.
\eeq
The physical heavy sneutrino mass-eigenstates are determined by
diagonalizing $\mathcal{M}^2_{\tilde\nu_h}$.  At leading order, the
mass-eigenstates are mass-degenerate sneutrino/antisneutrino pairs,
with masses and mixing angles (with respect to the basis in which $M$
is diagonal) determined by the diagonalization of $m_N^2$.  The
lepton-number violating off-block-diagonal matrix $m_B^2$ generates
sneutrino-antisneutrino mixing, and yields mass-splittings between
nearly degenerate heavy sneutrino pairs of order $\Delta m_{\tilde
\nu_h}\sim\mathcal{O}(m_B^2 M^{-1})$.

The complex elements of the sneutrino squared-mass matrix govern
CP-violating sneutrino phenomena, due to the non-degeneracy of masses
of the real and imaginary parts of the sneutrino fields.  Following
the discussion of the CP-properties of the sneutrino fields in
Section~\ref{sec:osc}, we find it convenient to define a new basis of
sneutrino interaction eigenstates of definite CP.  That is, we
decompose the complex sneutrino fields into real and imaginary parts:
\bea
\snl &=& \frac{1}{\sqrt{2}}\left[\snl^{(+)} + i\; \snl^{(-)}\right] \;, \\
\snr &=& \frac{1}{\sqrt{2}}\left[\snr^{(+)} + i\; \snr^{(-)}\right] \;,
\eea
where the $[+,-]$ superscripts indicate that the corresponding
sneutrino eigenstates are CP-even and CP-odd.  With respect to the
CP-basis,
\beq
-\mathscr{L}_{\rm mass}=\half(\snl^{(+)T},\,
\snl^{(-)T})\mathcal{P}^\dagger \mathcal{M}^2_{\tilde\nu_\ell}\mathcal{P}
 \left(\begin{array}{c} \snl^{(+)} \\ \snl^{(-)} \end{array}\right)+
 \half(\snr^{(+)T},\,  \snr^{(-)T})\mathcal{P}^T
\mathcal{M}^2_{\tilde\nu_h}\mathcal{P}^*
 \left(\begin{array}{c} \snr^{(+)} \\ \snr^{(-)} \end{array}\right)\,,
\eeq
where $\mathcal{P}$ is the $6\times 6$ unitary matrix introduced in
eq.~(\ref{peedef}).

That is, with respect to the CP-basis, the effective squared-mass
matrix for the light sneutrinos is given by:
\beq \label{cpmass}
\overline{\mathcal{M}}^{\,2}_{\tilde\nu_\ell}\equiv
\mathcal{P}^\dagger\mathcal{M}^2_{\tilde\nu_\ell}\mathcal{P}=
\left(\begin{array}{cc} \re(M_{LC}^2+M_{LV}^2) & \quad
-\im(M_{LC}^2+M_{LV}^2) \\
\im(M_{LC}^2-M_{LV}^2) & \quad \phm\re(M_{LC}^2-M_{LV}^2)\end{array}\right)\,.
\eeq
This is a \textit{real} symmetric matrix (which is easily checked by
recalling that $M^2_{LC}$ and $M^2_{LV}$ are, respectively, hermitian
and complex symmetric matrices), as the CP-basis consists of real
self-conjugate scalar fields.

If $\im M_{LC}^2=\im M_{LV}^2=0$, then the sneutrino mass-eigenstates
are also definite eigenstates of CP.  If in addition $\re M_{LV}^2\neq
0$, then the would-be sneutrino-antisneutrino pairs are organized into
CP-even/CP-odd pairs of nearly degenerate sneutrinos~\cite{Howie}.

Since $\overline{\mathcal{M}}^{\,2}_{\tilde\nu_\ell}$ is real
symmetric, it can be diagonalized by a $6\times 6$ real orthogonal
matrix, $\znu$ via:
\bea
\znu^T\, \overline{\mathcal{M}}^{\,2}_{\tilde\nu_\ell}
\znu = (m_{S_1}^2\,,\,m_{S_2}^2\,,\,\ldots\,,\,m_{S_6}^2) \,,
\label{eq:znudef}
\eea
and the corresponding physical sneutrino mass eigenstates, $S_k$
($k=1,\ldots,6$), can be identified as linear combinations of the
CP-even and the CP-odd sneutrino eigenstates:
\bea
\left(\begin{array}{c}
\snup_\ell\\
\snum_\ell
\end{array}\right)
=
\znu
\left(\begin{array}{c}
S_1\\
\vdots\\
S_6
\end{array}\right) \;.
\eea
Matching with the notation employed by our discussion of sneutrino
oscillations in Section~\ref{sec:osc}, we note that the sneutrino
interaction eigenstates, 
$\snu_\ell$, can be expressed in terms of the physical
(self-conjugate) sneutrino mass eigenstates $S_k$ via:
\bea
\snl^I = \frac{1}{\sqrt{2}}\sum_{k=1}^6
(\znu^{Ik} + i\znu^{I+3,k})S_k \;.   \label{rot1}
\eea
Comparing \eqs{essdef}{rot1}, we can identify:
\beq \label{xy}
X^{IK}=\frac{1}{\sqrt{2}}\left(\znu^{IK}+i\znu^{I+3,K}\right)\,,\qquad
Y^{IK}=-\frac{i}{\sqrt{2}}\left(\znu^{I,K+3}+i\znu^{I+3,K+3}\right)\,,\quad
(I,K=1,2,3)\,,
\eeq
which can be inverted to obtain:
\beq
\znu=\sqrt{2}\begin{pmatrix} \re~X & \,\,\,-\im~Y\\ \im~X & \,\,\,
\phm\re~Y\end{pmatrix}\,.
\label{eq:zxy}
\eeq
One can easily verify that the orthogonality of $\znu$ implies the
unitarity of $\mathcal{W}$ defined in \eq{wdef} [and vice versa].  In
particular, \eqs{wdiag}{eq:znudef} imply that
$\znu=\mathcal{P}^\dagger\mathcal{W}$, in which case
\beq \label{zzww}
\znu^T \znu=\mathcal{W}^T\mathcal{P}^* \mathcal{P}^\dagger
\mathcal{W}=\mathcal{W}^T
\binom{0\quad\mathds{1}}{\mathds{1}\quad 0}
\mathcal{W}= \mathcal{W}^\dagger \mathcal{W}=\mathcal{I}\,,
\eeq
after using the explicit forms for $\mathcal{W}$ and $\mathcal{P}$.

In summary, we have derived the light effective sneutrino squared-mass
matrix by exploiting the seesaw mechanism in the sneutrino as well as
in the neutrino sector.  Our calculation is quite general under the
parameter assumptions specified by \eqst{assume1}{assume5}.  We found
that $\mathcal{M}_{\snl}^2$ depends on two $3\times 3$ matrix blocks,
$M^2_{LC}$ and $M^2_{LV}$, given by \eqs{mlc}{mlv}, respectively.  In
particular, $M^2_{LV}$ is responsible for the splitting of the masses
of would-be sneutrino-antisneutrino pairs, or equivalently the
mass-splitting of CP-even/CP-odd sneutrino pairs, $\snu^{(\pm)}_\ell$,
in the CP-conserving limit.  As we shall see in Sections~\ref{sec:lc}
and \ref{sec:lv}, the matrices $M^2_{LC}$ and $M^2_{LV}$ provide a
convenient parameterization for a number of interesting physical
observables, such as neutrino masses and radiative lepton decays.

\setcounter{equation}{0}
\section{Constraints on lepton number conserving parameters}
\label{sec:lc}

The input parameters that govern sneutrino mixing phenomena and
sneutrino decays are encoded in matrices $M_{LV}^{2}$ and $M_{LC}^{2}$
given by \eqs{mlv}{mlc}, respectively [or, alternatively, in the
physical sneutrino masses and the orthogonal matrix $\znu$ defined
in~\eq{eq:znudef}].  At present, apart from neutrino oscillations,
only lepton number conserving processes are observed in current
experiments.  These processes constrain the entries of the lepton
number conserving matrix $M_{LC}^2$.  In this Section we investigate
bounds on the structure of $M_{LC}^2$ imposed by the measurements of
the muon magnetic moment anomaly, the $g_\mu-2$, the electric dipole
moment (EDM) of the electron and the radiative flavor changing charged
lepton decays, $\ell^{\,J}\ra \ell^{\,I}\gamma$.  The latter have
also been worked out in detail in \Ref{herrero}.  Additional constraints
due to $\ell_J^-\to\ell_I^-\ell_I^-\ell_I^+$ decays and $\mu$--$e$
conversion in nuclei are also relevant and have been analyzed in
\Rref{herrero,herrero2}.  These constraints can yield further
restrictions on the structure of $M_{LC}^2$, although we
shall not present this analysis here.

We briefly summarize the constraints from current experiments
relevant for the computations presented in this Section.  The
most recent experimental measurement of the muon anomalous magnetic
moment ($a_\mu^{\rm exp}$) exhibits a slight discrepancy \cite{expg2}
relative to the predicted value of the Standard Model ($a_\mu^{\rm
th}$).  A recent theoretical review of the computation of the Standard
Model prediction\cite{g2sigma} yielded $\delta a_{\mu} \equiv
a_\mu^{\rm exp}-a_\mu^{\rm th}=(2.94\pm 0.89)\times 10^{-9}$, where
all theoretical and experimental errors are added in quadrature,
corresponding to a $3.3\,\sigma$ effect.  Thus, we roughly expect that
the contribution to the muon anomalous magnetic moment from new
physics beyond the Standard Model to be no larger than $\delta a_{\mu}
\lesssim 3 \times 10^{-9}$.  There is no experimental evidence of an
nonzero EDM for the electron ($d_e$).  The most stringent upper bound,
obtained in \Ref{Regan:2002ta}, is $d_e\leq 1.6\times10^{-27}$~e~cm at
90\% CL.  Likewise, there is no experimental evidence for radiative
flavor-changing charged lepton decays.  The 90\% CL upper limits to
the branching ratios for the muon and tau-lepton radiative decays are
given by: $\BR(\mu \to e \gamma) \le 1.2 \times 10^{-11}$, $\BR(\tau
\to e \gamma) \le 1.1\times 10^{-7}$ and $\BR(\tau
\to \mu \gamma) \le 6.8 \times 10^{-8}$\cite{PDG}.

\subsection{Supersymmetric corrections to the lepton-photon vertex}
\label{sec:vertex}

\begin{figure}[t!]
\begin{center}
\begin{picture}(400,160)(0,0)
\ArrowLine(60,50)(120,50)
\ArrowLine(10,50)(60,50)
\ArrowLine(120,50)(170,50)
\Text(20,45)[t]{$\ell^{\,J}$}
\Text(160,45)[t]{$\ell^{\,I}$}
\Text(90,45)[t]{$f$}
\Text(58,90)[t]{$S$}
\Text(122,90)[t]{$S$}
\Text(43,70)[t]{$-k$}
\Text(142,70)[t]{$p-k$}
\Text(75,140)[t]{$\gamma^{\mu}$}
\Text(50,45)[t]{\bf 1}
\Text(130,45)[t]{\bf 2}
\Text(100,120)[t]{\bf 3}
\DashArrowLine(90,110)(50,50){4}
\DashArrowLine(130,50)(90,110){4}
\Photon(90,110)(90,145){4}{3}
\Text(20,135)[t]{$\mbox{\boldmath$(a)$}$}
\Text(90,25)[t]{\bf 1: $i(a^J P_L + b^J P_R)$}
\Text(90,10)[t]{\bf 2: $i(b^{I*} P_L + a^{I*} P_R)$}
\Text(90,-5)[t]{\bf 3: $-ieq_S(p-2k)^{\mu}$}
\DashArrowLine(250,50)(330,50){4}
\ArrowLine(210,50)(250,50)
\ArrowLine(330,50)(370,50)
\Text(220,45)[t]{$\ell^{\,J}$}
\Text(360,45)[t]{$\ell^{\,I}$}
\Text(290,45)[t]{$S$}
\Text(258,90)[t]{$f^C$}
\Text(322,90)[t]{$f^C$}
\Text(275,140)[t]{$\gamma^{\mu}$}
\Text(250,45)[t]{\bf 1}
\Text(330,45)[t]{\bf 2}
\Text(300,120)[t]{\bf 3}
\ArrowLine(250,50)(290,110)
\ArrowLine(290,110)(330,50)
\Photon(290,110)(290,145){4}{3}
\Text(220,135)[t]{$\mbox{\boldmath$(b)$}$}
\Text(290,25)[t]{\bf 1: $i(a^J P_L + b^J P_R)$}
\Text(290,10)[t]{\bf 2: $i(b^{I*} P_L + a^{I*} P_R)$}
\Text(290,-5)[t]{\bf 3: $-ieq_f\gamma^{\mu}$}
\end{picture}
\end{center}
\caption{One-loop SUSY diagrams contributing to radiative, $\ell^{\,J}\ra
\ell^{\,I}\gamma$, decays.  In (a), the scalar $S$ is a charged
slepton and the fermion $f$ is a neutralino.  In (b), the scalar $S$
is a sneutrino and the fermion $f$ [$f^C$] is a positively
[negatively] charged chargino ($q_f=1$).}
\label{fig:tridiags}
\end{figure}
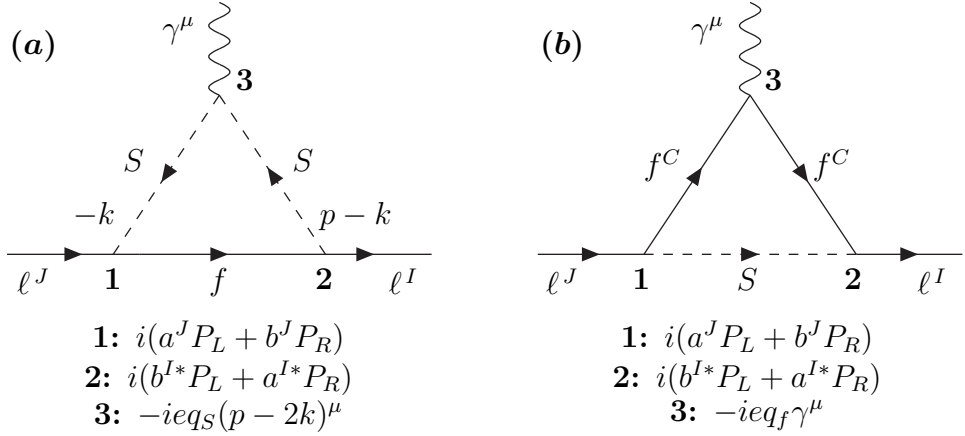

The amplitudes for the processes of interest are obtained by
evaluating triangle diagrams that contribute to the one-loop
correction to the lepton-photon $\ell^{\,J} \ell^{\,I} \gamma$ vertex.
Supersymmetric corrections to this vertex arise from the two
topologies of diagrams depicted in~\fig{fig:tridiags}.
The corresponding Feynman rules
required for the vertices are given in \eqs{B3}{B4}
of~\ref{app:feyrul}.
The anomalous magnetic moment and electric dipole moment (EDM)
of the leptons and the lepton flavor violating decays $\ell^{\,J}\ra
\ell^{\,I}\gamma$ are derived from the following terms of an effective
Hamiltonian:
\bea
\mathscr{H} \ = \ e \left ( C_{L}^{IJ} \:
\bar{\ell}^{\,I} \sigma^{\mu\nu} P_{L} \ell^{\,J}  \ + \ C_{R}^{IJ}\:
\bar{\ell}^{\,I} \sigma^{\mu\nu} P_{R} \ell^{\,J} \right ) \,
 F_{\mu \nu} \,,
\eea
which can be extracted from the computation of the effective one-loop
$\ell^{\,I} \ell^{\,J}\gamma$ vertex.

The computation of the Wilson coefficients $C_L,C_R$ is
straightforward.  After calculating the contributions of diagrams (a)
and (b) of~\fig{fig:tridiags} and expanding in momenta of external
particles, we find for their total Wilson coefficients
\bea
C_L^{iIJ} &=& C_{1L}^{iIJ} + m_{\ell^I} C_{4L}^{iIJ} + m_{\ell^J}
C_{4R}^{iIJ} \;, \nonumber\\[2mm]
C_R^{iIJ} &=& C_{1R}^{iIJ} + m_{\ell^I} C_{4R}^{iIJ} + m_{\ell^J}
C_{4L}^{iIJ} \;,
\label{clcr}
\eea
where the index $i$ labels the contribution of diagrams $i=a, b$ and
the $m_{\ell^I}$ ($I=1,2,3$) are the lepton masses.  For diagram (a)
we obtain,
\bea
C_{1L}^{aIJ} &=& \frac{1}{2(4\pi)^2} q_S b^{I*} a^J m_f
C_{12}(m_S,m_f) \;,
\hskip 5mm
%
C_{1R}^{aIJ} = \frac{1}{2(4\pi)^2} q_S a^{I*} b^J m_f C_{12}(m_S,m_f) \;,
\nonumber\\[2mm]
C_{4L}^{aIJ} &=& \frac{1}{2(4\pi)^2} q_S a^{I*} a^J C_{23}(m_S,m_f)\;,
\hskip 9.5mm
C_{4R}^{aIJ} = \frac{1}{2(4\pi)^2} q_S b^{I*} b^J C_{23}(m_S,m_f)\;,
\label{Ca}
\eea
and for the diagram (b),
\bea
C_{1L}^{bIJ} &=& \frac{1}{(4\pi)^2} q_f b^{I*} a^J m_f
C_{11}(m_f,m_S)\;,
\hskip 5mm
C_{1R}^{bIJ} = \frac{1}{(4\pi)^2} q_f a^{I*} b^J m_f
C_{11}(m_f,m_S)\;, \nonumber\\[2mm]
C_{4L}^{bIJ} &=& \frac{1}{2(4\pi)^2} q_f a^{I*} a^J C_{23}(m_f,m_S)\;,
\hskip 9.5mm
C_{4R}^{bIJ} = \frac{1}{2(4\pi)^2} q_f b^{I*} b^J C_{23}(m_f,m_S)\,,
\label{Cb}
\eea
where $m_f$ and $m_S$ are the masses of the fermion $f$ and scalar
$S$, respectively, and all other parameters are defined
in~\fig{fig:tridiags}.  The loop integrals appearing in~\eqs{Ca}{Cb}
are:
\bea
C_{11}(x,y) &=& - \frac{x^2 - 3 y^2}{4 (x^2 - y^2)^2} +
\frac{y^4}{(x^2 - y^2)^3}\log\frac{y}{x} \;, \nonumber\\
C_{12}(x,y) &=& - \frac{x^2 + y^2}{2 (x^2 - y^2)^2} - \frac{2 x^2 y^2}
{(x^2 - y^2)^3}\log\frac{y}{x}\;, \nonumber\\
C_{23}(x,y) &=& - \frac{x^4 - 5 x^2 y^2 - 2y^4}{12 (x^2 - y^2)^3} +
\frac{x^2 y^4}{(x^2 - y^2)^4}\log\frac{y}{x} \;.
\eea
The full Wilson coefficients $C_L$ and $C_R$ are obtained by summing
over all relevant triangle diagrams in the model.  In our case just
two of them contribute: diagram~(a) with charged
slepton and neutralino exchange and diagram~(b) with
sneutrino and chargino exchange.

\subsection{$\boldsymbol{(g-2)_\mu}$ and the electron EDM}

The formalism described above leads easily to expressions for the EDM
of the electron and for the muon magnetic moment anomaly
$(g_\mu-2)/2$.  For both processes $I=J$, so that the flavor-diagonal
piece of the effective Hamiltonian is given by
\bea
\mathscr{H} &=& e \: \bar \ell^{\,J}
\sigma_{\mu\nu} \biggl [ \re C_{1L}^{JJ} +
m_{\ell^J}(C_{4L}^{JJ} + C_{4R}^{JJ}) - i\im C_{1L}^{JJ}\gamma_5 \biggr ]
\ell^{\,J} F^{\mu\nu} \label{mom} \;,
\eea
where we used the relation $C_{1R}^{JJ} = C_{1L}^{JJ*}$.  By matching
to the standard form \cite{jdj,Feng:2001sq}:\footnote{In \eq{mdmedm},
the unit of electric charge $e$ is taken positive, so that the
electron charge is $-e$ (which also coincides with the convention
adopted by~\Refs{jdj}{Feng:2001sq}).  \Eq{mdmedm} is consistent with
the corresponding effective Lagrangian of \Ref{jdj}, by noting that
Commins \textit{et al.} define the anomalous magnetic moment of the
electron to be $\kappa=-a_e$ (J.D.~Jackson, private communication).}
\bea \label{mdmedm}
\mathscr{H} &=& -\frac{e}{4 m_{l^J}} a_J\bar \ell^{\,J}\sigma_{\mu\nu}
\ell^{\,J} F^{\mu\nu} + \frac{id_{\ell^{J}}}{2}\bar \ell^{\,J}
\sigma_{\mu\nu} \gamma_5 \ell^{\,J} F^{\mu\nu}\;,
\eea
where $a_J \equiv (g_J-2)/2$ is the magnetic moment anomaly and
$d_{\ell^J}$ is the EDM of the lepton, one can extract the expressions
for the electron EDM, $d_e$, and for $g_{\mu}-2$,
\bea
d_e &=& -2 e \: \im C_{1L}^{11}\;, \label{eedm} \\
a_\mu &=& -4 m_{\mu} \left [ \re C_{1L}^{22} \ + \ m_{\mu}(C_{4L}^{22} +
C_{4R}^{22})\right ] \;.
\eea

In principle, both quantities can be used to set bounds on parameters
such as $M$, $m_N^2$, $m_B^2$ and $X_\nu$ that govern the heavy
sneutrino sector.  However, the one-loop contribution to the
$C_{1L}^{11}$ from \fig{fig:tridiags}(b), which is sensitive to the
sneutrino sector, is real if the chargino parameters $\mu$ and
$M_2$ are real.  Hence, the electron EDM measurement does not yield
any constraints on sneutrino parameters at one loop.  However, there
can be sensitivity due to potentially large two-loop corrections; for
further details see Ref.~\cite{Farzan2}.  Similarly, the neutrino
magnetic and/or electric dipole moments\footnote{Note that for
Majorana particles only transition dipole moments can be nonzero.}
are also insensitive to the heavy sneutrino sector at one-loop, since
there is no possibility of attaching the photon to a one-loop graph
that involves the sneutrino-neutrino-neutralino vertex
(see~\ref{app:feyrul}).

The amplitudes displayed in \fig{fig:tridiags} can give sizable
contributions to the anomalous magnetic moment of the muon.  These
contributions are flavor diagonal and are sensitive mostly to the
overall mass scale of the sleptons, gauginos and light
sneutrinos---i.e.  to the diagonal entries of corresponding mass
matrices.  Thus, the measurement of $a_\mu$ can be used to set lower
bound on these SUSY masses.  Assuming that the discrepancy
between the experimentally observed
muon anomalous magnetic moment and the theoretical prediction of the
Standard Model, $\delta a_{\mu} \lesssim 3\times 10^{-9}$,
is due to new physics effects arising from the diagrams
of \fig{fig:tridiags}, one can deduce lower bounds on the magnitude of
slepton squared-mass parameter as a function of $M_2$ and $\tan\beta$.
Examples of such bounds are listed in Table~\ref{tab:g2}.
\begin{table}[tbp]
\begin{center}
\begin{tabular}{cp{1cm}cp{1cm}cp{1cm}c}
\hline
&& $M_2=100$ && $M_2=200$ && $M_2=300$\\
\hline
$\tan\beta$ && $(m_L)^{min}$  && $(m_L)^{min}$ && $(m_L)^{min}$ \\
\hline
5 && 170 && 110 && 70 \\
10 && 300 && 270 && 210 \\
15 && 420 && 420 && 370 \\
20 && 530 && 570 && 530 \\
25 && 650 && 740 && 700 \\
\hline
\end{tabular}
\end{center}
\caption{Lower bounds on the square root of $(m_L^2)_{22}$
from the measurement of $a_\mu$.  All masses are in GeV.}
\label{tab:g2}
\end{table}

Note that potential contributions to $M_{LC}^2$ [cf.~(\ref{mlc})] from
the terms containing the Dirac mass $m_D$ are suppressed by a quantity
of $\mathcal{O}(m_N^2 M^{-2})$.  As we will show in
Section~\ref{sec:llg}, this ratio can be at most of the order of
$10^{-2}$, otherwise the Dirac mass term $m_D$ would generate
unacceptably large contributions to rare $\ell^{\,J}\ra
\ell^{\,I}\gamma$ decays.   Thus, the muon anomalous magnetic moment
can be effectively used to set a lower bound on the diagonal $22$
element of the soft slepton squared-mass matrix $m_L^2$ and on the
gaugino mass parameter $M_2$, as specified in Table~\ref{tab:g2}.  The
dependence on $m_R^2$ and $\mu$ is significantly weaker.

\subsection{Radiative charged lepton decay: 
$\boldsymbol{\ell^{\,J}\ra \ell^{\,I}\gamma}$ }
\label{sec:llg}

The $\ell^{\,J} \rightarrow \ell^{\,I} \gamma$ decay width is given by
\bea
\Gamma(\ell^{\,J} \rightarrow  \ell^{\,I} \gamma) = \frac{e^2 m_{l^J}^3}{4\pi}
\left(|C_L^{IJ}|^2 + |C_R^{IJ}|^2\right) \;.   \label{eq58}
\eea
The corresponding branching ratio is obtained by dividing the result
of \eq{eq58} by the tree level decay width, $\Gamma(\ell^{\,J}
\rightarrow \ell^{\,I} \nu^J \bar{\nu}^I) = m_{\ell^J}^5 G_F^2/192
\pi^3$ (where we ignore $W$-propagator effects and a very small
correction due to the nonzero mass of the light final state charged
lepton).  In particular, the branching ratios for the experimentally
interesting decays $\mu \rightarrow e \gamma$ and $\tau
\rightarrow \mu \gamma$ are given by:
\bea
\BR(\mu\rightarrow e\gamma) = \frac{48\pi^2e^2}{m_{\mu}^2 G_F^2}
\left(|C_L^{12}|^2 + |C_R^{12}|^2\right) \;,
\label{BR1}
\eea
and
\bea
\BR(\tau \rightarrow \mu \gamma) = \frac{48\pi^2e^2}{m_{\tau}^2 G_F^2}
\left(|C_L^{23}|^2 + |C_R^{23}|^2\right) \label{BR2} \;.
\eea

At leading one-loop order, \fig{fig:tridiags}(a) yields an amplitude
that is proportional to the off-diagonal terms of the slepton soft
mass matrix $m_L^2$, and thus not relevant for setting bounds on heavy
sneutrino parameters\footnote{Of course this diagram is relevant when
$\ynu$-dependent corrections to $m_L^2$ entries are generated by the
renormalization group evolution of parameters.  This effect has been
studied extensively in the literature (see e.g., \Ref{Casas}), and we
will not repeat this discussion here.}.  The amplitude corresponding
to~\fig{fig:tridiags}(b) depends directly on the lepton flavor
conserving part of the light sneutrino mass matrix, $M_{LC}^2$.  This
can be verified by using the Feynman rules collected in
the~\ref{app:feyrul} and employing the mass insertion approximation
(MIA) expansion; for more details see e.g.~\Ref{MIPORO}.  Assume (at
least formally) that sneutrinos are closely degenerate in mass,
\bea
m_{S_k}^2 = m_0^2 + \delta m_{S_k}^2 \;,
\eea
and then expand the functions $C_L^{IJ}$ or $C_R^{IJ}$ [denoted
generically in \eq{taylor} by $f$], which depend on the squared-massed
$m_{S_k}^2$, up to the first order.  This results in
\bea
\!\!\!\!\!\!\!\!
f(m_{S_k}^2) \approx f(m_0^2) + (m_{S_k}^2 -m_0^2 )\left.\frac{\partial
f}{\partial m_{S_k}^2}\right|_{m_0^2} = f(m_0^2) -m_0^2\left.\frac{\partial
f}{\partial m_{S_k}^2}\right|_{m_{0}^2} + m_{S_k}^2 \left.\frac{\partial
f}{\partial m_{S_k}^2}\right|_{m_0^2},  \label{taylor}
\eea
where there is an implicit sum over $k$.  The advantage of this
procedure is that it allows one to perform the sum over the sneutrino
flavor index $k$ in evaluating \eqs{BR1}{BR2}.  For example, the
neutrino squared-masses always appear multiplied by a pair of
sneutrino mixing matrices (due to the form of the sneutrino couplings
given in~\ref{app:feyrul}).  Using the inverse of~\eq{eq:znudef}, one
obtains $\znu^{ik}\znu^{jk}m_{S_k}^2 =
(\overline{\mathcal{M}}^{\,2}_{\tilde\nu_\ell})^{ij}$.

It is possible to relax the assumption of approximately degenerate
sneutrino masses.  In particular, it can be shown diagrammatically
that it is better to use appropriate ratios in place of the
derivatives of~\eq{taylor} in the MIA expansion.  Thus, for $J>I$
(corresponding to the decay of a heavier lepton $\ell^{\,J}$ into a
lighter lepton $\ell^{\,I}$) and neglecting terms proportional to the
lighter lepton mass, one arrives at the simple result:
\bea
C_L^{IJ} &\simeq & 0 \;, \nonumber\\[2mm]
C_R^{IJ} &\simeq & C_{1R}^{bIJ}  + m_{\ell^J} \: C_{4L}^{bIJ} \nonumber\\
&\simeq & \frac{m_{\ell^J} }{(4\pi)^2}\frac{e^2}{2 s_W^2}
\left(M_{LC}^2\right)^{IJ} \left(
|Z_+^{1i}|^2 \left(\frac{\Delta C_{23}}{\Delta
m^2}\right)_{iIJ}
-\frac{\sqrt{2}}{\cos\beta} \frac{m_{\chi^+_i}}{M_W} \:
Z_+^{1i*} Z_-^{2i*} \left(\frac{\Delta C_{11}}{\Delta
m^2}\right)_{iIJ}
\right) , \nonumber \\[2mm]
\label{CRapp}
\eea
where the $Z_{\pm}$ are the chargino mixing matrices defined in
\Ref{Rosiek},
\bea
\left(\frac{\Delta C_{ij}}{\Delta m^2}\right)_{kIJ}  \equiv
\begin{cases}
\frac{\displaystyle C_{ij}(m_{\chi^+_k}, m_{\snl^I})
- C_{ij}(m_{\chi^+_k}, m_{\snl^J})}
{\displaystyle m_{\snl^I}^2 - m_{\snl^J}^2}\,, &  \text{for $I\neq J$}\,,\\[10pt]
\frac{\displaystyle\partial C_{ij}(m_{\chi^+_k}, m_{\snl^I})}
{\displaystyle \partial m_{\snl^I}^2}\,, &
\text{for $I=J$} \,.\end{cases}
\label{eq:cdiff}
\eea
and $ m_{\snl^I}$ are the three ``CP-averaged'' sneutrino masses,
given by the positive square roots of the eigenvalues of $M_{LC}^2$
[cf.~\eqs{eq:di_def}{essmass}].

Clearly, our approximate expression for $C^{IJ}_R$ given by
\eq{CRapp}, which enters the decay rates in \eq{eq58}, is proportional
to the lepton number conserving squared-mass matrix, $M_{LC}^2$,
defined in \eq{mlc}.  Even in the case where $m_L^2$ is diagonal,
contributions to radiative lepton decays arise from the off-diagonal
elements of $M_{LC}^2$ governed by the general form of the matrices
$m_D$ and $m_N^2$ [cf.  the third term in~\eq{mlc}].  Notice that the
flavor dependence disappears completely in the limit of diagonal
$m_{L}^{2}$ and $m^2_{N}= 0$ in which case $M_{LC}^2$ is diagonal.

The effect of the seesaw contribution to  the lepton number
conserving part of the sneutrino squared-mass matrix, $M_{LC}^2$, has
not been previously noticed in the literature.  This yields an extra
contribution to the decay branching ratios $\BR(\ell^{\,J}\ra
\ell^{\,I}\gamma)$.  Consequently, for a fixed set of chargino sector
parameters ($\mu$, $M_2$ and $\tan\beta$) and soft slepton
squared-mass matrix ($m_L^2$), the experimental bounds on the
radiative lepton branching ratios can be used [via
\eqss{BR1}{BR2}{CRapp}] to determine upper limits on the off-diagonal
matrix elements of $M_{LC}^2$.  Examples of such bounds for
$M_2=\mu=200$ GeV and two sets of $\tan\beta$ and $m_L^{min}$
(previously exhibited in Table~\ref{tab:g2}) are shown in
Table~\ref{tab:mlc}.  In obtaining these bounds, we assumed that
$m_L^2$ is diagonal so that \fig{fig:tridiags}(a) does not contribute
to the decay amplitude.\footnote{Non-vanishing off-diagonal elements
of $m_L^2$ should in most cases tighten the bounds on $M_{LC}^2$,
barring accidental cancellations between the amplitudes obtained from
\fig{fig:tridiags}(a) and (b).} We then varied the matrix elements of
$M_{LC}^{2}$ until the constraints from measurements were violated.
Moreover, we incorporated the full numerical one loop calculation for
$\ell^{\,J}\ra \ell^{\,I}\gamma$, presented in
Section~\ref{sec:vertex} rather than the approximate expressions
given, e.g., in \eq{CRapp}.  Notice that there exist lower bounds for
the diagonal elements of $M_{LC}^{2}$ from $(g-2)_{\mu}$, but upper
bounds for the off-diagonal elements of $M_{LC}^{2}$ from
$\BR(\ell^{\,J}\ra \ell^{\,I}+\gamma)$.

\begin{table}[tbp]
\begin{center}
\begin{tabular}{cp{1cm}cp{1cm}c}
\hline
$\tan\beta$ && $10$ && $20$ \\
\hline
$M_{LC}^2$ &&
$ \left (\begin{array}{ccc}
 \gtrsim 270^{2}  &  \lesssim 4^{2}  & \lesssim 11^{2} \\
 ...  & \gtrsim 270^{2} & \lesssim 31^{2} \\
 ...  & ...  & \gtrsim 270^{2} \end{array} \right )
\phantom{\begin{array}{c} 0 \\ 0 \\ 0\\ 0\\\end{array}}

$
&&
$ \left (\begin{array}{ccc}
 \gtrsim 570^{2}  &  \lesssim 8^{2}  & \lesssim 45^{2} \\
 ...  & \gtrsim 570^{2} & \lesssim 150^{2} \\
 ...  & ...  & \gtrsim 570^{2} \end{array} \right )
$
\\
\hline
\end{tabular}
\end{center}
\caption{Bounds on the structure of the matrix elements of
$M_{LC}^2$ for $M_2=\mu=200$ GeV.  All masses in the Table are given
in GeV.}
\label{tab:mlc}
\end{table}

The results of Table~\ref{tab:mlc} illustrate that the bounds on the
square roots of the off-diagonal elements of $M_{LC}^2$ are at least
10---100 times smaller than the square roots of the diagonal elements.
It is convenient to rewrite \eq{mlc} in the following form:
\bea \label{mlcnew}
\!\!\!\!\!\!\!\!
M_{LC}^2&= & m_L^2+\half M_Z^2\cos 2\beta+m_D^*M^{-1}m_N^2
(\mathds{1}+M^{-2}m_N^2)^{-1}M^{-1}m_D^T+\mathcal{O}(v^4 M^{-2})
\nonumber \\[6pt]
&=& m_L^2+\half M_Z^2\cos 2\beta+m_D^*M^{-1}m_N^2 M^{-1}m_D^T
+\mathcal{O}(v^4 M^{-2})+\mathcal{O}(v^2 m_N^4 M^{-4})\,,
\eea
where we have expanded out the quantity
$(\mathds{1}+M^{-2}m_N^2)^{-1}$ under the assumption that
$\|M^{-2}m_N^2\|<1$ (to be justified shortly).  \Eq{mlcnew} implies
that the off-diagonal elements of $M_{LC}^2$ are roughly of order
$m_D^2 m_N^2/M^2$ (barring any accidental cancellations).  If we
assume that $m_D$ is of order the electroweak scale, then the bounds
on the off-diagonal elements given in Table~\ref{tab:mlc} imply that
\bea \label{xdef}
x\equiv\frac{||m_N^2||}{||M^2||}\lesssim \mathcal{O}(10^{-2}) \;,
\eea
with the strongest bound given by $\mu \to e \gamma$ decay.
This result suggests that $\|m^2_{N}\|^{1/2}$ cannot be larger than
about 10\% of the Majorana mass scale $M$.  Hence, $M^{2} + m_{N}^{2}
\simeq M^{2}$ and for the estimates of the magnitude of the entries of
the lepton number violating mass matrix $M_{LV}^2$ in the next section
we henceforth set $m_{N}^{2}=0$.

\setcounter{equation}{0}
\section{Neutrino masses and the lepton number violating parameters}
\label{sec:lv}

In this section we examine the constraints on the lepton number
violating sneutrino squared-mass matrix $M_{LV}^2$ from our knowledge
of the physical (light) neutrino masses and mixing angles.

\subsection{One-loop contributions to neutrino masses}
\label{sec:self}

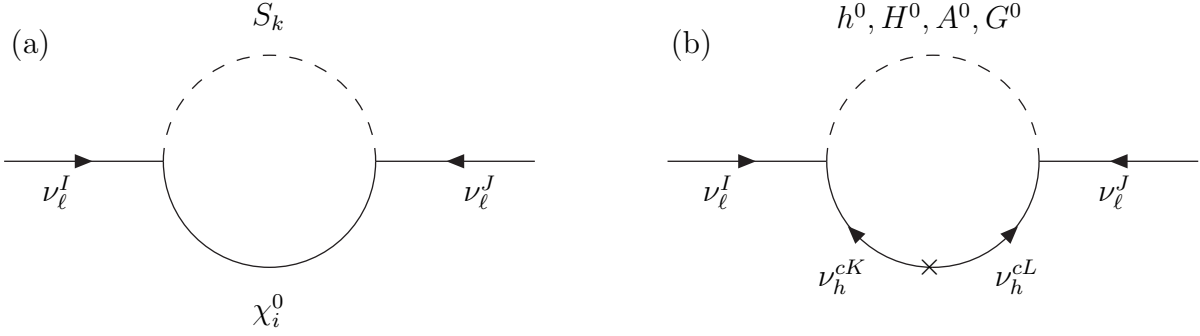
\begin{figure}[tb]
\begin{center}
\begin{picture}(200,110)(120,0)
\Text(0,100)[t]{(a)}
\ArrowLine(-10,50)(50,50)
\Text(10,45)[t]{$\nu_\ell^I$}
\ArrowLine(190,50)(130,50)
\Text(170,45)[t]{$\nu_\ell^J$}
\DashCArc(90,50)(40,0,180){5}
\Text(90,110)[t]{$S_k$}
\CArc(90,50)(40,-180,0)
\Text(90,0)[t]{$\chi^0_i$}
\Text(250,100)[t]{(b)}
\ArrowLine(240,50)(300,50)
\Text(260,45)[t]{$\nu_\ell^I$}
\ArrowLine(440,50)(380,50)
\Text(410,45)[t]{$\nu_\ell^J$}
\DashCArc(340,50)(40,0,180){5}
\Text(340,110)[t]{$h^0,H^0,A^0,G^0$}
\ArrowArcn(340,50)(40,-90,-180)
\ArrowArc(340,50)(40,-90,0)
\Text(340,10)[c]{$\times$}
\Text(307,13)[t]{$\nu_{h}^{cK}$}
\Text(373,13)[t]{$\nu_{h}^{cL}$}
\end{picture}
\end{center}\caption{One-loop corrections to light neutrino masses.
(a)~The loop consisting of light sneutrinos ($S_k, \; k=1\ldots 6$)
and neutralinos ($\chi^0_i, \; i=1\ldots 4$) is the dominant
contribution.  (b)~The loop consisting of a neutral Higgs (or
Goldstone) boson and a heavy neutrino contributes a relative
correction to the light neutrino mass of at most a few percent.  The
contributions of the corresponding graphs (not shown) in which the
light sneutrinos in~(a) are replaced by heavy sneutrinos and the heavy
neutrinos in~(b) are replaced by light neutrinos are suppressed by an
additional powers of $\mathcal{O}(v M^{-1})$ as explained in
\ref{app:oneloopnu}.}
\label{fignu}
\end{figure}

The effective operator that describes the light neutrino mass matrix
is given by:
\beq
-\mathscr{L}_{m_{\nu_\ell}}=\half M_{\nu_\ell}^{IJ}\nu_\ell^I
\nu_\ell^J+{\rm H.c.}
\eeq
Note that $\nu_\ell^I\nu_\ell^J$ is a $\Delta L=2$ operator, since it
changes lepton number by two units.  In \sect{sec:numass}, we
evaluated the tree-level contribution to $M_{\nu_\ell}$ [cf.
\eq{eq:mnu}].  However, one-loop contributions to the light neutrino
mass matrix can be significant, and in some cases these can be as or
more important than the tree-level contribution~\cite{Howie,mixmass}.
The dominant one-loop graph involves a loop containing neutralinos and
light sneutrinos, as shown in \fig{fignu}(a).  Due to the presence of
the lepton number-violating sneutrino squared-mass matrix $M^2_{LV}$,
which violates lepton number by two units, \fig{fignu}(a) can
contribute significantly to the light neutrino mass matrix.
Other one-loop contributions shown in~\fig{fignu}(b), yield
corrections to the light neutrino mass matrix of at most a few
percent, and thus can be neglected.

In order to establish the results just quoted, we begin by reviewing
the relevant interactions that govern the one-loop contributions to
the light neutrino masses.  The light neutrino couplings arise from
\eq{susyyuk} and the supersymmetric sneutrino-neutrino-neutral gaugino
interactions.  After isolating the interaction terms containing one
neutrino field, one arrives at
\bea
\mathscr{L}_{\nu} = -\ynu^{IJ}\left(\nu_L^I \nu_L^{cJ}H_2^2 +
\widetilde{H_2^2} \nu_L^I\tilde\nu_R^{J*}
+\widetilde{H_2^2} \nu_L^{cJ}\tilde\nu_L^{I*}\right)
+ \frac{i}{\sqrt{2}}(g_2 \widetilde W^3 - g_1 \widetilde B)\nu_L^I
\tilde\nu_L^{I*} + \Hc \;,
\label{eq:vlag0}
\eea
where $\widetilde W^3$ and $\widetilde B$ are the SU(2) and U(1)
neutral (two-component) gaugino fields, and $g_2$~and $g_1$ are the
corresponding gauge couplings.  Using \eqs{udef}{nutransform}, it
follows that $\nu_L\simeq \nu_\ell+m_D^* M^{-1}\nu_h^c$ and
$\nu_L^c\simeq \nu_h^c-M^{-1}m_D^T\nu_\ell$.
Likewise, it follows from \eqs{boldvee}{snutransform} that
\bea
\tilde\nu_L &\simeq &\tilde\nu_\ell+m_D^* M (M^2+m_N^2)^{-1}
\tilde\nu_h^*\,,\label{nulapp}\\[5pt]
\tilde\nu_R^* &\simeq &\tilde\nu_h^* -
(M^2+m_N^{2})^{-1}Mm_D^T\tilde\nu_\ell\,.\label{nurapp}  
\eea
Thus, the effective
interaction involving (at least) one light neutrino field is given by:
\bea
\!\!\!\!\!\!\!\!\!\!
\mathscr{L}_{\nu_\ell} & \simeq &
-\ynu^{IJ}\biggl\{\widetilde{H_2^2} \nu_\ell^I\tilde\nu_h^{J*}
+\nu_\ell^I \nu_h^{cJ}H_2^2 -(m_DM^{-1})^{KJ}
\left(\widetilde{H_2^2}\nu_\ell^K\tilde\nu_\ell^{I*}
+\nu_\ell^I\nu_\ell^K H_2^2\right)\nonumber \\
&& \qquad\qquad
-[(M^2+m_N^2)^{-1}Mm_D^T]^{JK}\widetilde{H_2^2}\nu_\ell^I\tilde\nu_\ell^K\biggr\}
\nonumber \\
&&\,\,\,
+ \frac{i}{\sqrt{2}}(g_2 \widetilde W^3 - g_1 \widetilde B)
\left[\nu_\ell^I\tilde\nu_\ell^{I*} + m_DM(M^2+m_N^{2\,*})^{-1}
\nu_\ell^I\tilde\nu_h^{I}\right]+ \Hc
\label{eq:vlag}
\eea

In order to perform the explicit loop computations, it is convenient
to rewrite \eq{eq:vlag} in terms of mass eigenstate fields.  The Higgs
field $H_2^2$ is expressed as \cite{gunhab1}:
\beq \label{h22}
H_2^2=\frac{1}{\sqrt{2}}\left[v_2+h^0\cos\alpha+H^0\sin\alpha
+i(\cos\beta A^0+\sin\beta G^0)\right]\,,
\eeq
in terms of the CP-even Higgs fields $h^0$ and $H^0$ (where
$m_{h^0}\leq m_{H^0}$), the CP-odd Higgs field $A^0$ and the Goldstone
field $G^0$, where $\tan\beta\equiv v_2/v_1$ and $\alpha$ is the
CP-even Higgs mixing angle.  We also define two-component
mass-eigenstate neutralino fields $\kappa_j^0$ ($j=1,\ldots,4$)
following \Ref{Rosiek} by
\beq \label{zndef}
\psi_i\equiv Z_N^{ij}\kappa_j^0\,,\qquad {\rm where} \qquad
\psi_i\equiv (-i\widetilde B\,,\,-i\widetilde W^3\,,\,\widetilde
H_1^1\,,\,\widetilde H_2^2)\,,
\eeq
and $Z_N$ is a unitary matrix that governs the Takagi-diagonalization
of the complex symmetric $4\times 4$ neutralino mass matrix,
$M_{\chi^0}$ via $Z_N^T M_{\chi^0} Z_N={\rm diag}(
M_{\chi_1^0}\,,\,\ldots\,,\,M_{\chi_4^0})$.

Before presenting the explicit computations, let us first estimate the
order of magnitude of the loop-contributions to the neutrino mass due
to the loop graphs of \fig{fignu}(a) and (b), and the corresponding
graphs (not shown) in which the light sneutrinos [heavy neutrinos] in
graph (a) [(b)] are replaced by heavy sneutrinos [light neutrinos].
This analysis is presented in~\ref{app:oneloopnu}---the results
obtained there imply that the graphs of \fig{fignu}(a) and (b) both
yield contributions to the one-loop light neutrino mass matrix of
order the tree-level light neutrino masses, multiplied by the
appropriate vertex couplings and a typical loop factor.  Other
one-loop contributions not shown in \fig{fignu} are suppressed by
additional powers of $\mathcal{O}(vM^{-1})$ and are utterly
negligible.

We begin with an examination of the loop amplitude of \fig{fignu}(b),
which is governed by the light neutrino-heavy neutrino-Higgs
interaction term of \eq{eq:vlag}.  The internal heavy neutrino line is
marked with an $\times$ to indicate the lepton-number violating
propagator proportional to its (diagonal) mass $M\delta^{KL}$.
Summing over all the internal neutral Higgs and Goldstone states, the
leading $\mathcal{O}(M)$ term vanishes, leaving a subleading term of
$\mathcal{O}(v^2 M^{-1})$, which is the magnitude of the
\textit{light} neutrino mass.  We find that~\fig{fignu}(b) yields a
leading contribution to the light neutrino mass that is proportional
to the tree-level light neutrino mass matrix [cf.~\eq{eq:mnu}]:
\bea
\delta M_{\nu_\ell}\approx
- \frac{M_{\nu_\ell}}{32\pi^2}\, \frac{g_2^2}{c_W^2}\,
\log\frac{\overline M}{\overline m_H} \;, \label{eq:higgs}
\eea
where $\overline M$ and $\overline m_H$ denote average heavy neutrino
and Higgs boson masses.  This correction turns out to be of the order
of at most few percent.  Additional corrections can also arise that
modify the flavor structure of $M_{\nu_\ell}$, but these are not
logarithmically enhanced and are thus even smaller.

Hence, the possibility of a \textit{significant} one-loop contribution
to the light neutrino mass matrix can only arise from
\fig{fignu}(a), which is governed by the light
sneutrino-neutrino-gaugino interaction term of \eq{eq:vlag}.  In the
following, we examine the corresponding loop graph in which the
external light neutrino fields are mass eigenstates $(\nu_\ell^J)^{\rm
phys}$ [cf.~\eq{MNS}].  Using four-component spinor methods, the
amplitude for this graph (with incoming four-momentum~$p$) will be
denoted by 
\beq
-i[ (\pslash \Sigma_V^{IJ} + \Sigma_S^{IJ}) P_L + (\pslash
\Sigma_V^{IJ*} + \Sigma_S^{JI*}) P_R]\,,
\eeq
where the generic self
energies $\Sigma^{IJ}_{V,S}(p^2)$ of the Majorana neutrino must be
symmetric in its indices $I,J$.  To evaluate this graph, we express
the neutrino-sneutrino-gaugino interaction Lagrangian in terms of the
four-component self-conjugate Majorana neutrino fields $\nu^I_M$ and
the Majorana neutralino fields~$\chi_i^0$
[cf. \ref{app:fermion}]:\footnote{More explicitly, the non-zero
components of $P_L\nu^I_M$ are the two-component neutrino fields
$(\nu^I_\ell)^{\rm phys}$, and the non-zero components of $P_L\chi^0$
are the two-component neutralino fields $\kappa_i^0$ introduced in
\eq{zndef}.}
\bea
\mathscr{L}_{\chi\nu\tilde\nu} = -\half \: (g_2 Z_N^{2i} - g_1 Z_N^{1i})(\znu^{Ik} -
i\znu^{(I+3)k})U_{MNS}^{IJ} \: \bar\chi^0_i P_L \nu_M^J S_k + \Hc\,,
\eea
where the neutralino mixing matrix $Z_N$ is defined in \eq{zndef}.
The resulting $\overline{\rm DR}$-renormalized neutrino mass matrix at
one-loop order is given by:
\bea
\!\!\!\!\!\!\!\!
(M^{(\rm 1-loop)}_{\nu_\ell})^{IJ} \ = \ m_{\nu_{\ell I}}(\mu_R) \,
\delta^{IJ} + \re \left [\Sigma_S^{IJ}(\overline{m}_{\nu_\ell}^2)
+ \half m_{\nu_{\ell I}}
\Sigma^{IJ}_V(\overline{m}_{\nu_\ell}^2) + \half
m_{\nu_{\ell J}} \Sigma^{JI}_V(\overline{m}_{\nu_\ell}^2) \right],
\label{numass}
\eea
where the loop diagrams are regularized by dimensional reduction and
the tree level diagonal mass, $m_{\nu_{\ell I}}$, is defined at the
renormalization scale $\mu_R$.  In addition,
$\overline{m}_{\nu_\ell}^2$, is some average neutrino mass scale,
which to a very good approximation can be taken to be zero in the
explicit loop calculations presented below.

In order to determine the masses of the light neutrinos at one-loop
accuracy, it is usually sufficient to calculate the diagonal matrix
elements of the self energies (i.e., by setting $I=J$ in \eq{numass}),
assuming that the tree-level neutrino masses are non-degenerate.
However, in some cases $\Sigma_{S,V}^{IJ}$ can be numerically large
for $I\ne J$.  If the latter holds, then one must re-diagonalize the
neutrino mass matrix, $(M_{\nu_\ell}^{\rm one-loop})^{IJ}$, in order
to obtain the loop-corrected physical neutrino masses and
corresponding mixing matrix $U_{\rm MNS}$ (more details of a similar
procedure in the context of $R$-parity violating models can be found,
e.g., in \Refs{Hempfling}{Rimmer}).

An explicit calculation of the diagram shown in \fig{fignu}(a), in the
limit $\overline{m}^2_{\nu_\ell}\rightarrow 0$, yields
\bea
\!\!\Sigma_S^{IJ} &\!\!=\!\!&  \frac{-m_{\chi^0_i}}{4(4\pi)^2} \:
{(g_2 Z_N^{2i} - g_1 Z_N^{1i})^2}\: (\znu^{Lk} - i\znu^{(L+3)k}) \:
(\znu^{Mk} - i\znu^{(M+3)k}) U_{\rm MNS}^{LI}U_{\rm MNS}^{MJ} \:
B_0(m_{\chi^0_i}, m_{S_k}), \nonumber
\\
\label{snu}
\\
\!\!\Sigma_V^{IJ} &\!\!=\!\!&  \frac{-1}{4(4\pi)^2}\:|g_2 Z_N^{2i} - g_1 Z_N^{1i}|^2\:
(\znu^{Lk} - i\znu^{(L+3)k}) \: (\znu^{Mk} + i\znu^{(M+3)k}) U_{\rm
MNS}^{LI}U_{\rm MNS}^{MJ*} \: B_1(m_{\chi^0_i}, m_{S_k}),\nonumber\\
\label{snuv}
\eea
with an implicit sum over repeated indices, where $m_{\chi^0_i}$ and
$m_{S_k}$ are the neutralino and sneutrino masses, respectively, and
$B_0$, $B_1$ are the standard 2-point loop-integrals \cite{passarino}
evaluated at $p^2=0$,
\bea
B_0(x,y) &=& \Delta - \log\frac{xy}{\mu_{R}^2} +1 - \frac{x^2 +y^2}{x^2
-y^2} \log\frac{x}{y} \;,\\
\label{eq:b0def}
B_1(x,y) &=& -\frac{1}{2}\Delta + \frac{1}{2}\log\frac{xy}{\mu_{R}^2}
-\frac{3}{4} - \frac{y^2}{2(x^2-y^2)} + \left(\frac{x^4}{(x^2-y^2)^2}
-\frac{1}{2} \right)\log\frac{x}{y} \;,
\label{eq:b1def}
\eea
with $\Delta \equiv 2/(4-d) -\gamma + \ln 4\pi$ set to $\Delta=0$ in
the minimal subtraction renormalization scheme.  Note that $\Sigma_S$
is finite, i.e.  in the sum over $k$ the dependence on $\Delta$ and
$\mu_R$ cancels exactly due to the orthogonality of $\mathcal{Z}$.
Likewise, $\Sigma_V^{IJ}$ is finite for $I\neq J$, which is easily
verified after using the orthogonality of $\mathcal{Z}$ and the
unitarity of $U_{\rm MNS}$.  This is to be expected since in the mass
basis there are (by definition) no tree-level off-diagonal neutrino
mass matrix elements.  In contrast, $\Sigma_V^{JJ}$ is divergent, and
after minimal subtraction it is here that the $\mu_R$ dependence
resides.

We now examine the relative magnitudes of the various contributions in
\eq{numass} to the loop-corrected neutrino mass.  First, we observe
that $\Sigma_{V}$ [given by \eq{snuv}] is dimensionless and has a
magnitude of the order of a typical electroweak correction (this has
been numerically confirmed).  Thus, the one loop contribution of the
terms proportional to the minimally subtracted $\Sigma_V$ in
\eq{numass} is at most a few percent of the tree-level neutrino mass.
Given the current experimental accuracy of neutrino data, this latter
correction can be neglected, as it does not provide any constraints on
sneutrino parameters.  Thus, we focus on $\Sigma_S$ [given by
\eq{snu}], which can be simplified by employing the MIA expansion
described in \sect{sec:llg}.  The end result is:
\bea
\!\!\!\!\!\!\!\!\!\!\!\!
\delta M_{\nu_\ell}^{IJ} & \equiv &(M^{\rm 1-loop}_{\nu_\ell})^{IJ}-m_{\nu_{\ell I}}
\delta^{IJ} \nonumber \\
& \simeq &
\frac{-1}{32 \pi^2} \: \sum_{i,K,M} m_{\chi^0_i}\re\left[(g_2 Z_N^{2i} - g_1
Z_N^{1i})^2 \: U_{\rm MNS}^{KI} U_{\rm MNS}^{MJ}
\left(M_{LV}^2\right)_{KM} \right]\ \:
\left(\frac{\Delta B_0}{\Delta m^2}\right)_{iKM}\!\!\!\!\!\!,
 \label{eq:vcorr}
\eea
where in analogy to~(\ref{eq:cdiff}) we define
\bea
\left(\frac{\Delta B_0}{\Delta m^2}\right)_{kIJ} \equiv
\begin{cases}\frac{\displaystyle B_0(m_{\chi^0_k}, m_{\snl^I})
- B_0(m_{\chi^0_k}, m_{\snl^J})} {\displaystyle m_{\snl^I}^2 -
m_{\snl^J}^2}\,, & \text{for $I\neq J$}\,,\\[10pt]
\frac{\displaystyle\partial B_0(m_{\chi^0_k}, m_{\snl^I})}
{\displaystyle \partial m_{\snl^I}^2}\,, &
\text{for $I=J$} \,.\end{cases}
\label{eq:bdiff}
\eea
and the CP-averaged sneutrino masses, $m_{\tilde\nu^I_\ell}$, are
defined below \eq{eq:cdiff}.  As expected, this contribution is finite
and is explicitly lepton number violating, as it is proportional to
the matrix $M_{LV}^2$.  \Eq{eq:vcorr} is a generalization of eq.~(7)
of~\Ref{Howie} to the 3-flavor seesaw model.\footnote{We correct here
a typographical in eq.~(7) of~\Ref{Howie} where $(g_2 Z_N^{2i} - g_1
Z_N^{1i})^2$ is incorrectly written as $|g_2 Z_N^{2i} - g_1
Z_N^{1i}|^2$.}

The results given in \sect{sec:self} can be used to estimate the
bounds on the heavy sneutrino soft parameters $m_N^2, m_B^2, X_{\nu}$
imposed by the current experimental measurements of neutrino masses
and mixing.  These bounds allow for a significant one-loop correction
to the light neutrino mass matrix, $\delta M_{\nu_\ell}^{IJ}$, which
could even compete with the corresponding tree-level masses.  Further
details will be given in Sections~\ref{sec:sugra} and \ref{sec:gen}.

\subsection{Radiative generation of neutrino masses and mixing}
\label{sec:mix}

It is very tempting to explain the characteristics of the neutrino
mass spectrum as a consequence of radiative corrections.  The most
economical possibility is one in which the pattern of neutrino masses
is entirely radiatively generated by the loop corrections.  However,
in the supersymmetric seesaw model this is not possible.  If one sets
$m_{\nu_{\ell I}}=0$ (for all $I$) in \eq{eq:vcorr}, then $m_D=0$ (or
equivalently, $Y_{\nu}=0$), in which case only the light
sneutrino-neutrino-gaugino interaction of \eq{eq:vlag} survives.
However, this interaction generates a one-loop neutrino mass that is
proportional to $M_{LV}^2$ [cf.~\eq{eq:vcorr}], which vanishes in the
limit of $m_D=0$.

Here, we shall be less ambitious and investigate whether the hierarchy
and/or the flavor mixing of neutrinos can be generated entirely by
loop effects.  As we shown below, such a scenario seems to be
possible.  However, in order to obtain the correct values of the light
neutrino mixing matrix elements, a fine-tuning of sneutrino parameters
may be required.

To be more specific, consider the following scenario.  At tree level
we assume the Yukawa coupling matrix $Y_\nu$
to be real, non-negative and flavor diagonal, i.e.  $Y_\nu^{IJ} =
Y_\nu^I\delta^{IJ}$ (with $Y_\nu^I\geq 0$).  
Consequently, the tree level neutrino mass
matrix [\eq{eq:mnu}] is also real, non-negative and diagonal so that
$U_{\rm MNS}^{\rm tree}=i\mathds{1}$.  Then, the one-loop correction
to the neutrino mass matrix [\eq{eq:vcorr}] is proportional to:
\bea
\alpha_{IJ} \equiv \frac{1}{32\pi^2} \sum_{i=1}^4 m_{\chi^0_i} (g_1 Z_N^{1i} -
g_2 Z_N^{2i})^2 \left(\frac{\Delta B_0}{\Delta m^2}\right)_{iIJ} \,.
\label{eq:a}
\eea
If one assumes that the flavor splitting of the light sneutrino masses
is small, then the ratio $\left({\Delta B_0}/{\Delta
m^2}\right)_{iIJ}$ is approximately constant with the respect to the
indices $I,J$, so that $\alpha_{IJ}\approx \alpha$ is roughly
constant.  Therefore, the one-loop corrected neutrino mass matrix
[\eq{numass}] can be written as
\bea
m_{\nu_\ell}^{(\rm 1-loop)} \simeq -m_D M^{-1} m_D +
\re\left(\alpha M_{LV}^2\right) \;.
\label{eq:vcorr_spec}
\eea
Since we have assumed above that $Y_\nu$ is diagonal, it follows that
$m_D\equiv v_2 Y_\nu/\sqrt{2}$ is also diagonal, 
in which case there is no need to
distinguish between $m_D$ and its transpose.
For simplicity, we shall further assume that $m_N^2\ll M^2$.  Then, using
\eq{mlv} for $M_{LV}^2$, in which only the leading $\mathcal{O}(v
M^{-1})$ terms are kept [under the assumption that
$m_B^2\sim\mathcal{O}(vM)$ as suggested by \eq{assume4}], we may
express \eq{eq:vcorr_spec} in the following form:
\bea
m_{\nu_\ell}^{(\rm 1-loop)} & \simeq & -[\mathds{1} -  \re(\alpha
X_{\nu})]\, m_D M^{-1} m_D \, [\mathds{1}- \re(\alpha
X_{\nu}^T)]\nonumber \\[3mm] &-& 2  \, m_D \frac{1}{M}\re(\alpha m_B^2)
\frac{1}{M} m_D
\ +  \re(\alpha X_{\nu})\, m_D M^{-1} m_D \, \re (\alpha X_{\nu}^T) \;.
\label{eq:vcorr_spec1}
\eea

To achieve the correct hierarchy of neutrino masses and mixings, one
possible strategy is to demand that the sum of the last two terms on
the right hand side of \eq{eq:vcorr_spec1} is negligible, in which
case the first term yields the correct physical neutrino masses and
the mixing matrix.  Then, using \eq{takagi}, we perform a
Takagi-diagonalization to identify the physical (loop-corrected)
neutrino masses and mixing matrix elements:
\bea
-[\mathds{1} - \re(\alpha  X_{\nu})] \, m_D M^{-1} m_D \, [\mathds{1}-
\re(\alpha  X_{\nu}^T)] \ = \ (U_{\rm MNS}^{\rm phys})^*\,
m_{\nu_\ell}^{phys} \, (U_{\rm MNS}^{\rm phys})^\dagger \;,
\label{eq:vmix2}
\eea
where $m_{\nu_\ell}^{phys}$ is the (non-negative) diagonal physical
neutrino mass matrix.  One can solve \eq{eq:vmix2} analytically for
$\re(\alpha X_\nu)$, which yields:
\bea
\re(\alpha X_{\nu}) = \mathds{1}-i (U_{\rm MNS}^{\rm phys})^*
(m_{\nu_\ell}^{phys})^{1/2}RM^{1/2}m_D^{-1}\;,
\label{radnu}
\eea
where $R$ is a complex orthogonal matrix, subject to the restriction
that the right hand side of \eq{radnu} is real.  Thus, starting from
\textit{any} hierarchy of the tree-level diagonal, non-vanishing
Yukawa couplings $Y_\nu^I$, the special choice of $X_\nu$ given in
\eq{radnu} allows us to reproduce the correct neutrino mass hierarchy
and the mixing matrix.

Clearly, the scenario just presented is not very realistic from the
phenomenological point of view.  To achieve the desired result, a
specific form of the $X_\nu$ parameter, very close to perturbativity
limit of $Y_\nu$ and the charged slepton masses is required, as well
as a rather precise cancellation between the last two terms of
\eq{eq:vcorr_spec1}.  Nevertheless, our example above provides an
analytical existence proof for a radiative mixing scenario.  In
general, for given $Y_\nu$ and $M$, many choices of sneutrino
parameters leading to the correct pattern of neutrino masses and
mixing at the one-loop level exist, but they need to be determined
numerically.  Presumably, all successful scenarios require a certain
degree of fine-tuning, but perhaps some solutions would be deemed
acceptable.

\subsection{Universal parameters at the scale $\boldsymbol{M}$}
\label{sec:sugra}

The magnitudes of the parameters $A_{\nu}$, $m^{2}_{B}$ and
$m^{2}_{N}$ that govern the behavior of the heavy sneutrino sector are
connected with the mechanism of supersymmetry breaking [cf.
\eq{lsoft}].  These parameters decouple at the scale $M\gg M_Z$ where
the sneutrino superfield $\wh{N}$ decouples.  If the scale $M$ is
close to the GUT scale then soft SUSY breaking parameters are
restricted by GUT symmetry considerations.  Further assumptions on the
minimality of the K{\"a}hler potential in supergravity simplify our
input parameters considerably, at the scale $M\sim M_{GUT}$,
\begin{eqnarray}
 A_{\nu} \ = \ A_{0} \, Y_{\nu}\quad , \quad
 m^{2}_{B} \ = \ m_{0} \, M  \quad , \quad
 m_{N}^{2} \ = \ x M^{2} \;, \label{univ}
 \end{eqnarray}
where $A_{0}$ is a complex number, $m_{0}$ and $x$ are real numbers,
$M$ is a diagonal $3\times 3$ Majorana neutrino matrix [cf.~\eq{takagi}]
and $Y_{\nu}$ is the neutrino Yukawa coupling [cf.~\eq{superpot}].

Under the universality assumptions of \eq{univ}, the matrices
$M_{LC}^{2}$ and $M_{LV}^{2}$ assume the following simple forms at the
GUT scale:
\begin{eqnarray}
M_{LC}^{2}  &=&  m_{L}^{2}   \ + \ \frac{1}{2} M_{Z}^{2} \cos 2\beta +
\frac{x}{1+x} m_{D}^{*} m_{D}^{T} \;, \label{umlc} \\[2mm]
M_{LV}^{2}    &=&  \frac{2 \, M_{\nu_{\ell}}}{1+x} \left (
A_{0} + \mu^{*}\cot\beta -  \frac{m_{0}}{1+x} \right ) \;,\label{umlv}
\end{eqnarray}
where the light tree-level neutrino mass matrix $M_{\nu_\ell}$ is
given in~\eq{eq:mnu}.  As parameters ``run'' from the GUT scale to low
energies, $m_{L}^{2}$ receives renormalization from other Yukawa and
gauge interactions.  In contrast, all the parameters associated with
the superfield $\wh{N}$ are hardly affected since $M\sim M_{GUT}$.
Moreover, the neutrino mass matrix $M_{\nu_{\ell}}$ and the
superpotential parameter $\mu$ are both multiplicatively renormalized.
Hence, just above the scale of low-energy supersymmetry breaking, the
low-energy value of $M_{LV}^2$ is still given by \eq{umlv}, with the
parameters on the right-hand side defined at the low scale.  At the
low-energy supersymmetry-breaking scale the $\overline{\rm DR}$
running neutrino mass matrix $M_{\nu_{\ell}}(\mu_{R})$ [or its
diagonal form $m_{\nu_{\ell}^{I}}(\mu_{R}$)] receives finite threshold
corrections from the neutralino--sneutrino loop in \fig{fignu}(a).
The one-loop correction to the neutrino mass matrix given in
\eq{eq:vcorr} is proportional to the diagonal tree-level neutrino mass
matrix.\footnote{Indeed, assuming universal parameters at the GUT
scale, and noting that $x\lesssim\mathcal{O}(10^{-2})$
[cf. \eq{xdef}], it follows that $M_{LC}^2\simeq m_{LC}^2\mathds{1}$
at the GUT scale, where $m_{LC}^2$ is one of the approximately
degenerate eigenvalues of $M_{LC}^2$.  The positive square roots of
the eigenvalues of $M_{LC}^2$, evaluated at the low-energy scale, are
identified as the three CP-averaged light sneutrino masses.  Although
$m_L^2$ is no longer proportional to the identity matrix at
low-energies, this latter effect is formally of higher order in the
loop expansion of $\delta M_{\nu_\ell}^{IJ}$ [cf. \eq{eq:vcorr}].
Consequently, we can neglect the flavor splitting of the CP-averaged
light sneutrino masses in the evaluation of the ratio $(\Delta
B_0/\Delta m^2)_{iKM}$, in which case this ratio is roughly constant
with respect to the indices $K$ and $M$ as discussed below \eq{eq:a}.}
Hence, the one-loop corrected neutrino masses assume the very simple
and suggestive form
\begin{eqnarray}
m^{(\rm 1-loop)}_{{\nu_\ell}^{I}} \ = \ m_{{\nu_\ell}^I} \, \left [
1 + 2{\rm Re}\frac{\alpha}{(1+x)} \left (A_{0} + \mu^{*}\cot\beta
- \frac{m_{0}}{1+x} \right ) \right ] \;, \label{eq:1loop}
\end{eqnarray}
where $\alpha$ is defined in \eq{eq:a} and all parameters are now
defined at the scale $\mu_{R}=M_{Z}$.

\begin{table}[t]
 \centering
   \begin{tabular}{@{} lp{1cm}cp{2cm}cp{1cm}r @{}} 
   \toprule
   \multicolumn{7}{c}{Input Parameters} \\
   \cmidrule(lr){1-7} 
 \multicolumn{3}{c}{Neutrino Sector} & & \multicolumn{3}{c}{SUSY Sector} \\
 \midrule
 $m^{\rm phys}_{\nu_{{\ell}^{1}}}$ && $10^{-14}$ && $A_{0}$ && 0  \\
 $m^{\rm phys}_{\nu_{{\ell}^{2}}}$ && $\sqrt{\Delta m^{2}_{{\rm sol}}}$ && $m_{0}$ && 0  \\
 $m^{\rm phys}_{\nu_{{\ell}^{3}}}$ && $\sqrt{\Delta m^{2}_{{\rm atm}}}$ && $\mu$ && 350  \\
 $\theta_{1}$ &&  0.2+0.1 i && $\tan\beta$ && 10  \\
 $\theta_{2}$ &&  0.3 && $M_{\tilde{B}}$ && 95 \\
 $\theta_{3}$ &&  0.1 + 0.5i && $M_{\tilde{W}}$ && 189  \\
 $M_{1}$      &&  $10^{14}$ && $x$ && 0.0 \\
 $M_{2}$      &&  $2\times 10^{14}$ && $m_{L}$ && 197  \\
 $M_{3}$      &&  $5\times 10^{14}$ && $m_{R}$ && 135  \\
 \bottomrule
\end{tabular}
\caption{If not otherwise indicated, the input parameters
that govern the neutrino and SUSY sectors listed above have been
employed in our numerical analysis.  We take
$\Delta m^{2}_{{\rm sol}}= (8.0^{+0.4}_{-0.3})\times 10^{-5}~{\rm eV}^2$ and
$\Delta m^{2}_{{\rm atm}}= (2.45\pm 0.55)\times 10^{-3}~{\rm eV}^2$ 
from Ref.~\cite{PDG}.  The values for $\theta_{1,2,3}$ above are 
representative choices (as these angles are not fixed by the light
neutrino data).
All mass parameters in the above table are in GeV units.  }
\label{tab2}
\end{table}

We next examine the light sneutrino mass difference.  
Since the results of Table~\ref{tab:mlc} imply that $M^2_{LC}$ is very
close to diagonal form, it follows that $Q_0\simeq 1$ (cf.~discussion
above \eq{qzero}].  Combining the results of
\eqss{eq:qdef}{beedef}{deltamnu},  we derive
\begin{eqnarray}
\left (\frac{\Delta m_{\tilde\nu_\ell}}{m_{\nu_{\ell}}} \right )\ls{I}
 \ =\ \frac{2}{m_{\tilde\nu_{\ell^{I}}}m_{\nu_{\ell^{I}}}}
\left|\frac{(M_{\nu_\ell})_{II}}
{1+x} \left ( A_{0} + \mu^{*} \cot\beta -
 \frac{m_{0}}{1+x} \right )\right|\;,
\label{dm2}
\end{eqnarray}
which is identical to the one flavor case found in \eq{eq:dmnu} and in
Ref.\cite{Howie} if the neutrino mass matrix $M_{\nu_\ell}$ 
is diagonal.  In the more general case of non-diagonal
$M_{\nu_\ell}$, the
diagonal elements of the neutrino mass matrix
do not coincide with the neutrino masses 
$m_{\nu_{\ell^I}}$.  Consequently, the quantity 
$(\Delta m_{\tilde\nu_\ell}/m_{\nu_{\ell}})\ls{I}$ exhibits non-trivial
dependence on the flavor index~$I$.

%

To produce quantitative results, we need to initialize the neutrino
Yukawa couplings in such a way that we always reproduce the
``observed'' MNS mixing matrix.  Using \eqs{eq:mnu}{eq:vmphys}, it
follows that
\begin{eqnarray}
m_{\rm D} = i U^*_{\rm MNS} \, (m_{\nu_\ell}^{\rm phys})^{1/2} \,
R^{T} \, M^{1/2} \;,
\label{mdd}
\end{eqnarray}
where $R$ is an arbitrary complex orthogonal matrix~\cite{Casas}, with
three (complex) angles, $\theta_{{1,2,3}}$.
(As the sign of $R$ is undetermined, one may choose ${\rm det}~R=1$
without loss of generality.)
In the plots that follow, we
assume a hierarchical spectrum for the neutrinos, and all relevant
input parameters are displayed in Table~\ref{tab2}.
The value for $m_{L}$ adopted in Table~\ref{tab2} is
consistent with a supersymmetric interpretation of the observed
experimental excess for $\delta a_{\mu}$.

\begin{figure}[t!] 
\centering
\includegraphics[width=6in]{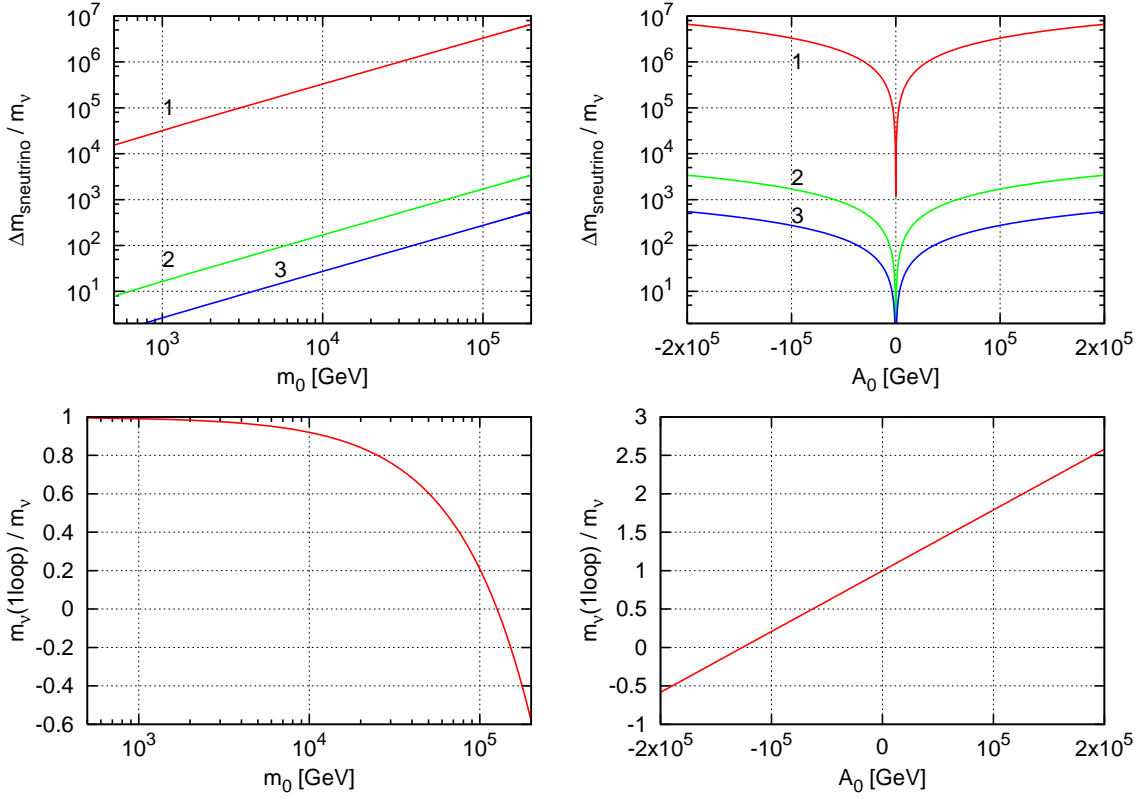}
\caption{Predictions for the ratios $(\Delta
m_{\tilde\nu_\ell}/m_{\nu_{\ell}})\ls{I}$ and $(m^{(\rm
1-loop)}_{{\nu_\ell}}/{m_{\nu_{\ell}}})\ls{I}$ for the three neutrino
states ($I=1,2,3$) as functions of the soft SUSY-breaking
parameters $m_{0}$ and $A_{0}$.
When varying $m_0$ [left panels]
we set $A_0=0$ and when varying $A_0$ [right panels] we set $m_0=0$.
\vspace{0.05in}}
\label{fig:massdif}
\end{figure}

In \fig{fig:massdif} we plot the ratios $(\Delta
m_{\tilde\nu_\ell}/m_{\nu_{\ell}})\ls{I}$ [upper panels] and $(m^{(\rm
1-loop)}_{{\nu_\ell}}/{m_{\nu_{\ell}}})\ls{I}$ [lower panels] as
functions of the SUSY-breaking 
parameters $m_{0}$ [left panels] and
$A_{0}$ [right panels].  
When varying $m_0$ we set $A_0=0$ and when varying $A_0$ we set $m_0=0$.
Otherwise, our input parameters are as specified in 
Table~\ref{tab2}.  
In obtaining these results, 
we have incorporated the full one-loop
contribution to the neutrino masses.  In the two lower panel plots, the 
ratios $(m^{(\rm
1-loop)}_{{\nu_\ell}}/{m_{\nu_{\ell}}})\ls{I}$ are nearly independent
of the flavor $I$, and thus only one curve is shown.
Our numerical results confirm our analytical approximate
formulae of \eqs{eq:1loop}{dm2} and demonstrate that one must have
$m_{0} \lesssim 10^{5}$ GeV ($|A_{0}| \lesssim 10^{5}$ GeV) to
guarantee that the radiative corrections to neutrino masses are less
than 80\% of the tree level neutrino mass.  In this case, the
sneutrino mass difference is at most $\Delta m_{\tilde\nu_\ell}
\lsim 300\,\Delta m_{atm}
\simeq 15$ eV. 

\begin{figure}[ht!] 
\centering
\hspace{-0.5in}
\includegraphics[width=3.4in]{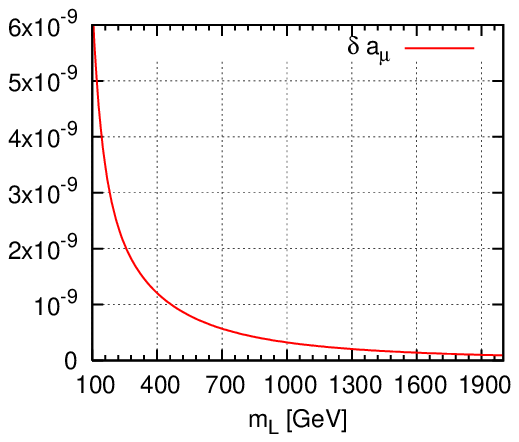}\includegraphics[width=3.4in]{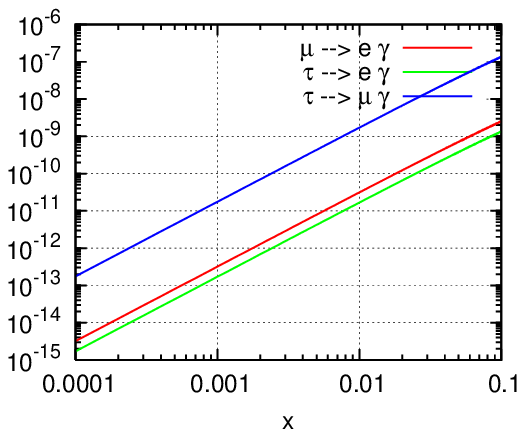}
\caption{(a) In the left panel, the contribution to the
muon anomalous magnetic moment from the diagrams in \fig{fig:tridiags}
as a function of $m_{L}=m_R$ is exhibited.  (b) In the right panel,
the prediction for 
BR($\ell^{\,J}\ra \ell^{\,I}\gamma$) is shown as a function of the
parameter $x=m_{N}^{2}/M^{2}$. 
The upper [lower] curves correspond to 
$\tau\to\mu\gamma$ [$\tau\to e\gamma$], 
and the middle curve to $\mu\to e\gamma$.
}
\label{fig:amu}
\end{figure}

For completeness, we plot in \fig{fig:amu} the results for $g_\mu-2$
anomaly and the branching ratios for the decays $\ell^{\,J}\ra
\ell^{\,I}\gamma$ in the case of universal parameters at the SUGRA
scale.  The results shown in \fig{fig:amu} confirm our choices of a
lower bound for $m_L$ [cf.~Table~\ref{tab:g2}] obtained in
\sect{sec:vertex} and an upper bound for~$x$ [cf.~\eq{xdef}] obtained in
\sect{sec:llg}.


\subsection{General case}
\label{sec:gen}

So far we have dealt with universal boundary conditions for the
supersymmetric parameters.  One can set general bounds for the lepton
number violating matrix elements of $M_{LV}^{2}$ from
\eq{eq:vcorr} and the ``naturalness'' assumption of $\delta
m_{\nu_{\ell}} \lesssim m_{\nu_{\ell}}$.  In the general case,
appropriate bounds can be derived only numerically and depend on the
particular form of the MNS matrix.  Analytical estimates can be
obtained using the following approach.  Let us require that the
one-loop corrections to the neutrino mass matrix do not significantly
affect the physical neutrino masses and their mixing.
Combining~\Eqs{eq:vmphys}{eq:vcorr}, one gets for any $I,J$:
\bea
|U_{\rm MNS}^{MI} \left(M_{\nu_\ell}\right)_{MN} U_{\rm MNS}^{NJ}| \geq
\left| \frac{m_{\chi^0_i}}{32 \pi^2} \: \re\left[(g_2 Z_N^{2i} - g_1
Z_N^{1i})^2 \: U_{\rm MNS}^{MI} U_{\rm MNS}^{NJ}
\left(M_{LV}^2\right)_{MN} \right]
\left(\frac{\Delta B_0}{\Delta m^2}\right)_{iMN}\right| \;.\nonumber\\
\label{eq:lvgen}
\eea
The structure of the $U_{\rm MNS}$ factors on both sides
of~\eq{eq:lvgen} is identical, so roughly [barring possible
cancellations between terms and the effects of truncating a potential
imaginary part\footnote{If the Higgsino mixing parameter $\mu$ and the
lepton trilinear coupling $A_{\ell}$ are real (the case of complex
$\mu$ and $A_{\ell}$ has been extensively discussed in the literature,
see e.g.~\cite{edm}) then there is no bound on the imaginary parts of
the matrices $M_{LC}^2$ and $M_{LV}^2$.}  of $U_{\rm MNS}^{MI}
\left(M_{\nu_\ell}\right)_{MN} U_{\rm MNS}^{NJ}$], the condition above
can be rewritten as:
\bea
|\left(M_{\nu_\ell}\right)_{MN} | = |\left(m_D M^{-1}
m_D^T\right)_{MN} | &\geq& \left| \frac{m_{\chi^0_i}}{32 \pi^2} \:
\re\left[(g_2 Z_N^{2i} - g_1 Z_N^{1i})^2 \: \left(M_{LV}^2\right)_{MN}
  \right] \left(\frac{\Delta B_0}{\Delta m^2}\right)_{iMN}\right|
\nonumber\\[8pt]
&\approx& |\alpha_{MN}\left(M_{LV}^2\right)_{MN}|\;,
\label{eq:lvtrunc}
\eea
with $\alpha_{MN}$ defined in~\eq{eq:a}.

Further estimates depend on the particular choice of the $m_D$ (or
$Y_\nu$) and $M$ and on the neutralino sector parameters.  For
example, using the parameters specified in Table~\ref{tab2}, one has
$\alpha_{MN}\approx \alpha \sim 4\times 10^{-6}~{\rm GeV}^{-1}$, so
that
\bea
|\left(M_{LV}^2\right)_{MN}|\leq 2.5\times 10^5~{\rm GeV}\,
|\left(M_{\nu_\ell}\right)_{MN}| \;.
\label{eq:genlv}
\eea
\Eq{eq:genlv} implies that in the general case one should expect the
entries of the matrix $M_{LV}^2$ to be no more than 5 or 6 orders of
magnitude larger then the typical scales in the effective neutrino
mass matrix; i.e.  of the order of a few MeV$^2$.  Bounds on
$M_{LV}^2$ can be also translated into bounds on $X_\nu$ and $m_B^2$.
{}From~\eq{mlv} one can see that, barring fine tuning, we have
approximate relations $M_{LV}^2\sim M_{\nu_\ell} X_\nu$ or
$M_{LV}^2\sim M_{\nu_\ell} m_B^2/M$.  Thus the rough estimates we made
above suggest that both $X_\nu$ and $m_B^2/M$ should be smaller than
approximately 100 TeV.

Stronger bounds on the matrix elements of $M_{LV}^2$ can be obtained
numerically after assuming some particular form of the MNS matrix.  As
an example, under the assumption of tri-bimaximal mixing
of~\Ref{Perkins} and the parameters given in Table~\ref{tab2},
\begin{eqnarray} \label{bmlv}
M_{LV}^{2} \ \lesssim \ \left (\begin{array}{ccc}  2\times 10^{-9} &
...  & ...   \\ ...  & \ 2\times 10^{-6} & ...  \\ ...  & ...  &
 10^{-5} \end{array} \right ) \;\; {\rm GeV}^{2} \quad,
\end{eqnarray}
where the dots indicate elements with similar bounds as the diagonal
ones.  The significant suppression of the lepton number violating
matrix elements of $M^2_{LV}$ relative to the lepton number conserving
matrix elements $M^2_{LC}\sim\mathcal{O}(v^2)$ is particularly
noteworthy.

\setcounter{equation}{0}
\section{Sneutrino Oscillations}
\label{sec:osc}

The theory behind sneutrino oscillations follows closely the very well
known theory of oscillations in the neutral Kaon-meson system.  The
light sneutrino state [cf.~\eq{nulapp}], 
$\widetilde\nu_\ell\simeq \widetilde\nu_L
-m_D^*M(M^2+m_N^2)^{-1}\widetilde\nu^*_R$ is to leading order in
$vM^{-1}$ the supersymmetric partner of left-handed neutrino
$\nu_L$, and therefore couples to the $W^\pm$ and $Z$ gauge bosons.
For the present discussion, it suffices to approximate:
$\widetilde\nu_\ell^I\simeq\widetilde\nu_L^I$, which we shall denote
simply by $\widetilde\nu_I$ in this Section.  The $\widetilde\nu_I$
can be produced, for example, in $e^+e^-$ annihilation via $s$-channel
$Z$ exchange:
\bea
e^+ + e^- \rightarrow \widetilde{\nu}_I +  \widetilde{\nu}_I^* \;.
\label{eq1}
\eea
When lepton number is conserved, the $\snu_I$ ($\snu_I^*$) possess a
definite lepton number equal to $-1$ ($+1$) and they are produced in
definite flavor eigenstates $I=1, 2, 3$.

It is convenient to introduce a two-dimensional complex vector space
spanned by a basis of vectors consisting of the sneutrinos states of a
given flavor $I$, $\ket{\snu_I}$ and $\ket{\snu_I^*}$.  Two important
operators that act on this state are:
\bea
\hat{L}\equiv\left ( \begin{array}{cc} -1 & \quad 0 \\ \phm 0 &\quad 1
\end{array} \right )\,,\qquad {\rm and} \qquad
CP \equiv \left ( \begin{array}{cc} 0 & \quad 1 \\ 1 &\quad 0
\end{array} \right )\,,
\eea
where $\hat{L}$ is the lepton number operator and $CP$ is the
CP-operator in the $\{\ket{\snu_I}, \ket{\snu_I^*} \}$ basis.  That
is, $\ket{\snu_I}$ and $\ket{\snu_I^*}$ are eigenstates of $\hat{L}$:
\bea \label{lepton}
\hat{L} \ket{\snu_I} = -\ket{\snu_I}\quad , \quad
\hat{L}\ket{\snu_I^*} = + \ket{\snu_I^*} \;,
\eea
and the charge-conjugate parity operator $CP$ transforms particle
states into antiparticle states:
\bea
CP \ket{\snu_I} = \ket{\snu_I^*}\quad , \quad CP
\ket{\snu_I^*} = \ket{\snu_I} \,.
\eea
The eigenstates of CP are given by
\bea
\ket{\snup_I} \equiv \frac{1}{\sqrt{2}} \left ( \ket{\snu_I}+\ket{\snu_I^*}
\right )
\quad , \quad  \ket{\snum_I} \equiv \frac{1}{i\sqrt{2}} ( \ket{\snu_I} -
\ket{\snu_I^*} )\;,
\eea
with definite eigenvalues
\bea
CP \ket{\snup_I } = + \ket{\snup_I} \quad , \quad CP \ket{\snum_I} = -
\ket{\snum_I} \;.
\eea
The CP-even sneutrino state of flavor $I$, $\ket{\snup_I}$, and the
CP-odd sneutrino state of flavor $I$, $\ket{\snum_I}$, are states of
indefinite lepton number.  Of course, these states are the real and
imaginary parts of the sneutrino field of definite lepton number,
\bea
\snu_I = \frac{1}{\sqrt{2}} (\snup_I + i \snum_I) \;.
\eea

Inevitably, in a supersymmetric model with a mechanism that yields
neutrino flavor oscillations, the sneutrino flavor states should
oscillate as well.  The sneutrino mass eigenstates, $S_k, (k=1,2...6)$
are linear combinations of the CP eigenstates $\ket{\snu^{(\pm)}_I}$,
and for a three flavor system ($I=1, 2, 3$) they are related by:
\bea
\ket{\snup_I} = \znu^{Ik} \ket{S_k} \quad , \quad \ket{\snum_I} =
\znu^{(I+3)k} \ket{S_k}
\;,
\eea
where the real orthogonal $6\times 6$ matrix with $\znu^{ij}$ has been
introduced in \eq{eq:znudef}.  The $\ket{S_k}$ are states of definite
CP unless the following CP-violating conditions hold:
\bea
\znu^{I(J+3)}\neq 0 \quad, \quad \znu^{(I+3)J} \ne 0\,,\qquad
I,J=1,2,3\,.
  \label{cpv}
\eea
In the presence of complex parameters in the Lagrangian (whose phases
cannot be absorbed by field redefinition), one expects the conditions
specified in \eq{cpv} to be satisfied (even in the case of a
one-generation model).

Let us initially focus our analysis on the CP-conserving
one-generation model.  Consider the time evolution of the sneutrino
states.  The time dependence of a sneutrino in the state
$\ket{\snupm}$ is governed by a definite frequency $\omega_\pm
=E_\pm/\hbar$ where $E_\pm = (p^2 c^2 + m_\pm^2 c^4)^{1/2}$.  where
$m_+$ and $m_-$ are the masses of $\ket{\snup}$ and $\ket{\snum}$
respectively.  If these masses are large compared to momentum $p$ then
the corresponding energies are $E_\pm \simeq m_\pm c^2$ (in which  
case, $\omega_\pm\simeq m_\pm$ in units where $\hbar=c=1$).  In addition
to the time-dependent phase, we must also account for the fact that
the sneutrinos decay exponentially (e.g. into a chargino and a lepton)
with a lifetime of $\tau_{\pm}$ (for $\widetilde\nu^{\pm}$
respectively).  We exhibit this time dependence explicitly by writing
\bea
\Psi_+(t) = e^{-i \omega_+ t - \frac{t}{2 \tau_+}} \ket{\snup}
\qquad , \qquad
\Psi_{-}(t) = e^{-i \omega_- t - \frac{t}{2 \tau_-}} \ket{\snum}\;,
\eea
where the $\snupm$ are time-independent state vectors, That is,
starting at $t=0$, the probability for finding particle in the
sneutrino state $\snup$ is given by $|\braket{\snup}{\Psi(t)} |^2
=e^{-t/\tau_+}$, as expected.

The well known striking effects of the K-system (e.g., $K$--$\overline
K$ mixing and regeneration) can also occur in the sneutrino system.
For example, we demonstrate how sneutrinos states $\ket{\snu}$ can
turn to states $\ket{\snu^*}$.  If we start off with a sneutrino state
that is $\Psi(0) = \ket{\snu} = \frac{1}{\sqrt{2}}(\ket{\snup}+i
\ket{\snum})$ at $t=0$, then it follows that at time $t$,
\bea
\ket{\Psi(t)} = \frac{1}{\sqrt{2}} \left [ e^{-i \omega_+ t -
\frac{t}{2 \tau_+}} \ket{\snup} + i e^{-i \omega_- t - \frac{t}{2
\tau_-}} \ket{\snum} \right ] \;.   \label{eq:11}
\eea
Then, the probability amplitude that the sneutrino $\ket{\snu}$ is in
state $\ket{\snu^*}$ is
\bea
P_{\snu\rightarrow \snu^*}(t) = |\braket{\snu^*}{\Psi(t)}|^2 =
\frac{1}{4} \left [ e^{-t/\tau_+}+e^{-t/\tau_-} - 2 \:
e^{-\frac{1}{2} \left (\frac{t}{\tau_+}+\frac{t}{\tau_-} \right )}
\cos[(\omega_+ - \omega_-)t] \right ] \;.\label{eq:13}
\eea
The quantum interference effects can only be seen if $t \simeq
\tau_+ \simeq \tau_-$ and $(m_+-m_-) \: t \equiv (\Delta m)t =
\mathcal{O}(1)$.  That is,
\bea
\frac{\Delta m}{\Gsnu} \simeq \mathcal{O}(1)\;, \label{gamma}
\eea
where $\Gsnu$ is an average decay rate for the sneutrino, and $\Delta
m$ is the mass difference of the CP-even and CP-odd sneutrino states.
\Eq{eq:13} describes the oscillations of sneutrinos into
antisneutrinos, or equivalently the oscillation between states of
definite CP quantum number.  We shall call this phenomena CP-driven
oscillations.

Similarly, one may compute the probability that the initial state
$\ket{\snu}$ is in the state $\ket{\snu}$ at time $t$.  We find
\bea
P_{\snu\rightarrow
\snu}(t) = |\braket{\snu}{\Psi(t)}|^2 = \frac{1}{4} \left [
e^{-t/\tau_+}+e^{-t/\tau_-} + 2 \: e^{-\frac{1}{2}
\left (\frac{t}{\tau_+}+\frac{t}{\tau_-} \right )} \cos[(\omega_+ -
\omega_-)t] \right ] \;.   \label{eq:14}
\eea
One can also easily verity that $P_{\snu^*\rightarrow \snu^*} =
P_{\snu\rightarrow \snu}$ and $P_{\snu^*\rightarrow \snu} =
P_{\snu\rightarrow \snu^*}$.  However, the probability
$P_{\snu\rightarrow \snu}$ is proportional to the number of negatively
charged leptons ($N_{l^-}$) due to the decay $\snu \rightarrow l^- +
\chi^+$ while $P_{\snu\rightarrow \snu^*}$ is proportional to the
number of positively charged leptons ($N_{l^+}$) due to the decay
$\snu^* \rightarrow l^+ + \chi^-$.  Then the asymmetry,
\bea
A_{l} = \frac{N_{l^-} - N_{l^+}}{N_{l^-} + N_{l^+}} \;,
\label{eq:asy}
\eea
is proportional to the quantum interference term $\cos(\Delta m \: t)$
in \eqs{eq:13}{eq:14}.  That is, the lepton charge asymmetry $A_l$
oscillates in time and provides a possible method for experimentally
determining the value of $\Delta m$.

The signal for sneutrino--antisneutrino oscillations can be
interpreted as the observation of a sneutrino that decays into a final
state with a ``wrong-sign'' charged lepton.  The phenomenological
implications of such wrong-sign charged lepton final states at future
colliders have been explored recently in Ref.~\cite{huitu}.

We now turn to the three-generation model (allowing for the
possibility of CP-violation) and consider the additional possibility
of flavor metamorphosis.  We pose the following question: Given the
state $\ket{\snu_I}$ at time $t=0$, what is the probability that the
sneutrino at time $t$ is in the state $\ket{\snu_J^*}$ or
$\ket{\snu_J}$?  Following the arguments given above \eq{eq:11}, we
find that a sneutrino wave function involves with time according to
\bea
\ket{\Psi_I(t)} = \frac{1}{\sqrt{2}}\: (\znu^{Ik} + i \znu^{(I+3)k} )
 \: e^{-i \omega_k t - \frac{t}{2 \tau_k}} \: \ket{S_k}\;.
\eea
Hence, the probabilities to be in the state $\ket{\snu_J^*}$ or
$\ket{\snu_J}$ at time $t$ are given by:
\bea
P_{\snu_I \rightarrow \snu^*_J}(t) &=&  P_{\snu^*_I\rightarrow\snu_J}(t)
=\frac{1}{4} \: \sum_{k,s=1}^6
\: e^{-t \left [ \frac{1}{2\tau_k} + \frac{1}{2\tau_s} \right ]} \cos
\left [(\omega_k-\omega_s) t \right ] \: \times \:
\nonumber \\
& & \left ( \znu^{Jk}\znu^{Ik}\znu^{Js}\znu^{Is} + \znu^{(J+3)k}
\znu^{(I+3)k} \znu^{(J+3)s} \znu^{(I+3)s} -2
\znu^{Jk}\znu^{Ik}\znu^{(J+3)s}\znu^{(I+3)s} \right.   \nonumber \\
&+& \left.  \znu^{Jk}\znu^{(I+3)k}\znu^{Js}\znu^{(I+3)s} +
\znu^{(J+3)k}\znu^{Ik}\znu^{(J+3)s}
\znu^{Is} +2 \znu^{Jk}\znu^{(I+3)k}\znu^{(J+3)s}\znu^{Is} \right )\;,
\nonumber \\ \label{eq:flos1} \\[4mm]
P_{\snu_I \rightarrow \snu_J}(t) &=&  P_{\snu^*_I\rightarrow \snu^*_J}(t)
=\frac{1}{4} \:
\sum_{k,s=1}^6 \: e^{-t \left [ \frac{1}{2\tau_k} + \frac{1}{2\tau_s}
\right ]} \cos \left [(\omega_k-\omega_s) t \right ] \: \times \:
\nonumber \\
& & \left ( \znu^{Jk}\znu^{Ik}\znu^{Js}\znu^{Is} + \znu^{(J+3)k}
\znu^{(I+3)k} \znu^{(J+3)s} \znu^{(I+3)s} +2
\znu^{Jk}\znu^{Ik}\znu^{(J+3)s}\znu^{(I+3)s} \right.   \nonumber \\
&+& \left.  \znu^{Jk}\znu^{(I+3)k}\znu^{Js}\znu^{(I+3)s} +
\znu^{(J+3)k}\znu^{Ik}\znu^{(J+3)s}
\znu^{Is} -2 \znu^{Jk}\znu^{(I+3)k}\znu^{(J+3)s}\znu^{Is} \right ) \;.
\nonumber \\
\label{eq:flos2}
\eea
Note that the probabilities in \eqs{eq:flos1}{eq:flos2} are unchanged
under the interchange of flavor indices $I$ and $J$, respectively.
The three-generation model possesses both flavor and CP-driven
oscillations.

In the supersymmetric seesaw model, neutrino mixing and masses are
governed by a variety of parameters that contribute to the tree-level
and one-loop neutrino mass matrix (cf. \sect{sec:mix}).  Some of these
parameters also are relevant for determining the structure of the real
orthogonal sneutrino mixing matrix $\znu^{ij}$, which controls the
properties of the sneutrino mixing as shown above.  Consequently, the
bounds on the model parameters discussed in Sections~\ref{sec:lc}
and~\ref{sec:lv} can be used to significantly constrain the general
form of~\eqs{eq:flos1}{eq:flos2}.

The mass splittings among sneutrinos of different flavors is typically
much larger than the sneutrino--antisneutrino mass splitting between
sneutrino states of a given flavor.  In particular, due to the
renormalization group evolution of parameters, $\Delta m_{{IJ}}^{2}$
is generally larger than few GeV$^{2}$, even in the case of
universality assumptions at the high scale, whereas
sneutrino--antisneutrino mass splittings are typically of order the
light neutrino masses.  The observability of oscillations depends on
the ratio $\Delta m/\Gamma$ [cf.~\eq{gamma}].  Because the total decay
width, $\Gamma$, is universal for a given sneutrino, whereas the
scales of the corresponding mass splittings are so different, it
follows that $\Delta m/\Gamma\sim\mathcal{O}(1)$ can be satisfied only
for one of the two oscillation phenomena.  That is, at most one
oscillation phenomenon, either flavor oscillations {\it or} CP-driven
oscillations, can be observed.

Consider first the CP-driven oscillations.  These oscillations can be
observed if the lifetime of the sneutrinos is sufficiently long (the
appropriate numerical requirements are given later in this section).
In this case, flavor-driven oscillations are much faster and have a
very short ``baseline'', so these oscillations are unobservable in
collider experiments.  Therefore, one can take a time average over
flavor-changing terms in the sums in~\Eqs{eq:flos1}{eq:flos2}, setting
them effectively to zero, and retain only those terms where the mass
splitting is CP-driven and not flavor-driven (i.e. keep only those
terms with $s=k$ or $s=k+3$).  Now, the sum over $s$ can be performed,
and~\eqs{eq:flos1}{eq:flos2} simplify to:
\bea
P_{\snu_I \rightarrow \snu^*_J} &=& \sum_{K=1}^3 \: \left(
e^{-t/\tau_{K_+}} \: \left|X^{IK}X^{JK}\right|^2 +
e^{-t/\tau_{K_-}} \: \left|Y^{IK}Y^{JK}\right|^2
\right)
\nonumber \\
&-& 2 \: \sum_{K=1}^3 \: e^{-t \left [ \frac{1}{2\tau_{K_+}} +
\frac{1}{2\tau_{K_-}} \right ]} \cos \left [\Delta_K t \right ] \:
\re\left(X^{IK}X^{JK}Y^{IK}Y^{JK}\right) \;,
\label{eq:flos3} \\
P_{\snu_I \rightarrow \snu_J} &=& \sum_{K=1}^3 \: \left(
e^{-t/\tau_{K_+}} \: \left|X^{IK}X^{JK}\right|^2 +
e^{-t/\tau_{K_-}} \: \left|Y^{IK}Y^{JK}\right|^2
\right)
\nonumber \\
&+& 2 \: \sum_{K=1}^3 \: e^{-t \left [ \frac{1}{2\tau_{K_+}} +
\frac{1}{2\tau_{K_-}} \right ]} \cos \left [\Delta_K t \right ] \:
\re\left(X^{IK}X^{JK}Y^{IK}Y^{JK}\right) \;, \label{eq:flos3a}
\eea
where $\Delta_K \equiv \omega_K-\omega_{K+3}$ and we have used
\eq{eq:zxy} to express the $6\times 6$ matrices $\znu$
in terms of the $3\times 3$ matrices $X$ and $Y$.

\Eqs{eq:flos3}{eq:flos3a} are easily interpreted.  For ``long
baseline'' oscillations, one needs first to project flavor $I$ onto
some $K$ (via the $X^{IK},Y^{IK}$ factors), then the CP-driven
oscillation takes place between the would-be sneutrino--antisneutrino
states $S_K$ and $S_{K+3}$, and finally the result is projected back
onto flavor $J$.

Further simplification is possible if we exploit the bounds on the
parameters due to the $\ell^{\,J}\ra \ell^{\,I}\gamma$ decays obtained
in \sect{sec:llg} to conclude that the matrix $M_{LC}^2$ is very close
to diagonal form.  In this case, the matrix $Q_0$ that diagonalizes
$M_{LC}^2$ [cf.~\eq{eq:di_def}] is close to the identity matrix.
Moreover, the matrix elements of $R$ [cf.~\eq{eq:xy}] are suppressed
by the ratio of $\Delta m_{\tilde\nu}/ m_{\tilde\nu}$, and are
therefore negligible.  It then follows that $X\simeq Y\simeq
T/\sqrt{2}$, where $T\equiv{\rm diag}(e^{-i\phi_1/2}\, , \,
e^{-i\phi_2/2}\, , \, e^{-i\phi_3/2})$ and $\phi_J\simeq\arg
(M^2_{LV})_{JJ}$ [cf.~\eq{eq:tdef}].  If we consider flavor conserving
(i.e.~$I=J$) sneutrino--antisneutrino oscillations, then there is one
large contribution in \eq{eq:flos3} in the sum over $K$ for $I=K$,
whereas the contributions of $I\neq K$ are strongly suppressed by the
squares of mixing angles.  Therefore, the dominant contribution to the
probability for sneutrino--antisneutrino oscillations is given by:
\bea
P_{\snu_I \rightarrow \snu^*_I} &\approx & \frac{1}{4} \left[
e^{-t/\tau_{I_+}} + e^{-t/\tau_{I_-}} - 2 \: e^{-t
\left [ \frac{1}{2\tau_{I_+}} + \frac{1}{2\tau_{I_-}} \right ]}
\cos(\Delta_I t) \,\cos (2\phi_I) \right] \;,
\label{eq:flos4}
\eea
which coincides exactly with the formula obtained previously for the
one generation case [cf.~\eq{eq:13}] in the CP-conserving limit (where
$M^2_{LV}$ is a real matrix so that $\cos 2\phi_I=1$).  Similarly, for
$P_{\snu_I \rightarrow \snu_I}$, one reproduces~\eq{eq:14} in the same
limiting case.

To complete the analysis of the sneutrino oscillation formulae, we
must compute the total sneutrino decay width, $\Gamma_k\equiv
\Gamma(S_{k}\ra {\rm anything})=1/\tau_{S_{k}}$.  Supposing that the
neutralino is the lightest supersymmetric particle (LSP), the
sneutrino decay width is the sum of the partial widths of the
following two kinematically available decay
chains,\footnote{$\Gamma(S_k\to\ell^{\mp\,I}+\chi_i^\pm)$ indicates
the sum of the sneutrino partial widths to the lepton--chargino and
its charge-conjugated final states.}
\begin{Eqnarray}
&&\!\!\!\!\!\!\!
\Gamma(S_k \rightarrow \ell^{\mp\,I}+\chi_i^\pm) = g_{2}^{2} \,
\frac{m_{S_k}}{32 \pi} \left(1-\frac{m_{\chi_i}^2}{m_{S_k}^2} \right)^{3/2}
 |Z_+^{1i}|^2 \left(|\znu^{Ik}|^2 + |\znu^{(I+3)k}|^2
\right)
 \;,\label{gam1}
\\
&&\!\!\!\!\!\!\!
\Gamma(S_k \rightarrow \nu^I+\chi^0_i) = \frac{g_{2}^{2}}{c_W^2}
\frac{m_{S_k}}{64 \pi} \left(1-\frac{m_{\chi_i^0}^2}{m_{S_k}^2} \right)^{3/2}
 |Z_N^{1i} s_W - Z_N^{2i} c_W|^2
\sum_{J=1}^3 \left| (\znu^{Jk}-i\znu^{(J+3)k})U_{\rm MNS}^{JI}
\right|^2\;.   \nonumber \\ \label{gam2}
\end{Eqnarray}%
In deriving the formulae above, we have used the Feynman Rules
\eqs{B1}{B4} from~\ref{app:feyrul} and have taken the lepton masses to
zero.  \Eqs{gam1}{gam2} agree with Ref.\cite{Howie} in the limit
$U_{\rm MNS} = \znu = 1$.
Writing $\mathcal{Z}_{\tilde\nu}$ in terms of $X$ and $Y$
[cf. \eq{eq:zxy}], it easily follows that the decay rates of the
sneutrinos $S_k$ with $k=1,2,3$ [$k=4,5,6$] depend on $X$ [$Y$] alone.
Since $X$ and $Y$ differ only by the ``small'' $R$ matrix
[cf. \eq{eq:xyqr}], it follows that $\tau_{I_+}\simeq\tau_{I_-}$,
which can be used to further simplify the expression given by
\eq{eq:flos4}.

The total sneutrino decay width is given by:
\begin{eqnarray}
\Gamma_k &=& \sum_{I=1}^3\sum_{i=1}^2
\Gamma(S_k \rightarrow \ell^{\mp\,I}+\chi_i^\pm)
+ \sum_{I=1}^3\sum_{i=1}^4 \Gamma(S_k \rightarrow \nu^I+\chi^0_i)
\nonumber\\
&=& g_{2}^{2} \, \frac{m_{S_k}}{32 \pi} \left[
\sum_{i=1}^2 \left(1-\frac{m_{\chi_i}^2}{m_{S_k}^2} \right)^{3/2}
|Z_+^{1i}|^2 + \frac{1}{2c_W^2} \sum_{i=1}^4
\left(1-\frac{m_{\chi_i^0}^2}{m_{S_k}^2}
\right)^{3/2} |Z_N^{1i} s_W - Z_N^{2i} c_W|^2 \right] \;,\nonumber\\
\label{gtot}
\end{eqnarray}
where the summation over the lepton indices can be performed in the
limit of vanishing lepton masses, with the use of the orthogonality
[unitarity] relations for the matrices $\znu$ [$U_{\rm MNS}$].

How can one observe sneutrino CP-oscillations?  Consider the following
scenario: suppose that the LHC finds sneutrinos with masses that are
accessible at a future International Linear Collider (ILC).  Then, at
the ILC, the sneutrinos are produced through the annihilation process
of \eq{eq1}, and subsequently decay into [leptons + charginos] and
[neutrinos + neutralinos] following the decay widths given by
\eqs{gam1}{gam2}, respectively.
Sneutrino CP-oscillations will then be observed only if the asymmetry
$A_l$ defined in \eq{eq:asy}, is appreciable, {\it i.e.,} $A_l \sim
{\mathcal O}(1)$, which can be realized if both $\Delta m_k$ is small
(providing a long enough oscillation base) and the sneutrino decay
rate is sufficiently slow such that $\Delta m_k/\Gamma_k\sim
\mathcal{O}(1)$.  This scenario is impossible if the sneutrinos are
sufficiently heavy compared to the neutralinos and/or charginos, in
which case (neglecting the phase space suppression in \eq{gtot} and
performing the summation over the chargino and neutralino indices) the
sneutrino decay rate is approximately given by:
\begin{eqnarray}
\Gamma_k &\approx&  g_{2}^{2} \, \frac{m_{S_k}}{32 \pi} \left[
\sum_{i=1}^2  |Z_+^{1i}|^2 + \frac{1}{2c_W^2} \sum_{i=1}^4
|Z_N^{1i} s_W - Z_N^{2i} c_W|^2 \right] = g_{2}^{2} \,
\frac{m_{S_k}}{32 \pi} \left( 1 + \frac{1}{2c_W^2} \right)
\;.\nonumber\\
\label{gtot1}
\end{eqnarray}
The expression above depends only on the sneutrino mass and cannot be
suppressed by a particular choice of mixing angles of the $\znu, Z_+$
or $Z_N$ matrices.  Thus, using the results of Section~\ref{sec:lv},
one can check that the ratio $\Delta m_k/\Gamma_k$ is always much too
small for the sneutrino oscillations to be observed.  As an example,
in the case of universal parameters discussed in
Section~\ref{sec:sugra}, for the lightest sneutrino and $m_{0}$,
$|A_{0}| \lesssim 10^{5}$ GeV we obtain
\begin{eqnarray}
 \frac{\Delta m_{S}}{\Gamma_{S}} \lesssim 2.7 \times 10^{{-6}} \;,
\end{eqnarray}
which is very far from the value $\mathcal{O}(1)$ required for the
observability of sneutrino oscillations.

In the case of 2-body decays, the decay width $\Gamma_k$ can be only
suppressed by choosing an appropriate hierarchy of particle masses.
Most of the decay channels in \Eqs{gam1}{gam2} would have to be closed
kinematically, with the open channels strongly suppressed either by
the very small phase space factors (which requires rather unnatural
degeneracy between sneutrino and neutralino or chargino masses), or by
sufficiently small mixing angles for the relevant channel.  An
alternative possibility is one where the sneutrinos are lighter then
all charginos and neutralinos, so that all 2-body decay channels are
closed, but heavier than some charged slepton.  In this case,
$\widetilde\nu\to\widetilde\ell^{\pm}W^{\mp}$, and assuming that the
$W$ is produced off-shell the end result is a 3-body decays that can
produce an observable charged lepton.  Three-body phase space
significantly suppresses the sneutrino decay rate (relative to the
two-body decay rates discussed above), and can yield observable
sneutrino--antisneutrino oscillations, as shown in ref.~\cite{Howie}.
However in such a scenario, either the charged slepton is the LSP,
which is strongly disfavored by astrophysical data, or the charged
slepton decays to some new lighter supersymmetric particle, which
requires extending the model beyond the seesaw-extended MSSM
considered in this paper~\cite{kchoi}.
As we have shown, the oscillations in the
three-generation case does not differ much from the one-generation
case, where the flavor indices are summed over
[cf.~\eqs{eq:flos4}{gtot}].  Thus, the results of ref.~\cite{Howie}
can also be used without significant changes in the three-generation
case discussed in this paper.

Finally, we discuss the case of sneutrino {\it flavor} oscillations.
These oscillations are described by~\eqs{eq:flos1}{eq:flos2} with
indices $I\neq J$.  For any choice of $I\neq J$, both equations can be
significantly simplified using the bounds on the structure of
sneutrino mixing matrices derived in Sections~\ref{sec:lc} and
\ref{sec:lv}.  These bounds imply that the off-diagonal elements of
matrices $Q$ and $R$ [defined in~\eqs{eq:qdef}{eq:rdef}] are small,
which then imply [via \eqs{eq:xyqr}{eq:zxy}] that the off-diagonal
elements of the matrices $X$, $Y$ and $\znu$ are likewise small.
Thus, to a good approximation one can keep in~\eqs{eq:flos1}{eq:flos2}
only terms at most quadratic in the non-diagonal elements of $\znu$.
For example, in the sum of the first term of the product of four
$\mathcal{Z}_{\tilde\nu}$'s in~\eq{eq:flos1}, it is sufficient to keep
only terms with $s,k=I,I+3,J,J+3$.  Assuming that the lifetimes of all
eigenstates are very similar (i.e., $\tau \simeq \tau_{k}$), all the
dominant terms can be summed to give a simple final expression valid
for $I\ne J$ ,
\bea
\!\!\!\!\!\!\!\!
P_{\snu_I \rightarrow \snu_J} &\approx& e^{-\frac{t}{\tau}} \:
\biggl\{ |Q^{IJ}Q^{JJ*}|^2 + |Q^{JI}Q^{II*}|^2 + 2 \re
\left(Q^{IJ}Q^{JJ*}Q^{JI*}Q^{II}\right) \cos\Delta m_{IJ}t \biggr\}\,,
\label{eq:flav2}
\eea
where $\Delta m_{IJ} \equiv m_{\snu_I} - m_{\snu_J}$.

The analogous expression for the sneutrino-antisneutrino oscillation
probability $P_{\snu_I \rightarrow \snu^*_J}$ is bilinear in the
matrix elements of $R$ [cf.~\eq{eq:rdef}].  The latter are at most of
${\cal O}(10^5m_\nu/v)\lsim 10^{-6}$ and thus lead to completely
negligible sneutrino--antisneutrino transition rates.\footnote{An
accurate estimate of $P_{\snu_I \rightarrow \snu^*_J}$ should also
take into account similarly small effects produced by the admixture of
the heavy sneutrino states in the definition of the $\snu_I$, which
were neglected in derivation of~\eqs{eq:flos1}{eq:flos2}. However,
given the extremely small transition probabilities, we do not present
the full analysis here.}

The form of \eq{eq:flav2} is explicitly invariant with respect to
rephasing, $Q^{IJ}\ra Q^{IJ}e^{i\phi_J}$.  Thus, without loss
of generality, we may replace $Q$ by $Q_0$ [cf.~\eq{eq:qdef}] in
\eq{eq:flav2}, where the off-diagonal matrix elements of the unitary matrix
$Q_0$ are given approximately by \eq{qzero} and the diagonal elements
of $Q_0$ are fixed by unitarity.  As $Q_0$ is close to the identity
matrix, the following approximations are valid: $Q_0^{JJ}\simeq 1$ and
$Q_0^{JI*} \simeq - Q_0^{IJ}$ for $I\neq J$.  In this approximation,
\eq{eq:flav2} simplifies for $I\neq J$ to:
\bea
P_{\snu_I \rightarrow \snu_J} &\approx& 2 e^{-t/\tau} \:
\left[ |Q_0^{IJ}|^2 - \re (Q_0^{IJ})^2 \cos\Delta m_{IJ}t \right]\,
.  \label{eq:flav2a}
\eea
If one uses the approximate expression given in \eq{qzero},
$Q_0^{IJ}\simeq (M^2_{LC})^{IJ}/(m_{\snu_J}^2 - m_{\snu_I}^2)$,
then~\eq{eq:flav2a} yields the oscillation probabilities directly in
terms of the sneutrino squared-mass matrix elements.  As expected, the
sneutrino flavor-transition depends on the flavor-conserving matrix
$M^2_{LC}$.

Defining the oscillation length by $L = c t$ we can write
\begin{eqnarray}
\Delta m_{IJ} t   \ = \ 5.06 \times \, \Delta m_{IJ}\, ({\rm GeV})\,
 L({\rm fm}) \;.
\label{Dks}
\end{eqnarray}
As in neutrino oscillations, it is useful to define $\Delta m_{IJ}\, L
= 2\pi L/L_{0}$ where $L_{0}$ is the characteristic length of the
oscillation~:
\begin{eqnarray}
 L_{0} \ = \ 1.24 \, {\rm fm} \times \frac{1}{\Delta m_{IJ}~({\rm GeV}) } \;.
\end{eqnarray}
If the sneutrino mass difference is of $\mathcal{O}$(1~GeV), the
characteristic oscillation length is of order 1~fm.  Of course, the
characteristic length of oscillation must be smaller than or at most
comparable to the decay length of the particle for oscillations to be
observable.  In the case of the sneutrino, the decay length is [using
\eq{gtot1}]:
\begin{eqnarray}
L_{\tilde{\nu}} = c \tau \simeq \frac{28~({\rm
fm})}{m_{\tilde{\nu}}~({\rm GeV})} \;.
\end{eqnarray}
Hence, the condition $L_{\tilde{\nu}} \gsim L_{0}$ requires that
\begin{eqnarray}
\frac{\Delta m_{IJ}}{m_{\tilde{\nu}}} \gtrsim \frac{1}{25} \;.
\end{eqnarray}
Such a mass splitting between the sneutrino states of different
flavors is sensible.  Thus, the likelihood of observing flavor
sneutrino oscillations at colliders depends primarily on the degree of
suppression caused by the mixing angles in the matrix $Q$.
It is instructive to input some representative numbers
in~\eq{eq:flav2}.  Thus, for $\Delta m_{12} = 10~{\rm GeV},
m_{\tilde{\nu}} = 270~\mathrm{GeV}$, $\tan\beta=10$ and taking into
account the bounds of Table~\ref{tab:mlc}, we obtain for
$\tilde\nu_{\mu} \ra \tilde\nu_{e}$ oscillations at time
$t=\tau=\Gamma^{-1}$ [cf. \eq{gtot}]:
\begin{eqnarray}
P_{\snu_\mu \rightarrow \snu_e} &\approx& 1.25 \times
10^{-5} \ [1 \ - \ \cos(\Delta m_{12}\tau) ] \;,
\end{eqnarray}
Thus, as a consequence of the bounds from neutrino masses and
radiative flavor changing decays obtained in Sections~\ref{sec:lc}
and~\ref{sec:lv}, we conclude that \textit{in the see-saw extended
MSSM, sneutrino flavor oscillations are difficult to observe at
colliders}.

If the bounds of Sections~\ref{sec:lc} and \ref{sec:lv} could be
avoided, say with some cancellation mechanism (which in the absence of
such a mechanism would appear unnatural), then it may be possible to
find regions of the supersymmetric parameter space where flavor
oscillations are observable.
Then, at the ILC, one can define a flavor asymmetry for the number of
muons vs. electrons in the final state, analogous to
\eq{eq:asy}.   A time-variation of this flavor asymmetry would indicate
the presence of flavor oscillations.

\setcounter{equation}{0}
\section{Conclusions}
\label{sec:conc}

In this paper, we have studied sneutrino mixing phenomena in the
seesaw-extended MSSM, allowing for the full complexity of the
three-generation model (which includes both flavor-changing and
CP-violating effects).  We have focused primarily on the
soft-SUSY-breaking matrix parameters $m_{N}^{2}, m_{B}^{2}$ and
$A_{\nu}$, which govern the structure of the sneutrino squared-mass
matrices.
We have found a convenient parameterization of the sneutrino sector,
where all relevant physical observables depend analytically on a pair
of $3\times 3$ mass matrices $M_{LV}^{2}$ and $M_{LC}^{2}$ given in
\eqs{mlv}{mlc}, respectively.  The elements of $M_{LV}^{2}$ violate
lepton number by two units, whereas elements of $M_{LC}^{2}$ are
lepton-number conserving parameters.

Within this framework, we have analyzed the constraints arising from
one-loop neutrino masses and mixings, from radiative flavor-changing
charged lepton decays, and from the electron electric dipole moment
(EDM).  We discovered new and potentially significant contributions to
radiative lepton decays $\ell^{\,J} \ra \ell^{\,I} + \gamma$ due to
the dependence of $m_N^2$ which modifies the MSSM value of
$M_{LC}^{2}$.  We also observed that although the $(g-2)_{\mu}$
measurement places non-trivial constraints on the SUSY-breaking
parameters, the electron EDMs do not yield any additional constraints
(at one loop) on the seesaw-extended MSSM parameters.  All conclusions
presented here are based on a complete numerical analysis of the
processes described above.\footnote{Fortran-77 and Maple-10 numerical
codes are available from the authors.}  In all cases, we have also
provided useful analytic approximations, which have served as a check
of our numerical work.

Sneutrino mixing phenomena takes on two different forms. The mixing of
sneutrinos and antisneutrinos violates lepton number by two units,
whereas sneutrino flavor mixing is a lepton-number conserving process.
Both forms of mixing are in present in principle in the
three-generation seesaw-extended MSSM.  In this paper, we have
generalized the sneutrino-antisneutrino mixing formalism, originally
presented in a one-generation model \cite{Howie}, to the
three-generation model.  This sneutrino-antisneutrino mixing then acts
back on the neutrino sector, and provides an important loop correction
to the neutrino mass matrix. In this paper, we examined the
possibility that starting from a diagonal neutrino mass matrix at
tree-level, the nontrivial flavor structure of the neutrino mass
matrix is generated entirely by the one-loop diagram that directly
involves the sneutrino--antisneutrino transition. Our analysis shows
that this is indeed possible, although in practice certain
fine-tunings among SUSY breaking parameters in the leptonic sector
seem to be unavoidable.

Returning to the sneutrino sector, we have derived analytical
expressions for both sneutrino-flavor oscillations and
sneutrino-antisneutrino oscillations in \eqs{eq:flos1}{eq:flos2}.  We
determined that if the constraints analyzed above are combined with
the assumption that sneutrinos can decay into two-body final states,
then sneutrino-antisneutrino oscillations are not observable at
colliders.  This is consistent with a similar result of the
one-generation model obtained in Ref.\cite{Howie}. This conclusion is
easily understood, by noting that the sneutrino-antisneutrino mass
difference, $\Delta m_{\tilde\nu}$, is proportional to the neutrino
mass and is at most of the order of 1 keV.  This is much smaller than
the corresponding width of the sneutrino, $\Gamma_{\tilde\nu}$, of
order 1 GeV or larger.  The observability of sneutrino-antisneutrino
oscillations at colliders requires that $\Delta m_{\tilde\nu}
\sim\Gamma_{\tilde\nu}$.  A sneutrino width of order 1 keV or less is
possible only if there are no kinematically allowed two-body final
states in sneutrino decay. In the seesaw-extended MSSM, this scenario
is possible only if a charged slepton is the lightest supersymmetric
particle, a possibility strongly disfavored by astrophysical data.
Other possibilities exist if one introduces new degrees of freedom
beyond the seesaw-extended MSSM, but this lies beyond the scope of
this paper.

Sneutrino flavor oscillations are more likely to be observable at
colliders, since the mass splitting between sneutrinos of difference
flavors can be of order 1 GeV or larger.  We have derived simple
approximate formulae for such oscillations and have estimated their
magnitudes.  Unfortunately, in the seesaw-extended MSSM, after
imposing bounds on bounding sneutrino mixing angles determined from
the analysis of radiative charged lepton decays, the resulting
probabilities for sneutrino flavor oscillations are likely to be too
small to be observed directly at colliders.

At present, within the seesaw framework for neutrino masses, few
handles exist for probing the physics at the seesaw scale.  At most,
one can hope to measure the MNS mixing angles, and determine neutrino
mass differences (and with a little luck, the absolute scale of
neutrino masses).  In the seesaw-extended MSSM, some of the physics of
the seesaw scale is imprinted on parameters that govern the properties
of the light sneutrinos.  With a precision program at future colliders
for measuring sneutrino observables, there are new opportunities to
explore the fundamental physics that is responsible for the origin of
neutrino masses.

\bigskip

\subsection*{Acknowledgements}

A.D. would like to thank the theory group at the Santa Cruz Institute
for Particle Physics for their kind hospitality.  A.D. acknowledges
K.~Sridhar and R.~Zwicky for enlightening comments and suggestions.
H.E.H. would like to thank the Institute for Particle Physics
Phenomenology for their kind hospitality during his visit to Durham.
H.E.H. is also grateful to the Aspen Center for Physics 
and the CERN Theory Group for providing
the time and space necessary for completing this project.  
In particular, conversations with Kiwoon Choi, Aaron Pierce, 
Elizabeth Simmons and Fabio Zwirner at CERN
were especially illuminating in contributing to our understanding
of the non-decoupling phenomenon
described in \ref{app:nondecoupling}.  J.R. is
especially grateful for a number of useful discussions with Borut
Bajc, Goran Senjanovic, Bohdan Grzadkowski and Jack Gunion during the
initial phases of this work.

A.D. and J.R. are partially supported by the RTN European Programme, 
MRTN-CT-2006-035505 (HEPTOOLS, Tools and Precision Calculations 
for Physics Discoveries at Colliders).
A.D also acknowledges support by the European Commission
for his participation in MRTN-CT-2006-035863-1
(UniverseNet).  H.E.H. is supported in part by U.S.  Department of
Energy grant number DE-FG02-04ER41268.  J.R. is also supported in part by
the Polish Ministry of Science and Higher Education Grant
No~1~P03B~108~30 for the years 2006-2008 and by the EC Project
MTKD-CT-2005-029466 (Particle Physics and Cosmology: the Interface).



\renewcommand{\thesection}{Appendix~\Alph{section}}
\renewcommand{\theequation}{\Alph{section}.\arabic{equation}}

\setcounter{equation}{0}  
\setcounter{section}{0}
\bigskip

\section{Notation for fermion fields}
\label{app:fermion}

Fermion fields in quantum field theory can be described by employing
either two-component or four-component fermion notation~\cite{dhm}.
In models where lepton number is not conserved, two-component fermion
notation is generally simpler and more efficient.  In this appendix,
we briefly discuss the relation between the two treatments.

In Table~\ref{tab:fields}, the fermionic fields associated with the
lepton and Higgs sectors of the seesaw-extended MSSM are listed.
These fermion fields can be viewed either as two-component fermion
fields or the left-handed projections of four-component fermion
fields, with $\Psi_L\equiv \half(1-\gamma\ls{5})\Psi$ and
\beq \label{ccpsi}
\Psi^c\equiv C{\overline\Psi}^{T}\,,\qquad \overline{\Psi^c}=-\Psi^T C^{-1}\,,
\eeq
where $\overline\Psi\equiv\Psi^\dagger\gamma^0$ and $C=-C^T$ is the
charge conjugation matrix.

For example, in four-component notation, given a four-component
(anticommuting) Dirac spinor $\nu_D$, we define the following
four-component spinors:
\beq
\nu_L\equiv P_L\nu_D\,, \qquad \nu_L^c\equiv
P_L\nu_D^c\,,\qquad \nu_R\equiv P_R\nu_D\,,
\quad {\rm and} \quad \nu_R^c\equiv P_R\nu_D^c\,,
\eeq
where $P_{L,R}\equiv\half(1\mp\gamma\ls{5})$, respectively.  The
corresponding two-component (anticommuting) fields are given by the
non-zero components of $\nu_L\equiv P_L\nu_D$ and $\nu_L^c\equiv
P_L\nu^c_D$.  Consequently, we shall use the same symbols $\nu_L$ and
$\nu_L^c$ for the corresponding two-component neutrino fields.
However, one must be careful to note that in our notation
\beq \label{clr}
\nu_L^c=C\overline{\nu_R}^T\,,\qquad \overline{\nu^c_R}
=-\nu_L^T C^{-1}\,,
\eeq
since, e.g., $\nu_L^c\equiv P_LC \overline \nu_D^T =
C(\overline{P_R\nu_D})^T$.  The same notation also applies to charged
fermion fields.  Our conventions for left and right-handed charged
conjugated fields follow those of~\Ref{langacker}.  Note that \eq{clr}
implies that anticommuting fermion fields satisfy:
\beq \label{nulr}
\overline{\nu_R^c}\nu_L^c=\overline{\nu_R}\nu_L\,,
\qquad
\overline{\nu_L^c}\nu_R^c=
\overline{\nu_L}\nu_R\,.
\eeq

In the text, the effective Lagrangians for fermion mass and
interaction terms are given in terms of two-component fermion fields.
These terms can be easily translated into the four-component spinor
notation .  As a first example, the dimension-five operator that
governs the standard seesaw mechanism [\eq{dim5}] contains a product
of two-component fermion fields, $L_i^I L_k^K$.  In terms of
four-component spinors, this product is given by $-(L^T)_i^I
C^{-1}L_k^K=(\overline{R^c})_i^I L_k^K$, where
$L_k^K\equiv(\nu_L^K\,,\,\ell^{\,K}_L)$ is now interpreted as a
doublet of four-component fermion fields as described above and
$(R^c)_i^I\equiv (\nu^{cI}_R\,,\,\ell^{\,cI}_R)$.

As a second example, we derive the four component version
of~\eq{eq:mterms} in the one-generation model.  One can redefine the
phases of the neutrino fields such that $m_D$ and $M$ are real and
non-negative.  The two-component spinor product $\nu_L\nu^c_L+{\rm
H.c.}$ translates to the product of four-component spinors: $-\nu_L^T
C^{-1}\nu_L^c+{\rm H.c.}=\overline{\nu_R}\nu_L
+\overline{\nu_L}\nu_R$, which is the usual Dirac mass term.
Similarly, the two-component spinor product $\nu_L^c\nu_L^c$
translates to the four-component spinor product
$-\nu_L^{c\,T}C^{-1}\nu_L^c=\overline{\nu_R}\nu_L^c$.  Hence, if the
Majorana mass term $M\neq 0$ in \eq{eq:mterms}, one cannot identify
the physical mass eigenstates as Dirac fermions.  For example, the
mass terms of the one-generation neutrino Lagrangian, which in terms
of two-component fermion fields is given by $-\mathscr{L}_{\rm
mass}=m_D\nu_L\nu_L^c +\half M\nu_L^c\nu_L^c+{\rm H.c.}$, translates
in four-component notation to
\bea
\hspace{-1in} -\mathscr{L}_{\rm mass}&=&\half
m_D(\overline{\nu_L}\nu_R+\overline{\nu_R}\nu_L
+\overline{\nu^c_L}\nu_R^c+\overline{\nu_R^c}\nu_L^c) +\half
M(\overline{\nu_R}\nu_L^c+\overline{\nu_L^c}\nu_R) \nonumber \\[8pt]
&=&\half\left(\begin{array}{cc}\overline{\nu_R^c} &
\overline{\nu_R}\end{array}\right)\,
\left(\begin{array}{cc} 0 & m_D\\ m_D & M
\end{array}\right)\,\left(\begin{array}{c} \nu_L \\ \nu_L^c\end{array}\right)
+ \half\left(\begin{array}{cc}\overline{\nu_L} &
\overline{\nu_L^c}\end{array}\right)\,\left(\begin{array}{cc} 0 & m_D\\ m_D & M
\end{array}\right)\,\left(\begin{array}{c} \nu_R^c \\ \nu_R\end{array}\right)
\nonumber \\[8pt]
&=&-\half\left(\begin{array}{cc}{\nu_L^T} &
{\nu_L^{c\,T}}\end{array}\right)\,C^{-1}
\left(\begin{array}{cc} 0 & m_D\\ m_D & M
\end{array}\right)\,\left(\begin{array}{c} \nu_L \\ \nu_L^c\end{array}\right)
+{\rm H.c.}\,,
\label{fn4}
\eea
where we have used \eq{nulr} to write the first line of \eq{fn4} in a
symmetrical fashion and \eq{clr} to obtain the final form above.

The Takagi-diagonalization of the neutrino mass matrix yields two
(self-conjugate) Majorana fermion mass-eigenstates.  This is
accomplished by introducing a unitary matrix~$\mathcal{U}$,
\beq
\left(\begin{array}{c} \nu_L \\ \nu_L^c\end{array}\right)=\mathcal{U}
\left(\begin{array}{c} P_L \nu_\ell \\ P_L\nu_h^c\end{array}\right)\,,
\eeq
such that
\beq
\mathcal{U}^T \left(\begin{array}{cc} 0 & m_D\\ m_D & M
\end{array}\right)\,\mathcal{U}=\left(\begin{array}{cc} m_{\nu_\ell} &
  0 \\ 0 & m_{\nu_h}\end{array}\right)\,,
\eeq
where $m_{\nu_\ell}\simeq m_D^2/M$ and $m_{\nu_h}\simeq M+m_D^2/M$.
The resulting neutrino mass Lagrangian is:
\beq \label{majmass}
-\mathscr{L}_{\rm mass}=-\half \left[m_{\nu_\ell}\nu_\ell^T C^{-1}P_L\nu_\ell
 +m_{\nu_h}\nu_h^{c\,T} C^{-1}P_L \nu_h^c\right]+{\rm H.c.}
\eeq
We can define four-component self-conjugate Majorana fields by:
\bea
\psi_M &\equiv &
P_L\nu_\ell+P_RC\overline\nu_\ell^T\,,
\qquad\quad\,\,\,
\overline\psi_M\equiv \overline\nu_\ell P_R
-\nu_\ell^T C^{-1}P_L\,, \label{lightmaj} \\
\Psi_M &\equiv &
P_L\nu_h^c+P_R C\overline\nu_h^{c\,T}\,,
\qquad\quad
\overline\Psi_M\equiv\overline\nu_h^c P_R
-\nu_h^{c\,T} C^{-1}P_L\,.\label{heavymaj}
\eea
Thus, \eq{majmass} reduces to the expected form:
\beq
-\mathscr{L}_{\rm mass} =
\half\left[m_{\nu_\ell}\overline\psi_M\psi\ls{M}
+m_{\nu_h}\overline{\Psi}_M\Psi_M\right]\,.
\eeq
\medskip

\setcounter{equation}{0}  
\section{A non-decoupling contribution to sneutrino masses when
$\boldsymbol{m_N^2\sim\mathcal{O}(M^2)}$}
\label{app:nondecoupling}

\noindent
{\bf B.1~~Non-decoupling effects when $\boldsymbol{m_N^2\gg v^2}$}

In \sect{sec:snumass}, we noted below \eq{mlv} 
 non-decoupling in the limit of $\|M\|\to\infty$ with
$\|m_N^2 M^{-2}\|$ fixed.  The lepton-number conserving 
$3\times 3$ squared-mass matrix of the light sneutrinos [\eq{mlc}]
can be written as:
\beq
M^2_{LC}= m_L^2+\half M_Z^2\cos 2\beta+m_D^*M^{-1}m_N^2 M^{-1}m_D^T
+\mathcal{O}(v^4 M^{-2})+\mathcal{O}(v^2 m_N^4 M^{-4})\,,
\eeq
after expanding the quantity
$(\mathds{1}+M^{-2}m_N^2)^{-1}$ under the assumption that
$\|M^{-2}m_N^2\|<1$.
Thus, we have a non-decoupling correction to the usual MSSM result 
of $\mathcal{O}(m_N^2 M^{-2})$ as previously noted.

To understand the origin of this non-decoupling phenomenon, we use
\eq{nurapp} which relates the
original right-handed sneutrino with the light and heavy sneutrino
states after block diagonalization of the sneutrino mass matrix.  To
formally integrate out the heavy sector and obtain the effective
theory of the light sneutrinos, we must write: 
\beq \label{e2} \wt N^{I}=\snu^I_h -
\epsilon_{kn} [(M^2+m_N^2)^{-1}MY^T_\nu]^{IJ} \wt L^{J}_n H^2_k\,,
\eeq 
before electroweak symmetry breaking, where we have used $\wt
N^I\equiv \snu_R^{I\,*}$.  Note that when $H_2^2$ is replaced by its
vacuum expectation value
$v_2/\sqrt{2}$, we recover \eq{nurapp} after using $m_D\equiv v_2
Y_\nu/\sqrt{2}$.  In addition, we have used $\wt L^J_1\simeq\snu^J_\ell
+\mathcal{O}(vM^{-1})$ and have worked consistently to leading order
in $vM^{-1}$.

Consider the contribution of $|dW/dN^J|^2$ to the scalar potential,
where $W$ is given by \eq{superpot}.  Then,
\beq
\frac{dW}{dN^J}=M^{JK}N^K+\epsilon_{ij} Y_\nu^{KJ} H^2_i L^K_j\,.
\eeq
After squaring, and including the soft-SUSY-breaking term
$\wt N^* m_N^2 \wt N$ (where $m_N^2$ is hermitian), we find:
\bea \label{e3}
\wt N^* m_N^2 \wt N &+&
\left(\frac{dW}{dN^J}\right)\left(\frac{dW}{dN^J}\right)^*=
\epsilon_{ij}\epsilon_{kn}Y_\nu^{KJ}Y_\nu^{IJ\,*}
H_i^2 H_k^{2\,*}\wt L_j^K \wt L_n^{I\,*}\nonumber \\[7pt]
&&\qquad +
\left[\epsilon_{ij}(Y_\nu M)^{KI} \wt N^{I\,*} 
H^{2}_i \wt L_j^{K}+\Hc\right]+(M^2+m_N^2)^{KJ}\wt N^{K\,*}\wt N^J\,.
\eea
To obtain the relevant operator that
survives in the low-energy effective theory, we insert \eq{e2} for
$\wt N^I$ in \eq{e3}, and then take the limit as $\|M\|\to\infty$,
In addition, we set $\snu_h=0$.  The end result is:
\beq \label{e4}
\epsilon_{kn}\epsilon_{ij}\left[Y^*_\nu Y_\nu^T
-Y^*_\nu M (M^2+m_N^2)^{-1}MY^T_\nu\right]^{JK}  
\wt L^{J\,*}_n \wt L_j^{K}H^{2\,*}_kH^{2}_i \,.
\eeq
Note that this is a dimension-4 (hard) SUSY-violating 
operator~\cite{spm} which vanishes
if $m_N^2=0$ [as $m_N^2$ is the only SUSY-breaking source
in \eq{e4}].  If $m_N^2<M^2$, one can
expand $(M^2+m_N^2)^{-1}$ in \eq{e4}, which yields:
\beq \label{e5}
\epsilon_{kn}\epsilon_{ij} [Y_\nu^* M^{-1}m_N^2 M^{-1}Y_\nu^T
+\mathcal{O}(m_N^4 M^{-4})]^{JK}
\wt L^{J\,*}_n \wt L_j^{K}H^{2\,*}_kH^{2}_i \,.
\eeq
We now replace $H_2^2\to v_2/\sqrt{2}$.  If $m_N^2\sim\mathcal{O}(v^2)$,
then the hard SUSY-breaking operator is of $\mathcal{O}(v^2 M^{-2})$,
which is the expected result.  Such corrections are extremely small, 
assuming that $v\ll \|M\|$, and can be be dropped from the low-energy
effective field theory of the light $\mathcal{O}(v)$ degrees of
freedom.   On the other hand, if $x\equiv \|m_N^2\|/\|M^2\|$
is held fixed to a finite positive value
as $M\to\infty$, then the hard SUSY-breaking
operator is of $\mathcal{O}(x)$, which must be kept in the low-energy
effective theory if $x$ is not too small.  

In the latter case, we see the presence of a non-decoupling effect in
the low-energy effective field theory of the $\mathcal{O}(v)$ degrees
of freedom as $M\to\infty$.  We identify this as a hard SUSY-breaking
effect described by the dimension-4 operator given by \eq{e5}.
Ultimately, this non-decoupling effect can be traced to the fact that
although $\nu_L$~[$\nu_L^c$] and $\snu_L$ 
[$\snu_R^*$] are superpartners, it is not quite true that
$\nu_\ell$ [$ \nu_h$] and $\snu_\ell$ [$\snu_h$]
are superpartners.  Explicitly [cf.~\eqs{nulapp}{nurapp}], whereas 
\beq
\nu_h^c\simeq \nu^c_L+M^{-1}m_D^T\nu_L\,,
\eeq
to leading order in $vM^{-1}$, we have:
\beq
\snu_h^*\simeq \snu_R^*+(M^2+m_N^2)^{-1}Mm_D^T\snu_L\,.
\eeq
Clearly, with $m_N^2\neq 0$, there is a slight discrepancy between
$\snu_h$ and the superpartner of $\nu_h$.
  
If we replace $H_2^2$ with its vacuum expectation value 
$v_2/\sqrt{2}$ in \eq{e4}
and again make use of  $\wt L^J_1\simeq\snu^J_\ell
+\mathcal{O}(vM^{-1})$, we obtain a contribution to $M_{LC}^2$:
Then \eq{e4} becomes:
\beq \label{e6}
[m_D^*m_D^T-m_D^* M (M^2+m_N^2)^{-1}Mm_D^T]^{JK}  
\snu^{J\,*}_\ell \snu_\ell^{K}\,,
\eeq
which correctly reproduces the last two terms of $M^2_{LC}$ given in
\eq{mlc}. 
Of course, the non-seesaw MSSM result of  $M^2_{LC}$
derives from the soft-SUSY-breaking term,
$\wt L_i^* m_L^2 \wt L_i$, and the $D$-term contribution,
$\half M_Z^2\cos 2\beta$.  As expected, in the $M\to\infty$ limit
(with $x\to 0$), the low-energy effective theory reproduces the 
non-seesaw MSSM result.  In this appendix, we have explained the origin of
the non-decoupling correction to the non-seesaw MSSM result in the
$M\to\infty$ limit with $x$ held fixed to a finite positive value.

Finally, we address the question of the allowed size of the matrix
parameter $m_N^2$.  Does it make sense to have $x$ close to
$\mathcal{O}(1)$?  In \Ref{cao}, it is
shown that for values of $x\sim 1$, there is a very large negative
shift in the mass of the lightest CP-even Higgs boson due to radiative
corrections from the heavy neutrino/sneutrino sector of the
seesaw-extended MSSM.  If we demand that there should be no unusually
large radiative correction to a \textit{physical observable} generated as a
result of $m_N\neq 0$, we can apply the 
results of \Ref{cao} for the radiatively-corrected physical 
Higgs masses to conclude
that $x\lsim 0.1$.  Note that this upper bound is less severe
than the bound of $x\lsim 0.01$ given in \eq{xdef}.  The latter was
obtained in \sect{sec:llg} from the bounds on rare charged lepton
radiative decay rates, which imply that the matrix $M_{LC}^2$ should
be close in form to a diagonal matrix. 

\medskip
\noindent
{\bf B.2~~Naturalness constraints on the magnitude of $\boldsymbol{m_N^2}$}

It seems that phenomenological constraints allow for the
possibility that $\|m_N^2\|$ is significantly larger than
$\mathcal{O}(v^2)$, in which case the non-decoupling contribution to
$M_{LC}^2$ may be significant (perhaps as large as a few percent of the
non-seesaw MSSM result).  However, if one imposes the usual
fine-tuning (or naturalness) 
requirements for the stability of the electroweak scale,
one can show that $\|m_N^2\|$ cannot be significantly larger than
$\mathcal{O}(v^2)$.  This can be verified by computing
the one-loop correction to the $H_2^2$ self-energy.  The computation
in the supersymmetric limit is performed explicitly in Appendix E,
section 7 of ref.~\cite{haberkane} for the Wess-Zumino model.  
This computation is easily adapted to the present case of interest 
(in which the
Higgs boson couples the the neutrino/sneutrino system).  We then
modify the supersymmetric computation in the case of 
the one-generation seesaw model by setting the boson (heavy sneutrino)
squared-mass to $M^2+m_N^2$ and the fermion (heavy neutrino) 
mass to $M$.  [Here, we are dropping terms of $\mathcal{O}(v^2)$.]
If $m_N^2\neq 0$ (which softly breaks the supersymmetry),
the quadratic divergence does not cancel exactly.  The
surviving contribution to the sqaured-mass term of $H_2^2$ is of the form
\beq \label{survive}
m_N^2 |Y_\nu|^2 \mathcal{I}(M^2,m_N^2)|H_2^2|^2\,,
\eeq
where $\mathcal{I}$ is a logarithmically divergent integral (that can
be regularized by dimensional reduction~\cite{dred}). 

We now add this one-loop result to the corresponding tree-level
contribution to the scalar potential:
\beq
(m_{H_2}^2+|\mu|^2)|H_2^2|^2\,.
\eeq
In order to achieve successful electroweak symmetry breaking with 
$v=246$~GeV, the complete coefficient multiplying $|H_2^2|^2$
must be of $\mathcal{O}(v^2)$.  
By assumption, we take $\mu\sim\mathcal{O}(v)$ [cf.~\eq{assumemu}].
If $m_N^2\gg v^2$, the correct scale of electroweak symmetry breaking can be
achieved only by an unnatural fine-tuning of the parameter
$m_{H_2}^2$.  Thus, naturalness requires that $m_N^2\sim v^2$.
We have not distinguished between $\mathcal{O}(v^2)$ and
$\mathcal{O}(M^2_{\rm SUSY})$ in the above discussion.  It is likely
that there is a slight separation of scales with $M_{\rm SUSY}\lsim
1~{\rm TeV}$.  By imposing the 
naturalness condition on the dynamics of electroweak symmetry
breaking (which ultimately is the motivation for TeV-scale
supersymmetry in the first place), we conclude that the expected natural
order of magnitude for $\|m_N^2\|$ is:
\beq
\|m_N^2\|\sim\mathcal{O}(M_{\rm SUSY})\,,
\eeq
as indicated by \eq{assume5}.

For completeness, we note that the same conclusion can be drawn by
considering the one-loop effective scalar potential, $V^{(1)}(\phi)$.
In particular, if we introduce a
hard momentum cutoff $\Lambda$, one obtains 
a one-loop contribution of~\cite{ewsbreview}
\beq \label{v1}
V^{(1)}(\phi) = \frac{\Lambda^2}{32\pi^2}\,\sum_i\, {\rm Str}\, M^2_i(\phi)
+ {1\over 64\pi^2}\,{\rm Str}\,\left\{ M^4_i (\phi)\left[
  \ln \frac{M^2_i(\phi)}{\Lambda^2} -\frac{1}{2}\right] \right\}\,,
\eeq
where
$M^2_i(\phi)$ are the contributing squared-mass matrices of particles
whose masses originate from their couplings to the Higgs boson,
with the vacuum expectation values 
replaced by the corresponding Higgs fields, $\phi$,
and
\beq \label{str}
  {\rm Str}\ \{\cdots\} = \sum_i(-1)^{2J_i} (2J_i+1) C_i\ \{\cdots\}\,.
\eeq
In \eq{str},
$C_i$ counts the electric charge and color degrees of freedom of
particle~$i$ (e.g., $C=2$ for the $W^\pm$ gauge boson and $C=6$ for a
colored quark, since we count both particle and antiparticle).
It is convenient to absorb the factor of $1/2$ in the last term on the
right hand side of \eq{v1}, by defining $\mu$ such that:
\beq \label{mudef}
\ln\frac{M_i^2(\phi)}{\Lambda^2}-\frac{1}{2}
\equiv \ln\frac{M_i^2(\phi)}{\mu^2}\,.
\eeq
Using the results of \eqss{heavynumass}{heavysnumatrix}{m2Hdef},
we focus on the contributions
to the supertraces from the heavy neutrinos and sneutrinos.  Indeed,
\beq
\sum_i {\rm Str}\, M^2_i(\phi)=2\,{\rm Tr}~m_N^2+\mathcal{O}(v^2)\,,
\eeq
although $m_N^2$ is field independent and thus contributes only to the
vacuum energy.  Here, we are interested in the implications of
naturalness associated with electroweak symmetry breaking (and not the
cosmological constant).  Thus we focus on the field-dependent part of
the scalar potential that is quadratic in the Higgs fields.  To do
this, we simply replace $m_D$ with $H_2^2 Y_\nu$.  For simplicity,
we shall examine the one generation seesaw model.  In this case, we
obtain the following scalar field-dependent squared-masses:
\bea
m^2_{\nu_h}&\simeq &M^2+2|Y_\nu|^2 |H_2^2|^2\,,\\
m^2_{\snu_h} &\simeq & M^2+m_N^2
+|Y_\nu|^2 |H_2^2|^2\left[1+\frac{M^2}{M^2+m_N^2}\right]\,.
\eea
Inserting these results into the last term on the right hand side of
\eq{v1}, and using \eq{mudef} to replace $\Lambda$ with $\mu$, we end
up with the following terms in $V^{(1)}(\phi)$ that contribute
to the coefficient of $|H_2^2|^2$
\bea
&& \hspace{-0.2in}
2\left\{(M^2+m_N^2)^2+2(2M^2+m_N^2)|Y_\nu|^2|H_2^2|^2\right\}
\ln\left[\frac{M^2+m_N^2+|Y_\nu|^2
    |H_2^2|^2\left(\frac{2M^2+m_N^2}{M^2+m_N^2}\right)}{\mu^2}\right]
\nonumber \\
&& \qquad -2\left\{(M^4+4M^2|Y_\nu|^2|H_2^2|^2\right\}
\ln\left[\frac{\displaystyle M^2+2|Y_\nu|^2
    |H_2^2|^2}{\mu^2}\right]\,,
\eea
where we have dropped terms of $\mathcal{O}(v^2|H_2^2|^2)$.
Expanding out the logarithms, the above expression reduces to 
\bea
&& \hspace{-0.2in}
 2\left\{(M^2+m_N^2)^2+2(2M^2+m_N^2)|Y_\nu|^2|H_2^2|^2\right\}
\left\{\ln\left[\frac{M^2+m_N^2}{\mu^2}\right]+|Y_\nu|^2|H_2^2|^2
\frac{2M^2+m_N^2}{(M^2+m_N^2)^2}\right\} \nonumber \\
&&
\qquad
-2\left\{(M^4+4M^2|Y_\nu|^2|H_2^2|^2\right\}
\left\{\ln\frac{M^2}{\mu^2}+\frac{2|Y_\nu|^2|H_2^2|^2}
{M^2}\right\}\,. 
\eea
If we keep only terms proportional to $|H_2^2|^2$, we end up with:
\beq
4|Y_\nu|^2|H_2^2|^2\left\{2M^2\ln\left(1+\frac{m_N^2}{M^2}\right)
+m_N^2\left[\ln\left(\frac{M^2+m_N^2}{\mu^2}\right)+
\frac{1}{2}\right]+\mathcal{O}(v^2)\right\}\,.
\eeq
One can check that the coefficient of $|Y_\nu|^2|H_2^2|^2$
is precisely $m_N^2\mathcal{I}(M^2,m_N^2)$, where $\mathcal I$ is
the integral appearing in \eq{survive} after $\overline{\rm DR}$
subtraction~\cite{dred}.

\medskip

\setcounter{equation}{0}  
\section{Feynman rules}
\label{app:feyrul}

We exhibit here the relevant Feynman rules for the calculation of
$\ell \rightarrow \ell^{\,\prime} \gamma$ presented in
Section~\ref{sec:llg}.  These rules are based on four-component
fermion notation (see~\ref{app:fermion}) and employ the conventions of
Ref.~\cite{Rosiek} for sfermion, chargino and neutralino masses and
mixing matrices.  The neutrinos $\nu^I$ are (self-conjugate) Majorana
fermions [cf. \eq{lightmaj}].  In the basis defined in
Section~\ref{sec:lagr} we obtain:

\vspace*{0.5cm}

\noindent
\begin{tabular}{ll}
\begin{picture}(150,70)(0,-10)
\DashLine(60,0)(10,0){5}
\Text(0,0)[c]{$S_k$}
\ArrowLine(60,0)(110,0)
\Text(120,0)[c]{$\chi^0_i$}
\ArrowLine(60,50)(60,0)
\Text(65,45)[l]{$\nu^I$}
\Vertex(60,0){2}
\end{picture}
&
\raisebox{35\unitlength}{
\begin{minipage}{5cm}
\lefteqn{
\frac{i}{2} \left [ (g_1 Z_N^{1i}  - g_2 Z_N^{2i} ) (\znu^{Jk}
- i \znu^{(J+3)k} )U_{MNS}^{JI} \: P_L \right.}
\vspace*{0.4cm}\hspace*{0.2cm}
\lefteqn{
+ \left .(g_1 Z_N^{1i*}  - g_2 Z_N^{2i*} ) (\znu^{Jk} + i
\znu^{(J+3)k} ) U_{MNS}^{JI*} \: P_R \right ]
\;,}\begin{equation}\label{B1}\end{equation}
\end{minipage}
}
\end{tabular}

\vspace*{1cm}

\noindent
\begin{tabular}{ll}
\begin{picture}(150,70)(0,-10)
\DashArrowLine(10,0)(60,0){5}
\Text(0,0)[c]{$L_k^+$}
\ArrowLine(60,0)(110,0)
\Text(120,0)[c]{$\chi_i$}
\ArrowLine(60,50)(60,0)
\Text(65,45)[l]{$\nu^I$}
\Vertex(60,0){2}
\end{picture}
&
\raisebox{35\unitlength}{
\begin{minipage}{5cm}
\lefteqn{
- i\left( g_2 Z_L^{Jk} Z_-^{1i} - Y_\ell^{\,J} Z_L^{(J+3)k}
Z_-^{2i} \right) U_{\rm MNS}^{JI} \: P_L
 \;, }\begin{equation}\label{B2}\end{equation}
\end{minipage}
}
\end{tabular}


\noindent
\begin{tabular}{ll}
\begin{picture}(150,70)(0,-10)
\DashArrowLine(10,0)(60,0){5}
\Text(0,0)[c]{$L_k^+$}
\ArrowLine(60,0)(110,0)
\Text(120,0)[c]{$\chi^0_i$}
\ArrowLine(60,50)(60,0)
\Text(65,45)[l]{$\ell^I$}
\Vertex(60,0){2}
\end{picture}
&
\raisebox{15\unitlength}{
\begin{minipage}{5cm}
\lefteqn{
i \left[ \left( \frac{g_2}{\sqrt{2}c_W} Z_L^{Ik} (Z_N^{1i} s_W +
Z_N^{2i} c_W) - Y_\ell^I Z_L^{(I+3)k} Z_N^{3i} \right) P_L \right.
}
\vspace*{0.4cm}
\lefteqn{
+ \left.\left( -g_1 \sqrt{2}\,  Z_L^{(I+3)k} Z_N^{1i*} -
Y_\ell^I Z_{L}^{Ik} Z_{N}^{3i*} \right) P_R \right]
\;,}\begin{equation}\label{B3}\end{equation}
\end{minipage}
}
\end{tabular}

\vspace*{1cm}

\noindent
\begin{tabular}{ll}
\begin{picture}(150,70)(0,-10)
\DashLine(60,0)(10,0){5}
\Text(0,0)[c]{$S_k$}
\ArrowLine(60,0)(110,0)
\Text(120,0)[c]{$\chi_i^C$}
\ArrowLine(60,50)(60,0)
\Text(65,45)[l]{$\ell^I$}
\Vertex(60,0){2}
\end{picture}
&
\raisebox{35\unitlength}{
\begin{minipage}{5cm}
\lefteqn{
-\frac{i}{\sqrt{2}} \left [  g_2 Z_+^{1i} (\znu^{Ik}
- i \znu^{(I+3)k} ) \: P_L
-Y_\ell^I Z_-^{2i*} (\znu^{Ik} - i \znu^{(I+3)k} ) \: P_R  \right ]
\;.  }\begin{equation}\label{B4}\end{equation}
\end{minipage}
}
\end{tabular}




\setcounter{equation}{0}  

\bigskip

\section{Order of magnitude estimates for contributions to
one-loop neutrino masses}
\label{app:oneloopnu}

In this appendix, we estimate the order of magnitude of the one-loop
contributions to the neutrino masses due to the graphs
of~\fig{fignu}(a) and (b), and the corresponding graphs (not shown) in
which the light sneutrinos [heavy neutrinos] in graph (a) [(b)] are
replaced by heavy sneutrinos [light neutrinos].

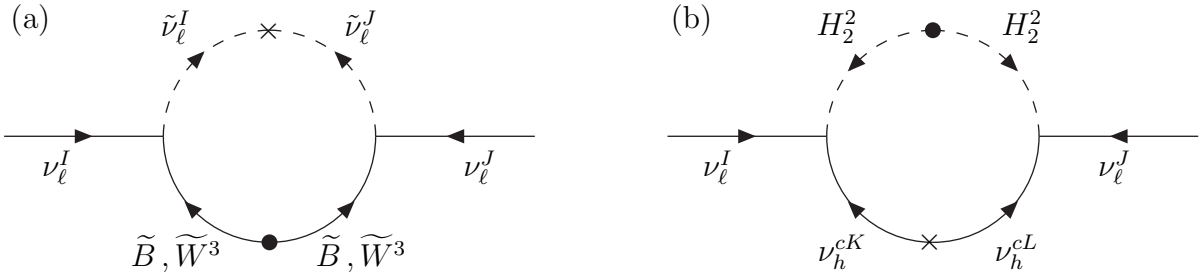
\begin{figure}[t]
\begin{center}
\begin{picture}(200,110)(120,0)
\Text(0,100)[t]{(a)}
\Text(90,90)[c]{$\times$}
\ArrowLine(-10,50)(50,50)
\Text(10,45)[t]{$\nu_\ell^I$}
\ArrowLine(190,50)(130,50)
\Text(170,45)[t]{$\nu_\ell^J$}
\DashArrowArc(90,50)(40,0,90){5}
\DashArrowArcn(90,50)(40,180,90){5}
\Text(55,98)[t]{$\tilde\nu_\ell^I$}
\Text(125,98)[t]{$\tilde\nu_\ell^J$}
\ArrowArcn(90,50)(40,-90,-180)
\ArrowArc(90,50)(40,-90,0)
\Vertex(90,10){3}
\Text(55,13)[t]{$\widetilde B$\,,\,$\widetilde W^3$}
\Text(125,13)[t]{$\widetilde B$\,,\,$\widetilde W^3$}
\Text(250,100)[t]{(b)}
\ArrowLine(240,50)(300,50)
\Text(260,45)[t]{$\nu_\ell^I$}
\ArrowLine(440,50)(380,50)
\Text(410,45)[t]{$\nu_\ell^J$}
\DashArrowArcn(340,50)(40,90,0){5}
\DashArrowArc(340,50)(40,90,180){5}
\Text(305,98)[t]{$H_2^2$}
\Text(375,98)[t]{$H_2^2$}
\ArrowArcn(340,50)(40,-90,-180)
\ArrowArc(340,50)(40,-90,0)
\Vertex(340,90){3}
\Text(340,10)[c]{$\times$}
\Text(307,13)[t]{$\nu_{h}^{cK}$}
\Text(373,13)[t]{$\nu_{h}^{cL}$}
\end{picture}
\end{center}\caption{One-loop corrections to light neutrino masses.
The $\times$ marks the location of the $\Delta L=2$ transition.
(a)~The loop consisting of light sneutrinos and gauginos.  The
$\times$ indicates the location of light sneutrino--antisneutrino
mixing, and the solid dot indicates a factor of the gaugino Majorana
mass in the numerator of the fermion-number-violating gaugino
propagator.  (b)~The loop consisting of the neutral Higgs field
$H_2^2$ and a heavy neutrino.  The $\times$ indicates the
lepton-number-violating heavy neutrino propagator, which is
proportional to $M\delta^{KL}$, and the solid dot indicates a mass
insertion of the form $(H_2^{2\,*})^2$.  The contributions of the
corresponding graphs (not shown) in which the gauginos in (a) are
replaced by the Higgsino $\widetilde H_2^2$, the light sneutrinos in
(a) are replaced by heavy sneutrinos, and the heavy neutrinos in (b)
are replaced by light neutrinos are all suppressed by an additional
powers of $\mathcal{O}(v M^{-1})$ as explained in the text.}
\label{app:fignu}
\end{figure}

In the case of graph (a), the dominant contribution involves the light
sneutrino--neutrino--gaugino interaction term\footnote{Of the three
light sneutrino-neutrino-neutralino interactions of \eq{eq:vlag}, the
two sneutrino-neutrino-higgsino interaction terms are suppressed by a
factor of $\mathcal{O}(m_D M^{-1})$ relative to the
sneutrino-neutrino-gaugino interaction, and can be neglected.}
of~\eq{eq:vlag}.  We can estimate the leading contribution of this
graph by replacing the internal lines by the interaction eigenstate
fields that appear in \eq{eq:vlag}, as depicted in~\fig{app:fignu}.
That is, we first replace the $S_k$ with the $\widetilde\nu_\ell^I$,
which must point \textit{away from both} external vertices, as shown
in \fig{app:fignu}(a).  The latter is possible only in the presence of
light sneutrino--antisneutrino mixing, which is indicated by the
$\times$ in \fig{app:fignu}(a).  Using the expected magnitudes of the
model parameters given by~\eqs{assume3}{assume4}, the $\times$ in
\fig{app:fignu}(a) produces a factor $\Delta m^2_{\tilde\nu_\ell}\sim
\mathcal{O}(v^3 M^{-1})$.  The neutralino line can be treated
perturbatively.  In the lowest order approximation, we take the
neutralino to be a gaugino (either $\widetilde B$ or $\widetilde W^3$,
with Majorana masses $M_1$ and $M_2$, respectively), and we treat the
mixing of the gauginos with the neutral higgsino states ($\widetilde
H_1^1$ and $\widetilde H_2^2$) as a perturbation.  The corresponding
gaugino propagators (with internal four-momentum $q$) shown in
\fig{app:fignu}(a) are fermion-number-violating propagators (indicated
by the clashing arrows), and are given by $iM_k/(q^2-M_k^2)$ for
$k=1,2$.  We denote the presence of the gaugino mass [which is of
$\mathcal{O}(v)$] in the numerator by the solid dot in
\fig{app:fignu}(a).  Not including this explicit factor of the gaugino
mass, the loop in graph (a) then consists of two massive scalar
propagators [with mass of $\mathcal{O}(v)$] and one
fermion-number-violating propagator; hence the loop integral has a
mass dimension of $-2$.  Thus, the corresponding loop integral is of
$\mathcal{O}(v^{-2})$.  Combining the above results, the order of
magnitude of the contribution of graph~(a) is:
\beq
C_L\frac{v^3}{M}\cdot \frac{1}{v^2}\cdot v=C_L\frac{v^2}{M}\,,
\eeq
which is indeed of order the tree-level neutrino mass multiplied by
the product of the relevant vertex coupling constants and a typical
loop factor of $1/16\pi^2$ (denoted by $C_L$ above).

Suppose we replace the light sneutrinos of graph (a) with heavy
sneutrinos.  In this case, the effect of heavy
sneutrino--antisneutrino mixing is $\Delta m^2_{\tilde\nu_h}\sim
\mathcal{O}(m_B^2)\sim\mathcal{O}(vM)$.  {}From \eq{eq:vlag}, we see
that there are potentially two contributions---one involving the
gauginos and one involving the higgsino $\widetilde H_2^2$.  In the
case of the gaugino loop graph, each vertex introduces a
$\mathcal{O}(vM^{-1})$ suppression.  Thus, following the analysis
above, we conclude that the order of magnitude of the heavy-sneutrino
loop is suppressed by a factor of $\mathcal{O}(v^2 M^{-2})$ as
compared with the light-sneutrino loop.  In the case of the loop graph
involving $\widetilde H_2^2$, we note that there is no diagonal
Majorana mass term for this higgsino field.  Moreover, $\widetilde
H_1^1$ does not couple to the external neutrinos, so we cannot use the
off-diagonal Majorana mass term $\mu \widetilde H_1^1 \widetilde
H_2^2$ for the fermion-number-violating neutralino propagator.
Therefore, the heavy-sneutrino loop can be neglected.

In the case of graph (b), the propagator of the heavy neutrino (with
internal four-momentum $q$) is given by $iM\delta^{KL}/(q^2-M^2)$, due
to the presence of the lepton-number violating mass $M$ (indicated by
the $\times$).  Since the loop integral is dimensionless, it naively
appears that the resulting loop integral should be of
$\mathcal{O}(M)$.  However, an explicit computation of the graph of
\fig{fignu}(b) demonstrates that the coefficient of the leading
$\mathcal{O}(M)$ term vanishes exactly after summing over the internal
neutral Higgs and Goldstone states.  The subleading term does not
vanish and is of $\mathcal{O}(v^2 M^{-1})$, which is the magnitude of
the \textit{light} neutrino mass.  This cancellation can be easily
understood by noting that the two vertices of~\fig{fignu}(b) arise
from interactions of \eq{eq:vlag} that involve $H_2^2$.  Thus we
replace the neutral Higgs and Goldstone lines of \fig{fignu}(b) by the
$H_2^2$ field [cf. \eq{h22}].  According to the interaction Lagrangian
of \eq{eq:vlag}, the $H_2^2$ field must point \textit{into both}
external vertices, as shown in~\fig{app:fignu}(b).  This requires a
mass insertion on the $H_2^2$ line of the form $(H_2^2)^2+{\rm H.c.}$
In fact, such a term exists in the MSSM Higgs potential \cite{gunhab1}
after shifting the neutral field $H_2^2\to H_2^2+v_2/\sqrt{2}$, which
results in a term of the form $\frac{1}{4} m_Z^2\sin^2\beta
(H_2^2)^2+{\rm H.c.}$ Thus, in the mass insertion approximation, graph
(b) consists of the lepton-number-violating heavy neutrino propagator,
two massive scalar field lines\footnote{In the MSSM Higgs sector,
after shifting the neutral Higgs fields by their vacuum expectation
values and applying the potential minimum conditions, there is a mass
term of the form $(\half m_Z^2\sin^2\beta+m_A^2\cos^2\beta)|H_2^2|^2$,
where $m_A^2\equiv -m_{12}^2/\sin\beta\cos\beta$ [and $m_{12}^2$
defined in \eq{lsoft}].  In evaluating graph (b) of \fig{app:fignu},
we treat the $|H_2^2|^2$ mass term exactly, and incorporate the
$(H_2^2)^2+{\rm H.c.}$ and $H_1^1 H_2^2 +{\rm H.c.}$ mass terms
perturbatively (via the mass insertion approximation).}  and an
insertion of $\mathcal{O}(v^2)$.  After extracting the factor of $M$
from the numerator of the heavy neutrino propagator, the remaining
loop integral now has a mass dimension of $-2$, which yields a result
of $\mathcal{O}(M^{-2})$.  Combining these result, the order of
magnitude of the contribution of \fig{app:fignu}(b) is given by:
\beq
C^{\,\prime}_L\frac{1}{M^2}\cdot M\cdot v^2 =
C^{\,\prime}_L\frac{v^2}{M}\,,
\eeq
which is again of order the tree-level neutrino mass multiplied by the
product of the relevant vertex coupling constants and a typical loop
factor (denoted above by $C^{\,\prime}_L$).  This result confirms our
previous argument above.  A careful evaluation of the leading behavior
of the loop integral (in the limit of $M\gg v$) then reproduces the
result obtained in \eq{eq:higgs}.  Note that the factor of $\sin^2
\beta\equiv v_2^2/v^2$ that arises in the mass insertion on the
$H^2_2$ line cancels out a similar factor of $v_2^2$ that appears in
$C^{\,\prime}_L\propto Y_\nu^2$.

If the heavy neutrinos in~\fig{app:fignu}(b) are replaced by light
neutrinos, the resulting contribution is suppressed by an additional
factor of $\mathcal{O}(v^2 M^{-2})$ due to the suppression of the
$\nu_\ell^I\nu_\ell^K H_2^2$ interaction of \eq{eq:vlag}.

\end{document}